\documentclass[12pt]{article}

\usepackage{xr-hyper}
\usepackage{hyperref}
\usepackage{makecell}
\usepackage{verbatim,color,amssymb,epsfig}
\usepackage{graphicx,setspace,float,rotating,longtable,amsmath,amssymb,epsfig,color,lscape,subfigure,url}
\usepackage{epstopdf,multirow,lscape,comment}
\usepackage{multirow}

\usepackage{tabularx}
\usepackage{minitoc}
\usepackage{caption}
\usepackage{booktabs}
\usepackage{array,mathabx}
\usepackage[normalem]{ulem}
\usepackage[percent]{overpic}
\usepackage{pdfpages}
\usepackage{colortbl}
\usepackage{tikz}
\usepackage{placeins}

\usepackage{url}
\usepackage{booktabs}
\usepackage{amsthm}
\usepackage{setspace}
\usepackage[table]{xcolor}
\usepackage{algorithm}
\usepackage{algpseudocode}
\usepackage{subcaption} 
\usepackage{fancybox}

\usepackage{xr}
\usepackage{enumitem}
\setlist{  
  listparindent=0pt,
  parsep=0.1in,
}

\renewcommand{\baselinestretch}{1}

\definecolor{babyblue}{rgb}{0.63, 0.79, 0.95}

\newcolumntype{H}{>{\setbox0=\hbox\bgroup}c<{\egroup}@{}}

\usepackage[normalem]{ulem}

\def\boxit#1{\vbox{\hrule\hbox{\vrule\kern6pt
          \vbox{\kern6pt#1\kern6pt}\kern6pt\vrule}\hrule}}

\def\sumi{\sum_{i=1}^n}
\def\diag{\hbox{diag}}
\def\wh{\widehat}

\def\diag{\hbox{diag}}
\def\log{\hbox{log}}

\def\var{\hbox{var}}

\def\Normal{\hbox{Normal}}

\def\bse{\begin{eqnarray*}}
\def\ese{\end{eqnarray*}}
\def\be{\begin{eqnarray}}
\def\ee{\end{eqnarray}}
\def\bq{\begin{equation}}
\def\eq{\end{equation}}
\def\bse{\begin{eqnarray*}}
\def\ese{\end{eqnarray*}}

\def\wh{\widehat}

\def\b1e{{\mathbf e}}

\def\bO{{\mathbf O}}

\def\bS{{\mathbf S}}

\def\bzero{{\mathbf 0}}

\def\bPhi{{\boldsymbol{\Phi}}}

\def\btheta{{\boldsymbol{\theta}}}

\def\bxi{{\boldsymbol{\xi}}}

\def\btau{{\boldsymbol{\tau}}}

\def\bgamma{{\boldsymbol{\gamma}}}

\newcommand{\balpha}{\mbox{\boldmath $\alpha$}}

\def\bA{{\mathbf A}}

\def\bB{{\mathbf B}}

\def\bB{{\mathbf B}}

\def\bS{{\mathbf S}}

\def\bz{{\mathbf z}}
\def\bZ{{\bf Z}}

\def\bW{{W}}

\def\bZ{{\mathbf Z}}
\def\bzero{{\mathbf 0}}
\def\0{{\mathbf 0}}

\def\Normal{\hbox{Normal}}

\def\Bernoulli{\hbox{Bernoulli}}

\usepackage{xr-hyper}
\usepackage{makecell}

\usepackage{hyperref}
\hypersetup{
    colorlinks=true,
    allcolors  =blue,
    pdftitle={Overleaf Example},
    pdfpagemode=FullScreen,
    }

\newcommand{\ny}[1]{\cellcolor{gray!20}{#1}} 
\def\expit{\text{expit}}

\newtheorem{Prop}{Proposition}

\usepackage[
  backend=bibtex,
  citestyle=ext-authoryear,
  bibstyle=ext-authoryear,
  giveninits=true,
  maxnames = 5, 
  articlein=false,
  maxcitenames=2,
	uniquename=init,   
  natbib=true
]{biblatex}

\usepackage{setspace}

\makeatletter
\input{ext-authoryear.bbx}
\makeatother

\DeclareNameAlias{sortname}{family-given}

\DeclareNameAlias{author}{sortname}
\DeclareNameAlias{editor}{sortname}
\DeclareNameAlias{translator}{sortname}

\DeclareFieldFormat[article,periodical]{volume}{\mkbibbold{#1}}
\DeclareFieldFormat[article,periodical]{number}{\mkbibparens{#1}}
\DeclareFieldFormat[article,periodical]{pages}{#1}

\DeclareSourcemap{
  \maps{
    \map{
      \pertype{article}
      \pertype{periodcial}
      \step[fieldset=number, null]
    }
  }
}

\renewbibmacro{volume+number+eid}{%
    \printfield{volume}%
        \printfield{number}%
    \setunit{\addcomma\space}%
    \printfield{eid}}
    
\def\spacingset#1{\renewcommand{\baselinestretch}%
{#1}\small\normalsize} \spacingset{1}

\makeatletter
\newcommand*{\addFileDependency}[1]{
\typeout{(#1)}
\@addtofilelist{#1}
\IfFileExists{#1}{}{\typeout{No file #1.}}
}\makeatother

\def\boxit#1{\vbox{\hrule\hbox{\vrule\kern6pt
          \vbox{\kern6pt#1\kern6pt}\kern6pt\vrule}\hrule}}

\begin{document}

\pagenumbering{arabic}
\baselineskip=28pt

\begin{center}
{\LARGE{\bf Federated Learning with Incomplete Data: When to Use Complete Cases and When to Weight}}
\end{center}

\baselineskip=12pt

\vskip 2mm
\begin{center}
Jesus E. Vazquez$^{*1}$,
Yicheng Shen$^{1}$, 
Jason Akulian$^2$,
Chad Hochberg$^3$,
Theodore J. Iwashyna$^3$,
Elizabeth A. Stuart$^{1}$, 
Jiayi Tong$^{1}$ \\ \vspace{0.2in}
$^1$Department of Biostatistics, Johns Hopkins University, Baltimore, MD 21231\\
$^2$Division of Pulmonary and Critical Care, University of North Carolina, \\ Chapel Hill, NC 27516 \\
$^3$Pulmonary and Critical Care Medicine, Johns Hopkins University, Baltimore, MD 21231\\
\vspace{1em}
*jvazqu18@jh.edu
\end{center}

\begin{center}
{\Large{\bf Summary}}
\end{center}
\baselineskip=12pt


Privacy constraints have driven the rise of federated learning (FL), which enables multi-site analyses without sharing individual participant data. Existing FL estimators largely assume complete data, whereas multi-site studies often face missingness. We develop a framework for FL with missing data, identifying conditions under which the complete case (CC) estimator is preferred over the inverse probability weighting (IPW) estimator. For settings where the CC estimator leads to bias, we introduce a calibrated weight estimation approach that combines candidate weighting models across sites and remains consistent if at least one is correctly specified at each site; we further show that pooling many weighting candidate models with redundant information degrades the calibrated estimator, so a small set is preferable. Consistency conditions are stated at the site level, ensuring that the federated estimator inherits validity from site-level properties. We prove consistency and derive a sandwich variance estimator that accounts for uncertainty in the outcome model, and in both the estimated weighting models and the calibration step. Additionally, we show that all estimators require only one or a few communication rounds, making them practical under real-world data-governance constraints. We illustrate the framework by evaluating risk factors for 90-day mortality among patients with pleural infections treated with intrapleural enzyme therapy.

\baselineskip=12pt
\par\vfill\noindent
\underline{\bf Some Key Words}: Federated Learning; Missing Data; Site Heterogeneity; Inverse Probability Weighting; Calibrated weights; 
$M$-estimator
\par\medskip\noindent
\underline{\bf Short title}: Federated Learning With Incomplete Data
\clearpage\pagebreak\newpage

\newlength{\gnat}
\setlength{\gnat}{22pt}
\baselineskip=\gnat

\clearpage\newpage
\spacingset{2}

\section{Introduction}
\label{sec:fl-intro}

Federated learning (FL) enables multi-site collaboration without pooling individual participant data (IPD), addressing privacy regulations and institutional policies that increasingly prohibit data sharing \citep{malpetti2025technical}. A growing body of work develops FL methods for regression estimation that address three key constraints: (i) \emph{privacy}---only summary statistics may leave a site \citep{xu2021federated}, (ii) \emph{communication efficiency}---one or few exchanges of data or information between sites, and (iii) \emph{heterogeneity}---allowing site-specific covariate distributions \citep{jordan2019communication, duan2021leverage, li2024developing}. However, most methods assume complete data---an assumption rarely met in practice \citep{li2023federated}.

Recent work has begun adapting missing data approaches to FL settings, primarily through imputation-based approaches: e.g., the communication-efficient surrogate likelihood framework \citep{jordan2019communication}, distributed multiple imputation \citep{chang2020federated}, and the FedIMPUTE family \citep{li2025fedimpute}. These approaches typically assume data are missing at random (MAR), with limited approaches for missing not at random (MNAR) settings \citep{galimard2018heckman}. While flexible, imputation approaches are often computationally intensive, with multiple imputation approaches at times requiring \textit{thousands} of communication rounds \citep{chang2020federated}, which may limit practicality in real-world settings.

The inverse probability weighting (IPW) estimator offers an alternative by up-weighting complete observations rather than imputing missing values, creating a pseudo-sample that approximates the full data-generating process \citep{robins1994estimation}. While the IPW estimator has been extended to federated causal inference for confounding adjustment in fully observed data \citep{yin2025federated}, weighted approaches for missing data in FL remain underdeveloped. Key challenges in implementing the IPW estimator in FL include: (i) how to leverage information across the network to estimate the weights when IPD cannot be pooled, and (ii) how to properly quantify uncertainty when the weighting model---a nuisance component required for estimation---is fitted across the network.

In this paper, we develop a framework for federated estimation with incomplete data. Our approach is organized around a decision-oriented question: given what is missing and what the missingness depends on at each site, what is the simplest valid estimator? We make three contributions. First, we characterize when the CC estimator is consistent in FL, establishing site-level conditions for consistency of the federated estimator and enumerating the combinations of missing variables and missingness mechanisms under which they hold, thereby extending the work of \cite{hughes2019accounting} and \cite{little2024comparison}. Second, for settings in which the CC estimator is biased, we extend the IPW estimation framework to FL through two strategies: \emph{site-specific} weighting, in which each site estimates its weighting model independently, and \emph{calibrated} weighting, in which sites share estimated weighting model parameters and combine candidate model predictions via calibration. We show that the calibrated estimator is consistent provided at least one candidate model is correctly specified at each site. Third, we derive distinct sandwich variance estimators for both IPW strategies that account for the missingness together with the uncertainty in the estimated weights, including the parameters of the calibration step, and we show that ignoring such uncertainty yields invalid inference even when the point estimates are unbiased. All proposed estimators are one-shot (a single round of communication) or few-shot, making them practical under real-world data-governance constraints.

We organized the paper as follows. Section~\ref{sec:fl-notation-assumptions} introduces notation  and Section~\ref{sec:fl-estimators} develops the estimators: we begin with the CC estimator and establish when it is valid, then present federated IPW under site-specific and calibrated weight estimation, and provide the form of the sandwich variance estimator. Section~\ref{sec:fl-simulation} evaluates performance through simulations. Section~\ref{sec:application} applies the estimators to a multi-site study of risk factors for 90-day mortality among patients with pleural infections treated with intrapleural enzyme therapy. Section~\ref{sec:fl-discussion} concludes with a discussion of practical recommendations and future directions.

\section{Notation and setup}
\label{sec:fl-notation-assumptions}

Let $\btheta$ denote the parameters of the pooled data outcome model
\be
\label{eqn:outcome-model}
E_{Y\mid X, \bZ}\{ g(Y) \} = m(X,\bZ; \btheta),
\ee
where $Y$ is the outcome of interest, $g(\cdot)$ is a known link function (e.g., identity link), $m(\cdot)$ is a mean function of the covariates $(X,\bZ)$ and $\btheta$, and is linear in $\btheta$. Capital letters denote random variables and lowercase letters denote their realizations; $f_{Y|X,\bZ}(y_i,x_i,\bz_i;\btheta)$ is the conditional density function of $Y$ given $(X,\bZ)$, evaluated at $(y_i,x_i,\bz_i)$, and indexed by $\btheta$. 

Let $R=1$ when $(Y, X)$ is fully observed and $R=0$ otherwise. Each site $k$ contributes $n^{(k)}$ observations, with $n = \sum_{k=1}^K n^{(k)}$ total. The observed data at site $k$ is denoted $\bO^{(k)}$; for example, when $X$ is subject to missingness, $\bO^{(k)} = (Y^{(k)}, X_R^{(k)}, R^{(k)}, \bZ^{(k)})$ where $X_R^{(k)} = \diag\{R^{(k)}\}X^{(k)}$. While we lead with the missing covariate problem, the methodology extends to settings where $Y$ or $(Y,X)$ are subject to missingness by redefining $R$ and $\bO^{(k)}$. Our goal is to estimate $\btheta$ using the $K$ partitioned incomplete datasets $\bO^{(k)}$ while preserving privacy and communication efficiency. We establish consistency and robustness properties via $M$-estimation theory, showing that the expectation of the estimating equation equals zero under the stated conditions \citep{tsiatis2006semiparametric}. Here, robustness refers to the ability of the estimator to remain consistent under misspecification (or absence) of the weighting model.

\section{Estimation framework}
\label{sec:fl-estimators}
This section introduces the CC and IPW estimators within the $M$-estimation framework, each defined as the solution $\wh\btheta$ to estimating equations, and shows how these equations adapt to the federated setting with missing data. Because the federated estimating equation decomposes additively across sites, conditions that hold at each site are sufficient for consistency of the pooled estimator. We therefore organize the development around a single site-level criterion: does the oracle probability of a complete observation depend on $Y$? When it does not, the CC estimator is consistent; when it does, the IPW estimator provides a consistent alternative (Proposition~\ref{prop:fl-consistency}).

\subsection{Complete case estimator}
\label{sec:fl-cc}

The CC estimator restricts the analysis to complete cases at each site---a 
direct application of standard federated algorithms to the complete cases, 
requiring no modification to the underlying estimation machinery 
\citep{chang2020federated, wu2025cola}. We present it here to establish 
consistency and to make explicit that the count aggregation representation 
for generalized linear models (GLM) enables a one-shot lossless federated algorithm not previously 
stated in the missing data context. Formally, $\wh\btheta_\text{CC}$ solves
\be
\label{eqn:cc-ee}
\sum_{i=1}^n \bPhi_\text{CC}(\bO_i;\btheta) = \sum_{k=1}^K \sum_{i=1}^{n^{(k)}} r_i^{(k)} \bS_\btheta(y_i^{(k)}, r_i^{(k)} x_i^{(k)}, \bz_i^{(k)};\btheta) = \bzero,
\ee
where $\bS_{\btheta}(y_i, r_i x_i, \bz_i;\btheta) = \partial \log f_{Y|X,\bZ}(y_i, r_i x_i, \bz_i;\btheta) / \partial \btheta$ is the score function. The double summation partitions the estimating equation into $K$ site-specific contributions. Depending on the outcome model---and by extension the link function $g(\cdot)$---different estimation strategies yield a one-shot algorithm. Under a linear regression model with normally distributed errors with variance $\sigma^2$, the sufficient information (SI) approach enables one-shot estimation \citep{chang2020federated}. When $m(X,\bZ;\btheta) = (X,\bZ)\btheta$, the OLS solution requires each site to transmit the sufficient statistics $\{X_R^{(k)}, \bZ^{(k)}\}^\top \diag\{R^{(k)}\} \{X_R^{(k)}, \bZ^{(k)}\}$ and $\{X_R^{(k)}, \bZ^{(k)}\}^\top \diag\{R^{(k)}\} Y^{(k)}$. An estimate is then obtained by
\bse
\wh\btheta_\text{CC} 
&=& \left\{(X_R,\bZ)^\top \diag(R) (X_R,\bZ)\right\}^{-1} (X_R,\bZ)^\top \diag(R) Y  \\
&=& \left[\sum_{k=1}^K \{X_R^{(k)},\bZ^{(k)}\}^\top \diag\{R^{(k)}\} \{X_R^{(k)},\bZ^{(k)}\}\right]^{-1}
\left[\sum_{k=1}^K \{X_R^{(k)},\bZ^{(k)}\}^\top \diag\{R^{(k)}\} Y^{(k)}\right].
\ese
The first line is the standard OLS formula; the second decomposes it into site-specific matrix summands. To estimate the error variance $\sigma^2$, $\wh\btheta$ is shared back to the sites, which compute and return their sum of squared residuals; these are then aggregated at the coordinating site (Algorithm~\ref{alg:fl-cc}).

When a GLM is of interest, a one-shot lossless algorithm can also be 
applied \citep{wu2025cola}. Here we present a generalization of this 
algorithm that will later support the formulation of the IPW estimator. 
Among the complete data, for site $k$ with $n^{(k)}$ observations, suppose 
there is a finite number of $q << n^{(k)}$ distinct combinations such that 
$u_j = (y,x,\bz)$ with corresponding counts $n_j^{(k)}$. The site-level data 
can be represented as
\be
\label{eqn:cc-u-glm}
\bO^{(k)} = 
\begin{pmatrix}
u_1, & w_1^{(k)} = n_1^{(k)}  \\
\vdots & \vdots\\
u_q, & w_q^{(k)} = n_q^{(k)} \\
\end{pmatrix}.
\ee
Under \eqref{eqn:cc-u-glm}, the estimating equation \eqref{eqn:cc-ee} becomes
$\sum_{i=1}^n \bPhi_{\text{CC}}(\bO_i;\btheta,\balpha) =\sum_{k=1}^K \sum_{j=1}^q w_j^{(k)} \bS_\btheta(u_j;\btheta)$. Because this representation depends only on the distinct outcome-covariate combinations and their aggregated counts, each site can share these summaries once, enabling computation of $\wh{\btheta}$ in a one-shot, lossless, federated manner without transmitting IPD (Algorithm~\ref{alg:fl-cc-glm}). It is worth noting that each combination must comply with minimum cell size requirements to satisfy suppression policies which may differ by data privacy governing body \citep{malpetti2025technical}.

\subsubsection{Consistency conditions}
\label{sec:consistency-cc}

In the federated setting, the CC estimator in~\eqref{eqn:cc-ee} is a sum of $K$ site-specific estimating equations. A sufficient condition for consistency is that each site-specific contribution has expectation zero at the true parameter value $\btheta_0$. Proposition~\ref{prop:fl-consistency} (i) characterizes when this condition holds; a proof is provided in Supplementary Material (Section~\ref{sec:supp-proposition-proof}).

\begin{Prop}
\label{prop:fl-consistency}
Assume that $Y$, $X$, or $(Y,X)$ are subject to missingness but $\bZ$ is 
fully observed, and that standard regularity conditions hold 
(Supplementary Material, Section~\ref{sec:supp-proposition-proof}). Under the site-level conditions below, the federated estimator is consistent. 
\begin{enumerate}
    \item[(i)] The complete case estimator is consistent if, at each site, 
    the oracle probability of a complete observation does not depend on 
    the outcome $Y$. This condition may be satisfied by different 
    missingness mechanisms across sites---including MCAR, MAR, and certain 
    MNAR mechanisms---so long as it holds site-by-site 
    (Table~\ref{tab:missingness-mcar-mar-mnar}). 
    \item[(ii)] The inverse probability weighting estimator is consistent if,  at each site, the weighting model is correctly specified. When the oracle probability does not depend on $Y$, correct specification is not required, but $Y$ must be excluded from the weighting model when $X$ is missing to avoid bias.
    \item[(iii)] The calibrated inverse probability weighting estimator is consistent if, at each site, at least one of the $J$ candidate weighting models is correctly specified. The correctly specified model need not be the one estimated at that site.
\end{enumerate}
\end{Prop}

\begin{table}[h]
\centering
\resizebox{0.8\textwidth}{!}{%
\begin{tabular}{lccc}
\toprule
& $Y$ & $X$ & $(Y,X)$ \\
\midrule
MCAR & \ny{$\Pr(R=1)$} & \ny{$\Pr(R=1)$} & \ny{$\Pr(R=1)$} \\[6pt]
MAR  & \ny{$\Pr(R=1 \mid X)$} & $\Pr(R=1 \mid Y)$ & \ny{$\Pr(R=1 \mid \bZ)$} \\
     & \ny{$\Pr(R=1 \mid \bZ)$} & \ny{$\Pr(R=1 \mid \bZ)$} &  \\
     & \ny{$\Pr(R=1 \mid X,\bZ)$} & $\Pr(R=1 \mid Y,\bZ)$ &  \\[6pt]
MNAR & $\Pr(R=1 \mid Y)$ & \ny{$\Pr(R=1 \mid X)$} & $\Pr(R=1 \mid Y,X)$ \\
     & $\Pr(R=1 \mid Y, X)$  & $\Pr(R=1 \mid Y, X)$ & $\Pr(R=1 \mid Y,X,\bZ)$ \\
     & $\Pr(R=1 \mid Y, \bZ)$  & \ny{$\Pr(R=1 \mid X,\bZ)$} & $\Pr(R=1 \mid Y,\bZ)$ \\
     & $\Pr(R=1 \mid Y, \bZ, X)$ & $\Pr(R=1 \mid Y,X,\bZ)$ & \ny{$\Pr(R=1 \mid X,\bZ)$} \\
     & & & \ny{$\Pr(R=1 \mid X)$} \\
     & & & $\Pr(R=1 \mid Y)$ \\
\bottomrule
\end{tabular}%
}
\caption{\textbf{Missingness mechanisms by missing variable(s).} Missing completely at random (MCAR), missing at random (MAR), and missing not at random (MNAR). Cells shaded in gray illustrate scenarios in which the complete case estimator remains consistent.}
\label{tab:missingness-mcar-mar-mnar}
\end{table}

Table~\ref{tab:missingness-mcar-mar-mnar} enumerates the missingness mechanisms under which Proposition~\ref{prop:fl-consistency}(i) holds, organized by what is missing and what explains it; shaded cells indicate settings where the CC estimator is consistent. The single-site literature characterizes when complete-case analysis is valid for an individual study, but federated learning introduces a consideration it does not address: sites may exhibit different missingness mechanisms. Does site-level validity imply validity at the pooled-level? Proposition~\ref{prop:fl-consistency}(i) establishes that it does---the federated CC estimator is consistent whenever each site's mechanism falls within the shaded cells, regardless of site differences. The determining condition is whether the probability of a complete observation depends on $Y$, not the MCAR-, MAR-, or MNAR-label. This distinction is often obscured in practice, where MNAR is conflated with \textit{non-ignorability} \citep{heitjan1991ignorability} and presumed to invalidate the CC estimator. Several MNAR settings preserve consistency while certain MAR settings do not \citep{bartlett2014improving, hughes2019accounting, little2024comparison, vazquez2026estimators}. The proof appears in the Supplementary Material 
(Section~\ref{sec:supp-proposition-proof}).

\subsection{Inverse probability weighting estimator}
\label{sec:fl-ipw}

The IPW estimator accounts for missingness by up-weighting observations with a low probability of being completely observed. Implementations, however, differ in how weights are estimated. We consider two strategies: (i) \emph{site-specific weighting}, in which weights are estimated independently at each site (Section~\ref{sec:fl-local}), and (ii) \emph{calibrated weighting}, in which sites share coefficients from their weighting models to calibrate weights across sites (Section~\ref{sec:fl-mr}). We then describe how each strategy impacts variance estimation (Section~\ref{sec:robust-sandwhich-estimator}). 

Generally, the IPW estimator is defined as the solution to 
\be
\label{eqn:ipw-ee}
\sum_{i=1}^n \bPhi_\text{IPW} (\bO_i;\btheta,\balpha)  = \sum_{k=1}^K \sum_{i=1}^{n^{(k)}} r_i^{(k)} \frac{\bS_\btheta(y_i^{(k)},r_i^{(k)}x_i^{(k)},\bz_i^{(k)};\btheta)}{\pi(y_i^{(k)}, r_i^{(k)}x_i^{(k)}, \bz_i^{(k)}; \balpha)} = \bzero,
\ee
where $\pi(\cdot)$ is a placeholder for the probability of a complete 
observation (Table~\ref{tab:missingness-mcar-mar-mnar}). The parameter 
$\balpha$ indexes the weighting model used to estimate this probability, 
which we assume is finite-dimensional. Under a linear regression model, an 
estimate for $\btheta$ can be obtained via weighted least squares (WLS). 
Define $\bW = \text{diag}\{r_1/\pi(y_1,r_1x_1,\bz_1;\balpha),\ldots,r_n/\pi(y_n,r_n x_n,\bz_n;\balpha)\}$, with $\bW^{(k)}$ denoting the block-matrix for site $k$. The WLS solution decomposes as
\bse
\wh\btheta_\text{IPW}
&=&
\left\{(X_R,\bZ)^\top \bW (X_R,\bZ)\right\}^{-1}
(X_R,\bZ)^\top \bW Y \\
&=& \left[\sum_{k=1}^K \{X_R^{(k)},\bZ^{(k)}\}^\top \bW^{(k)} \{X_R^{(k)},\bZ^{(k)}\}\right]^{-1}
\left[\sum_{k=1}^K \{X_R^{(k)},\bZ^{(k)}\}^\top \bW^{(k)}  Y^{(k)}\right].
\ese
Each site transmits the weighted sufficient statistics to obtain $\wh\btheta_\text{IPW}$; the error variance $\sigma^2$ is estimated as in the CC case (Algorithms~\ref{alg:fl-local-ipw}--\ref{alg:fl-mr-ipw}).

When a GLM is of interest, the one-shot algorithm described for the CC 
estimator can be extended. Instead of reporting the count $n_j^{(k)}$ as 
the weight, the weight $w_j^{(k)}$ corresponds to the sum of the inverse 
probabilities among observations within the $u_j$ combination. The 
site-level data to be communicated can be represented as
\be
\label{eqn:ipw-u-glm}
\bO^{(k)} = 
\begin{pmatrix}
u_1, & w_1^{(k)} = \sum_{i \in u_1} 1/\pi(y_i^{(k)}, r_i^{(k)}x_i^{(k)}, \bz_i^{(k)}; \balpha) \\
\vdots & \vdots\\
u_q, & w_q^{(k)} = \sum_{i \in u_q} 1/\pi(y_i^{(k)}, r_i^{(k)}x_i^{(k)}, \bz_i^{(k)}; \balpha)
\end{pmatrix}.
\ee
Under this formulation, \eqref{eqn:ipw-ee} becomes
$\sum_{i=1}^n \bPhi_{\text{IPW}}(\bO_i;\btheta, \balpha) = \sum_{k=1}^K 
\sum_{j=1}^q w_j^{(k)} \bS_\btheta(u_j;\btheta)$ 
(Algorithm~\ref{alg:fl-ipw-glm}). This expression generalizes the CC representation: when each observation receives unit weight, $w_j^{(k)}$ collapses to $n_j^{(k)}$ and \eqref{eqn:ipw-u-glm} reduces to 
\eqref{eqn:cc-u-glm}. The representation applies regardless of whether the weights are estimated via the site-specific or calibrated approach.

\subsubsection{Site-specific weight estimation}
\label{sec:fl-local}

Suppose the probability of observing a complete observation at site $k$ is modeled as $\pi_{Y,\bZ}(y_i^{(k)}, \bz_i^{(k)};\balpha_k) = \expit\{ m_{\balpha}(y_i^{(k)}, \bz_i^{(k)}; \balpha_k)\}$, where $m_{\balpha}(\cdot)$ is a known function of $(Y,\bZ)$ indexed by a finite-dimensional parameter vector $\balpha_k$ which may differ across sites. An estimate of $\balpha_k$ can be obtained at site $k$ by solving the $M$-estimating equation
\bse
\sum_{i=1}^{n^{(k)}} \bPhi_\text{Nuisance} (\bO^{(k)}_i;\balpha_k) 
&=& \sum_{i=1}^{n^{(k)}} \left\{ r^{(k)}_i - \pi_{Y,\bZ}(y_i^{(k)}, \bz_i^{(k)}; \balpha_k) \right\} \frac{\partial}{\partial \balpha_k^\top} m_{\balpha}(y_i^{(k)}, \bz_i^{(k)}; \balpha_k) = \bzero.
\ese
Because this $M$-estimating equation depends only on data available at site $k$, no communication is required to estimate the weights. Each site computes $\wh{\balpha}_k$ locally, constructs the weights $1/\wh{\pi}_{Y,\bZ}(y_i^{(k)}, \bz_i^{(k)}; \wh{\balpha}_k)$, and transmits the needed information to the coordinating site as described in Section~\ref{sec:fl-ipw}. Estimation of $\btheta$ then proceeds via the WLS approach or using the summary of outcome and covariate combinations and weights.

\subsubsection{Calibrated weight estimation}
\label{sec:fl-mr}

Site-specific weight estimation relies on a single weighting model at each site. If this model is misspecified, the resulting IPW estimator may be inconsistent (Proposition~\ref{prop:fl-consistency}(ii)). Calibrated weight estimation addresses this limitation by combining $J$ candidate models contributed from \textit{donors} across the network. Consistency is a site-level property: the IPW estimator is consistent at site $k$ if at least one of the $J$ candidate models correctly captures the missingness mechanism at site $k$---and this model need not have been estimated at site $k$. Pooling candidates therefore expands the set of models available locally, increasing the chance that at least one is correctly specified at each site.

We construct the set of candidate models by requiring some or all sites to share the estimated parameters from their site-specific weighting models. Let $\wh\balpha = (\wh\balpha_1^\top, \ldots, \wh\balpha_J^\top)^\top$ denote the collection of $J$ candidate parameter vectors, where each $\wh\balpha_j$ indexes a weighting model contributed by a participating site. Note that $J$ may exceed $K$ if any site contributes more than one candidate model. For each observation at site $k$, we define the vector of candidate probabilities for a complete observation as
\bse
\bgamma(y_i^{(k)}, r_i^{(k)}x_i^{(k)}, \bz_i^{(k)}; \wh\balpha) &=& \left\{\pi(y_i^{(k)}, r_i^{(k)}x_i^{(k)}, \bz_i^{(k)}; \wh\balpha_{1}), \ldots, \pi(y_i^{(k)}, r_i^{(k)}x_i^{(k)}, \bz_i^{(k)}; \wh\balpha_J) \right\}^\top.
\ese
The calibrated probability of a complete observation is then constructed as a linear combination of these candidate probabilities:
\bse
\pi^\text{cal}(y_i^{(k)}, r_i^{(k)}x_i^{(k)}, \bz_i^{(k)}; \wh\balpha) = \wh\btau_k^\top \bgamma(y_i^{(k)}, r_i^{(k)}x_i^{(k)}, \bz_i^{(k)}; \wh\balpha),
\ese
where $\wh\btau_k$ is a vector of calibration coefficients that combine the $J$ candidate probabilities at site $k$. Throughout, we reserve \emph{weight} for the inverse-probability weight $1/\pi^{\text{cal}}(\cdot)$ applied to an observation, and refer to $\wh\btau_k$ as calibration coefficients. The coefficients are chosen so that the combined probability reproduces the observed missingness indicator in expectation; specifically, $\wh\btau_k$ solves
\bse
\wh\btau_k &=& \left\{ \sum_{i=1}^{n^{(k)}} \bgamma(y_i^{(k)}, r_i^{(k)}x_i^{(k)}, \bz_i^{(k)}; \wh\balpha) \bgamma(y_i^{(k)}, r_i^{(k)}x_i^{(k)}, \bz_i^{(k)}; \wh\balpha)^\top \right\}^{-1} \\
&& \times \left\{ \sum_{i=1}^{n^{(k)}} \bgamma(y_i^{(k)}, r_i^{(k)}x_i^{(k)}, \bz_i^{(k)}; \wh\balpha) r_i^{(k)} \right\}.
\ese
The calibration coefficients $\wh{\btau_k}$ have two interpretations: as the solution to a least squares projection, or as the result of enforcing a calibration constraint via Lagrange multipliers \citep{han2014multiply, chen2023unified}. Because $\wh\btau_k$ is computed locally at each site using only the shared parameters $\wh\balpha$ from the network and local data, the calibration step requires no additional communication beyond the initial exchange of weighting model coefficients. After constructing $\pi^\text{cal}(\cdot)$, estimation proceeds as described in Section~\ref{sec:fl-ipw}. 

\subsubsection{Consistency conditions}
\label{sec:consistency-ipw}

Proposition~\ref{prop:fl-consistency}(ii) characterizes when the IPW estimator is consistent: when the oracle probability of a complete observation depends on $Y$, consistency requires that the weighting model be correctly specified at those particular sites. The calibrated weight estimation framework facilitates correct specification over the site-specific weights by combining $J$ candidate models across sites (Proposition~\ref{prop:fl-consistency}(iii)), achieving robustness through multiplicity \citep{han2014multiply}. Beyond the robustness that multiplicity affords, the calibration framework lets sites with insufficient sample sizes borrow strength from larger sites. Such borrowing matters in practice because federated networks are often dominated by small sites whose limited data may not resolve a weighting model on their own; calibrating against the better-estimated models of larger sites stabilizes their weights. More broadly, the value of calibration is not confined to sites with a correctly specified candidate: in Section~\ref{sec:sim-pool} we find that calibration can lower estimation error even when no candidate model is correct. 

Moreover, when the oracle probability does not depend on $Y$, correct specification of the weighting model is not required; however, including $Y$ in the weighting model may introduce bias in certain settings, such as when $X$ is missing. Empirically, $Y$ may appear associated with missingness in $X$ due to the structure of the outcome model~\eqref{eqn:outcome-model} rather than the true oracle mechanism \citep{bartlett2014improving, vazquez2026estimators}. Distinguishing true from empirical dependence requires subject-matter knowledge, which may be  formalized using missing data directed acyclic graphs. The distinction matters because including $Y$ when the oracle probability does not depend on $Y$ can introduce bias through a collider structure (Supplementary Material, Section~\ref{sec:supp-consistency-cc}); we illustrate the bias numerically in Section~\ref{sec:fl-simulation}.

\subsection{Robust sandwich estimator}
\label{sec:robust-sandwhich-estimator}


Standard sandwich variance estimation applies to the CC estimator since no weighting model parameters are involved; we describe the steps in Algorithms~\ref{alg:fl-cc} (sufficient information; few-shot) and \ref{alg:fl-cc-glm} (aggregated counts; one-shot). For the IPW estimator, however, estimating the weighting model parameters $\balpha$ (and by extension the weights) introduces uncertainty that propagates into the estimator of $\btheta$ \citep{robins1995analysis}. If this nuisance-parameter uncertainty is ignored, standard error estimates may be biased, resulting in under- or over-coverage of nominal confidence intervals and misleading inference. This issue is distinct from bias in $\widehat{\btheta}$: even when $\widehat{\btheta}$ is consistent, inference can be invalid if the variance is misestimated.

\subsubsection{Site-specific weight estimation}

To correctly adjust for $\balpha$ in the variance estimation of $\btheta$, a common remedy is to use a stacked $M$-estimating equation approach, in which the estimating equations for the outcome model and the weighting model are solved jointly \citep{tsiatis2006semiparametric}. Let $\bxi = (\btheta^\top,\balpha^\top)^\top$ denote the stacked parameter vector. Then
\be
\label{eqn:estimating-equation-stacked}
\sumi\bPhi_\text{Stacked}(\bO_i ;\bxi)
=
\sumi
\left\{
\begin{matrix}
\bPhi_\text{IPW}(\bO_i;\btheta,\balpha) \\
\bPhi_\text{Site 1, Nuisance}(\bO_i;\balpha_1) \\
\ldots \\
\bPhi_\text{Site J, Nuisance}(\bO_i;\balpha_J) 
\end{matrix}
\right\}
=
\sumi
\left\{
\begin{matrix}
\bPhi_\text{IPW}(\bO_i;\btheta,\balpha) \\
\bPhi_\text{Nuisance}(\bO_i;\balpha)
\end{matrix}
\right\}
= \bzero 
\ee
and the solution is denoted by $\wh\bxi = (\wh\btheta^\top,\wh\balpha^\top)^\top$. Following the construction of the stacked estimating equation approach, and under standard regularity conditions, $\wh\bxi$ has the asymptotic linear representation
\bse
n^{1/2}(\wh{\bxi}-\bxi_0) \rightarrow_d \Normal(\bzero, \bA_\text{Stacked}^{-1}\bB_\text{Stacked}\bA_\text{Stacked}^{-\top}), 
\ese
where an estimate of $\var(\bxi) = \bA_\text{Stacked}^{-1}\bB_\text{Stacked}\bA_\text{Stacked}^{-\top}$ is obtained by replacing population-known quantities with their sample estimated analogs:
\bse
\wh\bA_\text{Stacked} &=& n^{-1} \sum_{k=1}^K \frac{\partial}{\partial \bxi^\top} \left\{ \sum_{i=1}^{n^{(k)}} \bPhi_{\rm Stacked}(\bO_i^{(k)}; \bxi) \right\} \bigg|_{\bxi = \wh\bxi}, \ \wh\bB_\text{Stacked} = n^{-1} \sum_{k=1}^K \sum_{i=1}^{n^{(k)}} \left\{ \bPhi_{\rm Stacked}(\bO_i^{(k)}; \wh\bxi)^{\otimes2} \right\},
\ese
where $v^{\otimes 2} = vv^\top$ for a column vector $v$. Standard errors for $\btheta$ are extracted from the corresponding diagonal elements of $\wh\var(\wh\bxi)$. We also derive the asymptotic normality of $\wh\btheta$ from a representation that conditions on the $M$-estimator of the weighting-model parameters. The representation yields the same asymptotic variance as the stacked one but isolates the term contributed by estimating $\balpha$, clarifying how that uncertainty enters the variance estimator.

Under site-specific weighting, weighting model parameters are estimated separately within each site, so no cross-site dependence is created. As a result, the estimating equations can be decomposed by site, and both the stacked Jacobian matrix $\wh\bA_{\text{Stacked}}$ and the stacked score covariance matrix $\wh\bB_{\text{Stacked}}$ have a block-structured form. Under the assumption that the weighting-model parameters are estimated independently across sites, each of the sandwich matrices decomposes into a sum of $K$ site-specific contributions, a structure that limits the number of communication rounds and the information each site must transmit.

Let $\bA^{(k)}_{\btheta\btheta} = E\left\{ \partial \bPhi_\text{IPW}(\bO; \btheta, \balpha) / \partial \btheta^\top \right\}$, $\bA^{(k)}_{\btheta\balpha_k} = E\left\{ \partial \bPhi_\text{IPW}(\bO; \btheta, \balpha_k) / \partial \balpha_k^\top \right\}$, \\ $\bA^{(k)}_{\balpha_k \btheta} = E\left\{ \partial \bPhi_\text{Nuisance}(\bO, \balpha_k) / \partial \btheta^\top \right\}$, and $\bA^{(k)}_{\balpha_k \balpha_k} = E\left\{ \partial \bPhi_\text{Nuisance}(\bO; \balpha_k) / \partial \balpha_k^\top \right\}$. Because the weighting models indexed by $\balpha$ do not depend on $\btheta$, we have $\bA^{(k)}_{\balpha_k \btheta} = \bzero$. In a network with $K$ sites, the stacked Jacobian matrix takes the form
\bse
\wh\bA_{\text{Stacked}}
=
\begin{pmatrix}
\sum_{k=1}^K \wh\bA^{(k)}_{\btheta\btheta}
&
\wh\bA^{(1)}_{\btheta\balpha_1}
&
\wh\bA^{(2)}_{\btheta\balpha_2}
&
\cdots
&
\wh\bA^{(K)}_{\btheta\balpha_K}
\\
\bzero
&
\wh\bA^{(1)}_{\balpha_1\balpha_1}
&
\bzero
&
\cdots
&
\bzero
\\
\bzero
&
\bzero
&
\wh\bA^{(2)}_{\balpha_2\balpha_2}
&
\cdots
&
\bzero
\\
\vdots
&
\vdots
&
\vdots
&
\ddots
&
\vdots
\\
\bzero
&
\bzero
&
\bzero
&
\cdots
&
\wh\bA^{(K)}_{\balpha_K\balpha_K}
\end{pmatrix}.
\ese
Similarly, the stacked score covariance matrix has the block form
\bse
\wh\bB_{\text{Stacked}}
=
\begin{pmatrix}
\sum_{k=1}^K \wh\bB^{(k)}_{\btheta\btheta}
&
\wh\bB^{(1)}_{\btheta\balpha_1}
&
\wh\bB^{(2)}_{\btheta\balpha_2}
&
\cdots
&
\wh\bB^{(K)}_{\btheta\balpha_K}
\\
\wh\bB^{(1)}_{\balpha_1\btheta}
&
\wh\bB^{(1)}_{\balpha_1\balpha_1}
&
\bzero
&
\cdots
&
\bzero
\\
\wh\bB^{(2)}_{\balpha_2\btheta}
&
\bzero
&
\wh\bB^{(2)}_{\balpha_2\balpha_2}
&
\cdots
&
\bzero
\\
\vdots
&
\vdots
&
\vdots
&
\ddots
&
\vdots
\\
\wh\bB^{(K)}_{\balpha_K\btheta}
&
\bzero
&
\bzero
&
\cdots
&
\wh\bB^{(K)}_{\balpha_K\balpha_K}
\end{pmatrix},
\ese
where $\bB^{(k)}_{\btheta\btheta} = E\left\{ \bPhi_\text{IPW}(\bO; \btheta, \balpha)^{\otimes 2}  \right\}$, $\bB^{(k)}_{\btheta\balpha_k} = E\left\{ \bPhi_\text{IPW}(\bO; \btheta, \balpha_k) \bPhi_\text{Nuisance}(\bO; \balpha_k)^\top \right\}$, and $\bB^{(k)}_{\balpha_k \balpha_k} = E\left\{  \bPhi_\text{Nuisance}(\bO; \balpha_k)^{\otimes 2}  \right\}$. The block structure motivates a federated strategy: each site computes and sends out only its local covariance components, which are then summed centrally (Algorithm~\ref{alg:fl-local-ipw} and \ref{alg:fl-ipw-glm}). Computation of the robust sandwich variance estimator therefore parallels the SI approach: sites transmit only summary statistics, the block covariance components above, rather than IPD.

\subsubsection{Calibrated weight estimation}

The formulation in~\eqref{eqn:estimating-equation-stacked} would treat calibration coefficients $\wh\btau = (\wh\btau_{1}^\top, \dots, \wh\btau_{K}^\top)^\top$ as fixed, accounting only for the uncertainty in the candidate model parameters $\wh\balpha$; a better solution would propagate the uncertainty in estimating $\btau$. We append its estimating equation, $\sum_i \bgamma(\bO_i; \balpha)\{r_i - \bgamma(\bO_i; \balpha)^\top \btau\} = \bzero$, to the stacked system, yielding a joint $M$-estimator for $\bxi = (\btheta^\top, \balpha^\top, \btau^\top)^\top$. This distinguishes the calibrated estimator from the site-specific one in two 
ways. First, because each site constructs its weights from candidate models contributed across the network, the contribution of $\wh\bA_\text{Stacked}$ for a given 
site depends on multiple $\balpha_j$ rather than a single $\balpha_k$. Second, the calibration 
weights $\btau$ enter the variance as an additional estimated parameter, 
introducing blocks in $\wh\bA_\text{Stacked}$ and $\wh\bB_\text{Stacked}$ 
that the site-specific formulation does not contain. The resulting 
cross-site structure is given explicitly in the Supplementary Material 
(Section~\ref{sec:supp-variance-calibrated}); each site still transmits 
only summary contributions, preserving the federated and privacy-aware 
nature of the estimator.

\subsection{Communication requirements}
\label{sec:fl-communication}

Table~\ref{tab:communication-rounds} summarizes the communication requirements for each estimator. The CC estimator is the most communication-efficient and should be preferred when Proposition~\ref{prop:fl-consistency}(i) holds. The site-specific IPW estimator adds no communication rounds beyond the CC estimator, since weight estimation is carried out within each site, but transmits additional components of the sandwich variance estimator (e.g., $\bA^{(k)}_{\btheta\balpha_k}$). The calibrated IPW estimator requires one further round to share the weighting-model coefficients $\wh\balpha$ and transmits the most information of the three. Under finite outcome-covariate combinations, the CC estimator permits one-shot estimation, whereas the IPW estimators require additional rounds (Supplementary Material, Section~\ref{sec:supp-algorithms}).

\begin{table}[h!]
\centering
\small
\begin{tabular}{lccc}
\toprule
Step & CC & IPW (site-specific) & IPW (calibrated) \\
\midrule
1. Estimate weights locally & --- & $\times$ & $\times$ \\
2. Share weighting model coefficients & --- & --- & \checkmark \\
3. Calibrate weights locally & --- & --- & $\times$ \\
4. Transmit sufficient statistics & \checkmark & \checkmark & \checkmark \\
5. Compute $\wh\btheta$; return residuals & \checkmark & \checkmark & \checkmark \\
6. Transmit variance components & \checkmark & \checkmark & \checkmark \\
\midrule
Total rounds & 3 & 3 & 4 \\
\bottomrule
\end{tabular}
\caption{\textbf{Communication requirements by estimator.} ``---'': not applicable; ``$\times$'': local computation (no communication); ``\checkmark'': communication required. See Algorithms~\ref{alg:fl-cc}--\ref{alg:fl-mr-ipw} in Supplementary Material.}
\label{tab:communication-rounds}
\end{table}

\section{Simulation study}
\label{sec:fl-simulation}

Our simulations proceed in three parts: we determine how many candidate models to pool when correct specification is unlikely and the models from the donors are misspecified (Section~\ref{sec:sim-pool}), then fix that pool size to study a correctly specified weighting model where valid inference hinges on correcting the variance for the estimated weights (Section~\ref{sec:sim-homogeneous}). We close with a network in which a correctly specified model exists only at another site, and show that calibration borrows it to correct a site's own misspecification (Section~\ref{sec:sim-hetero}).

Data were generated under the linear regression model
\bse
Y = \beta_0 + \beta_1 X + \beta_2 Z_1 + \beta_3 Z_2 + \epsilon, \quad \epsilon \sim \Normal(0,\sigma^2),
\ese
with $(\beta_0,\beta_1,\beta_2,\beta_3,\sigma) = (1,1,1,1,5)$. Covariates were drawn as $Z_1 \sim \Bernoulli(0.5)$, $Z_2 \sim \Normal(Z_1,1)$, and $X \sim \Normal(Z_1 Z_2, 1)$, where $X$ is the covariate subject to missingness. The number of sites $K$ ranged from $5$ to $50$, and each site size $n^{(k)}$ was drawn from $\{30,100,1000\}$ with equal probability, so the network mixed small and large sites. The simulation study considered $2{,}000$ replications. 

Performance was summarized by percent bias, coverage of the nominal $95\%$ confidence intervals, and the ratio of the estimated to the empirical standard errors, for which a value above one indicates overestimation and below one underestimation. Standard errors were computed with and without correcting for the estimation of $\balpha$ and/or $\btau$ to isolate the consequence of ignoring uncertainty in the estimated weighting-model parameters. For the calibrated weighting estimator we parametrized the  weights to be non-negative and normalized, following \cite{chen2023unified}. We report results for the intercept $\beta_0$, which exhibits the most pronounced bias in the missing covariate setting \citep{hughes2019accounting}; the remaining coefficients, a logistic-regression analysis based on count aggregation, and further comparisons between the calibrated and site-specific estimators are deferred to Supplement Section~\ref{sec:supp-simulations}.

\subsection{Choosing the calibration pool}
\label{sec:sim-pool}

To choose the number of candidate weighting models to pool, we stressed the calibration step in a heterogeneous network: each site was assigned a missingness mechanism drawn at random from a library of ten logistic forms for the probability of a complete observation in $(Y, Z_1, Z_2)$, each with a distinct interaction structure, giving a subject-level missingness rate in $X$ of roughly $60\%$, and each site fit a working model whose form was likewise drawn at random from the library (details in Supplement Section~\ref{sec:supp-pool-forms}). The working model therefore rarely coincided with the mechanism that generated its data. Under this misspecification neither the site-specific nor the calibrated estimator is consistent; our aim was to show that borrowing models across the network may improve estimation over the site-specific estimator. We varied the number of sites $K$ from $10$ to $50$ and the number of donor models---the external models a site borrows---from one to nine, taking donors to be the largest sites.

We assessed the calibrated estimator by the percent bias and root mean square error (RMSE) of the intercept, and by the conditioning of the calibration step, summarized by the condition number $\kappa$ of the candidate matrix \citep{belsley2005regression}. The ideal is $\kappa = 1$, where the candidate predictions are mutually orthogonal and each donor contributes independent information; as the candidates grow redundant, $\kappa$ rises, the projection that defines $\wh\btau$ becomes unstable (details in Supplement Section~\ref{sec:supp-kappa}). 

\begin{figure}[h!]
    \centering
    \includegraphics[width=\linewidth]{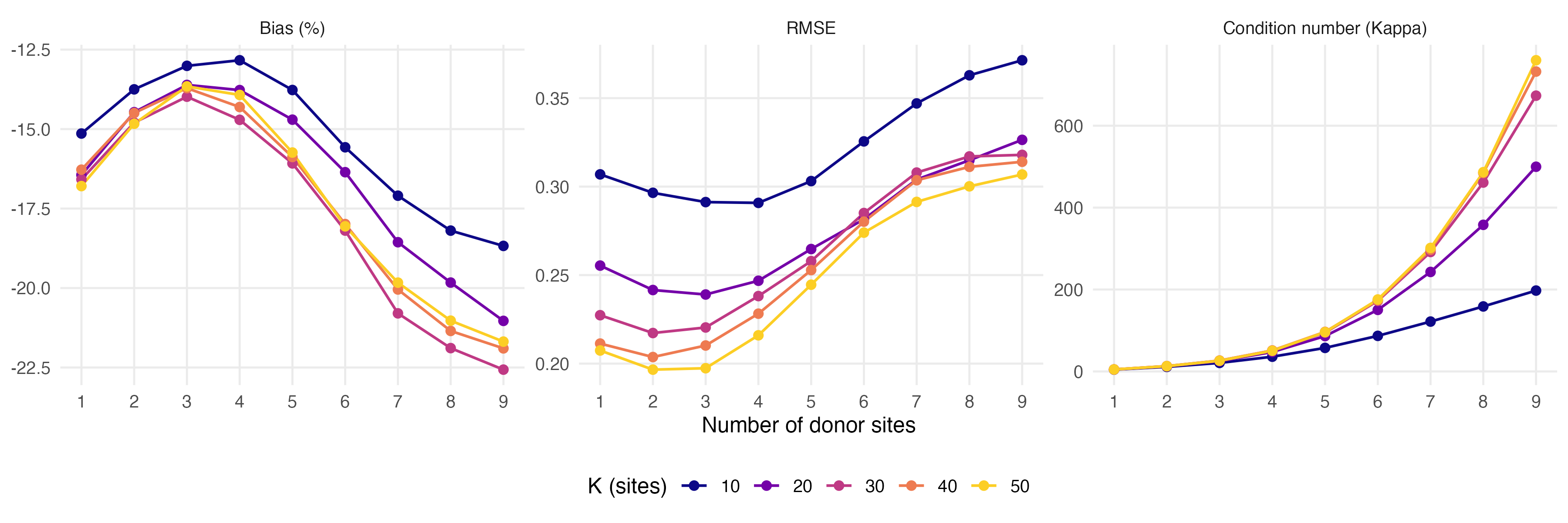}
    \caption{\textbf{Performance of the IPW estimator with calibrated weights by donor pool size.} Percent bias and RMSE of the intercept, and the condition number $\kappa$ of the candidate matrix, against the number of donor models, by network size $K$.}
    \label{fig:rmse-kappa}
\end{figure}

Both bias and RMSE behaved non-monotonically with the number of donors: each fell as the first few donors were added, reached a minimum at a small pool, and rose as the pool grew further. At $K=30$, the RMSE bottomed near $0.22$ at two donors before climbing to roughly $0.32$ by nine donors (Figure~\ref{fig:rmse-kappa}). The rise accompanied a sharp increase in $\kappa$, which stayed in the single-to-double digits through about four donors and then grew into the hundreds. The smallest network, $K=10$, carried the highest bias and RMSE across all pool sizes: with only ten sites, each added donor consumes a larger share of the network's independent information.

\begin{figure}[h!]
    \centering
    \includegraphics[width=\linewidth]{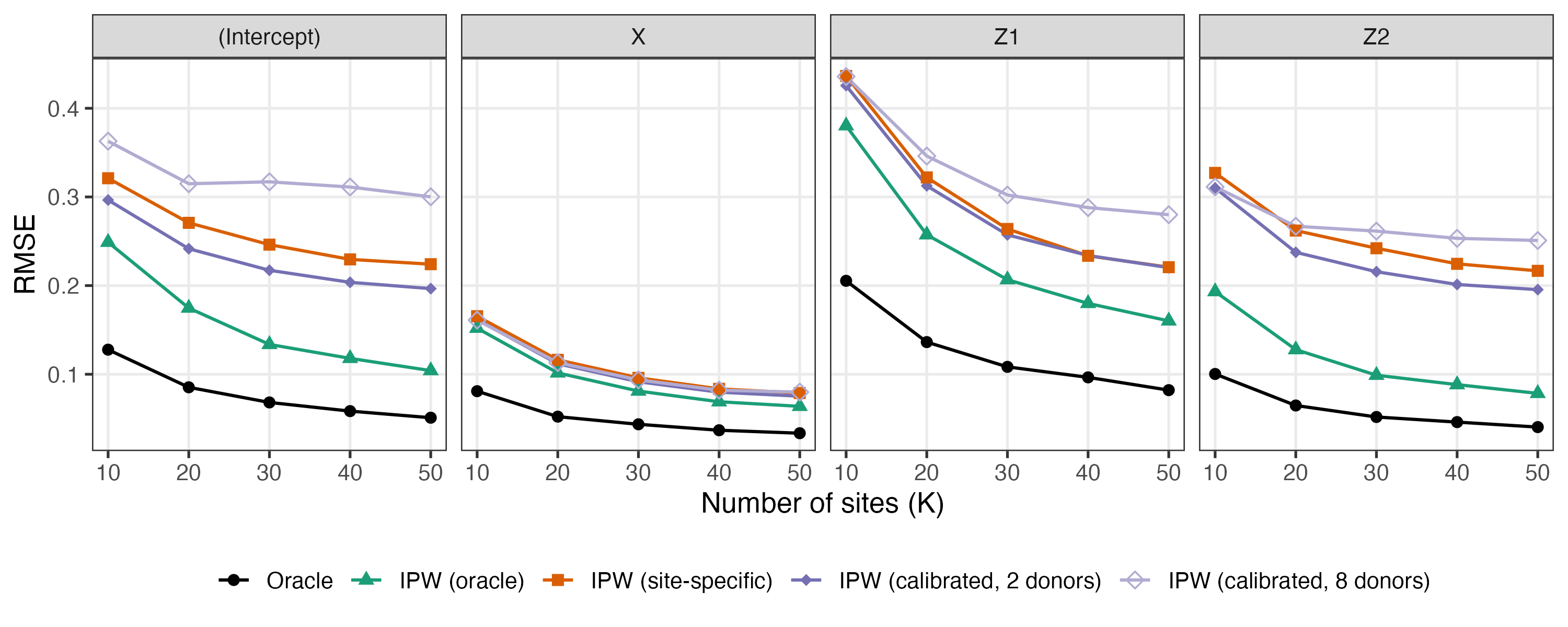}
    \caption{\textbf{Performance of the IPW estimator by weight choice and pool size.} RMSE for each regression coefficient against the number of sites $K$, comparing the oracle and oracle-weighted IPW benchmarks with the site-specific estimator and calibrated weights at two and eight donors.}
    \label{fig:pool-rmse}
\end{figure}

Figure~\ref{fig:pool-rmse} places the calibrated estimator against its benchmarks across the four coefficients. The oracle estimator, free of missingness, attained the lowest bias and RMSE, followed by the IPW estimator with oracle weights (no weight estimation). Among the estimated-weight strategies, the calibrated estimator with two donors matched or slightly improved on the site-specific estimator---most visibly for the intercept and $Z_2$---while the calibrated estimator with eight donors had the highest RMSE of the estimated-weight curves at every network size. The ordering reflects the ill-conditioning of Figure~\ref{fig:rmse-kappa}: borrowing a few external models sharpens the weights, whereas borrowing too many integrates redundant information that degrades the estimator. Taken together, a small pool captured most information in this simulation study, while additional donors contribute redundant predictions that destabilize the weights and can leave the estimator worse than the site-specific one. Therefore, we use the two largest sites as donors in the simulation scenarios that follow.

\subsection{Homogeneous missingness}
\label{sec:sim-homogeneous}

In a more controlled setting, we considered a network in which the missingness mechanism was shared across sites to evaluate Proposition~\ref{prop:fl-consistency} and showcase the importance of variance correction in FL. The subject-level missingness in $X$ was approximately $60\%$ and was introduced under MNAR as $\Pr(R=1\mid X,\bZ)=\text{expit}(-0.1+0.2X+0.2Z_1+0.2Z_2)$ and under MAR as $\Pr(R=1\mid Y,\bZ)=\text{expit}(-0.1+0.1Y+0.2Z_1+0.2Z_2)$, separately. We compared five weighting strategies: the oracle (true weights), pooled (weights from the pooled data), site-specific (local estimation only), calibrated (sharing $\wh\balpha$ from the two largest sites), and uniform (weights drawn from the inverse of a $\text{Uniform}(0.1,0.9)$ draw, included to probe sensitivity to weight misspecification). Each weighting model was fit by logistic regression for $\Pr(R=1\mid Y,\bZ)$ adjusting for $(Y,Z_1,Z_2)$, an intentional choice that introduced misspecification under the MNAR mechanism. 

The simulation results supported Proposition~\ref{prop:fl-consistency}. Under MNAR, where the probability of a complete observation did not depend on $Y$, the CC estimator and the IPW estimator with oracle or uniform weights all exhibited minimal bias in $X$ (Table~\ref{tab:numerical-results}); for instance, the percent bias of the CC estimator was below $1\%$. Forcing $Y$ into the weighting model, by contrast, introduced bias regardless of how the weights were estimated, and the IPW estimator using pooled weights led to roughly $11\%$ bias. Under MAR the pattern reversed: the CC estimator and the uniform-weight IPW estimator were substantially biased, near $118\%$ across $K$, whereas the IPW estimator with a correctly specified weighting model---oracle, pooled, site-specific, or calibrated---removed the bias. The contrast in results showcases that the IPW estimator can introduce or remove bias depending on which variables enter the weighting model relative to the oracle missingness mechanism.

\begin{table}[!h]
\centering
\resizebox{0.95\linewidth}{!}{
\begin{tabular}[t]{llrrrrrrrrrrrrrr}
\toprule
\textbf{Estimator} &  &  \textbf{Bias} & \textbf{SE} & \textbf{SD} & \textbf{Cov} & \textbf{Bias} & \textbf{SE} & \textbf{SD} & \textbf{Cov} &  \textbf{Bias} & \textbf{SE} & \textbf{SD} & \textbf{Cov} \\
\midrule
\addlinespace[0.3em]
\multicolumn{2}{l}{\textbf{Missing not at random}} 
&\multicolumn{4}{c}{\textbf{K=10}} &\multicolumn{4}{c}{\textbf{K = 30}} &\multicolumn{4}{c}{\textbf{K = 50}}\\
\addlinespace
\hspace{1em}Oracle &  & -0.40 & 21.87 & 23.27 & 94.98 & -0.38 & 11.07 & 11.51 & 94.55 & -0.05 & 8.33 & 8.39 & 95.22 \\
\addlinespace
\hspace{1em}CC &  & 0.29 & 30.21 & 31.91 & 94.93 & -0.51 & 15.33 & 16.19 & 94.31 & 0.08 & 11.54 & 11.76 & 94.98 \\
\addlinespace
\hspace{1em}IPW & Oracle & 0.39 & 30.40 & 32.12 & 95.03 & -0.58 & 15.42 & 16.30 & 94.26 & 0.10 & 11.61 & 11.77 & 95.22 \\
 & Pooled & 11.11 & 30.44 & 28.82 & 95.60 & 10.77 & 15.42 & 14.34 & 91.25 & 11.24 & 11.60 & 10.21 & 86.14 \\
 & Site-specific & 10.98 & 32.37 & 30.25 & 96.22 & 10.61 & 16.36 & 14.76 & 92.78 & 11.12 & 12.32 & 10.44 & 89.10 \\
 & Calibrated & 11.08 & 31.63 & 30.57 & 95.65 & 10.80 & 15.90 & 14.95 & 91.16 & 11.13 & 11.94 & 10.56 & 87.24 \\
 & Uniform & 0.05 & 35.99 & 39.56 & 93.50 & -0.50 & 18.55 & 19.18 & 94.60 & -0.01 & 13.98 & 14.26 & 94.93 \\
 \addlinespace
 & & \multicolumn{12}{c}{\textit{Robust sandwich estimator}}\\
 \addlinespace 
 & Pooled & 11.11 & 26.65 & 28.82 & 91.83 & 10.77 & 13.49 & 14.34 & 85.52 & 11.24 & 10.15 & 10.21 & 78.78 \\
 & Site-specific & 11.09 & 26.99 & 30.58 & 91.25 & 10.79 & 13.79 & 14.92 & 85.45 & 11.14 & 10.40 & 10.56 & 79.20 \\
 & Calibrated $(\balpha)$ & 10.98 & 30.61 & 30.25 & 93.74 & 10.61 & 18.29 & 14.76 & 93.12 & 11.12 & 14.50 & 10.44 & 91.25 \\
 & Calibrated $(\balpha,\btau)$ & 10.98 & 28.13 & 30.25 & 92.35 & 10.61 & 14.48 & 14.76 & 88.53 & 11.12 & 11.01 & 10.44 & 82.22 \\
\addlinespace
\multicolumn{2}{l}{\textbf{Missing at random}} 
&\multicolumn{12}{c}{} \\
\addlinespace
\hspace{1em}Oracle &  & -0.40 & 21.87 & 23.27 & 94.98 & -0.38 & 11.07 & 11.51 & 94.55 & -0.05 & 8.33 & 8.39 & 95.22 \\
\addlinespace
\hspace{1em}CC &  & -118.37 & 30.24 & 31.77 & 8.75 & -118.56 & 15.32 & 16.07 & 0.00 & -117.98 & 11.51 & 11.65 & 0.00 \\
\addlinespace
\hspace{1em}IPW & Oracle & -1.83 & 34.10 & 36.93 & 94.17 & -0.69 & 17.41 & 18.45 & 93.74 & -0.12 & 13.13 & 13.37 & 94.84 \\
 & Pooled & -1.56 & 34.22 & 32.47 & 97.51 & -0.63 & 17.43 & 15.84 & 97.37 & -0.06 & 13.14 & 11.51 & 97.28 \\
 & Site-specific  & -3.94 & 35.11 & 35.05 & 96.65 & -2.16 & 18.00 & 16.53 & 96.56 & -1.23 & 13.66 & 12.27 & 96.99 \\
 & Calibrated & -3.57 & 33.37 & 34.64 & 96.22 & -2.18 & 16.95 & 16.55 & 95.70 & -1.44 & 12.75 & 12.11 & 96.18 \\
 & Uniform & -118.33 & 36.01 & 38.42 & 14.15 & -118.19 & 18.52 & 19.28 & 0.14 & -117.83 & 13.94 & 14.26 & 0.00 \\
\addlinespace
 & & \multicolumn{12}{c}{\textit{Robust sandwich estimator}}\\
 \addlinespace 
 & Pooled & -1.56 & 29.28 & 32.47 & 94.65 & -0.63 & 14.92 & 15.84 & 93.88 & -0.06 & 11.25 & 11.51 & 94.55 \\
 & Site-specific & -3.96 & 29.61 & 35.07 & 92.77 & -2.11 & 15.45 & 16.55 & 93.40 & -1.25 & 12.06 & 12.26 & 93.88 \\
 & Calibrated $(\balpha)$ & -3.57 & 33.29 & 34.64 & 95.17 & -2.18 & 19.86 & 16.55 & 96.32 & -1.44 & 15.92 & 12.11 & 97.56 \\
 & Calibrated $(\balpha,\btau)$ & -3.57 & 31.15 & 34.64 & 94.22 & -2.18 & 16.38 & 16.55 & 94.46 & -1.44 & 12.56 & 12.11 & 94.60 \\
\addlinespace
\bottomrule
\end{tabular}}
\caption{\textbf{Simulation results of regression analysis with a missing covariate by number of sites.} Percent bias (\%), mean estimated standard error (SE), empirical standard deviation (SD), and $95\%$ confidence interval coverage (Cov), for the intercept coefficient ($\beta_0$) under homogeneous missingness. Calibrated $(\balpha)$ corrects the variance for the estimated weighting-model parameters only; Calibrated $(\balpha,\btau)$ additionally accounts for the estimated calibration coefficients. \label{tab:numerical-results}}
\end{table}

\begin{figure}
    \centering
    \includegraphics[width=\linewidth]{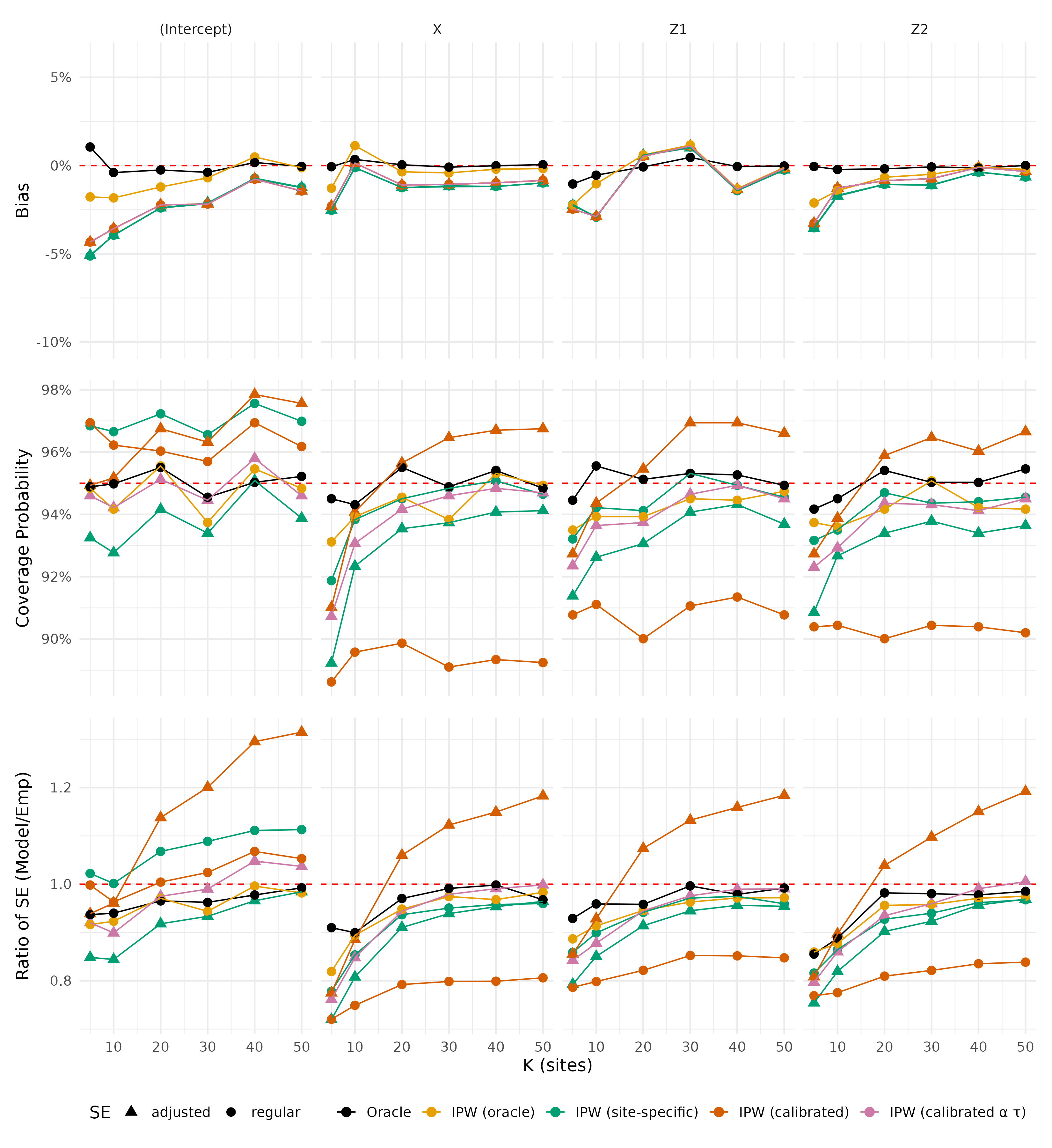}
    \caption{\textbf{Federated regression with covariate missingness under MAR: Comparison of inverse probability weighting strategies as the number of participating sites increases.} 
    Top row: percent bias; middle row: empirical coverage of 95\% confidence intervals; bottom row: ratio of model-based to empirical standard errors. 
    Columns correspond to regression coefficients for the intercept, partially observed covariate ($X$), and fully observed covariates ($Z_1$, $Z_2$). 
    \textit{Oracle}: no missingness; \textit{IPW (oracle)}: known true weights; \textit{IPW (pooled)}: weights estimated from stacked data; \textit{IPW (site-specific)}: weights estimated locally at each site; \textit{IPW (calibrated)}: weights calibrated using federated external models. 
    Standard errors are unadjusted (circles) or adjusted for estimation of nuisance parameters  $\balpha$  or both $\balpha$ and $\btau$ (triangles).}
    \label{fig:mar-ipw}
\end{figure}

Beyond bias, valid inference required correcting the variance for the estimated weighting model. Figure~\ref{fig:mar-ipw} reports the ratio of model-based to empirical standard errors across the weighting strategies under MAR. Ignoring the estimation of the weighting model, the standard errors were underestimated for $X$, $Z_1$, and $Z_2$ and overestimated for the intercept, producing under- and over-coverage of the nominal level, respectively, with the distortion increasing as $K$ grew. Correcting for the uncertainty in the weighting model parameters $\wh\balpha$ returned the ratios to near one and held them stable across network size for the site-specific weight estimator. The calibrated weighting estimator carries a second source of uncertainty beyond $\wh\balpha$---the estimated calibration coefficients $\wh\btau$---so correcting for $\wh\balpha$ alone is insufficient. Figure~\ref{fig:mar-ipw} traces the three levels for the coefficient of $X$. Without any correction, the ratio of model-based to empirical standard errors sat below one ($\approx 0.8$), producing under-coverage ($\approx 90\%$). Correcting for $\wh\balpha$ alone moved the ratio above one as the network grew ($\approx 1.1$ at $K=30$), turning coverage conservative ($\approx 96\%$). Correcting jointly for $(\wh\balpha^\top, \wh\btau^\top)^\top$ returned the ratio to near one and brought coverage to the nominal $95\%$. Similar results were observed for the rest of the coefficients. 

\subsection{Heterogeneous missingness}
\label{sec:sim-hetero}

To isolate the calibrated weighting estimator's ability to correct site-level misspecification, we constructed a controlled setting in which the missingness mechanism differed across sites. Half the sites followed a main-effects MAR mechanism, $\Pr(R=1\mid Y,\bZ)=\text{expit}(-0.2+0.1Y+0.1Z_1+0.1Z_2)$, and the other half a MAR mechanism carrying a $YZ_1$ interaction, $\Pr(R=1\mid Y,\bZ)=\text{expit}(-0.2+0.1Y+0.05Z_1+0.05Z_2+0.1YZ_1)$. The site-specific and pooled IPW estimators used a single working model that omitted the interaction, leaving the interaction-mechanism sites misspecified. The calibrated estimator instead drew candidate models from the two largest sites---one per mechanism, with the interaction-mechanism donor including the $YZ_1$ term---so its library contained a correctly specified model for each mechanism. By simulation design, then, only the calibrated estimator could remain unbiased.

Across the four coefficients, the oracle and oracle-weighted IPW estimators were unbiased $<0\%$, and the calibrated estimator tracked them closely, with absolute percent bias of at most about $3\%$ (largest for $Z_1$). The complete case and uniform-weight estimators were biased throughout---near $113\%$ for the intercept, $65\%$ for $Z_1$, and $9$--$12\%$ for $X$ and $Z_2$---since neither addresses a mechanism that depends on $Y$. The pooled and site-specific estimators, which share a working model that omits the $YZ_1$ interaction, were biased, roughly $26$--$28\%$ for the intercept and $56$--$67\%$ for $Z_1$, while leaving $X$ and $Z_2$ near zero (Table~\ref{tab:supp-numerical-results-hetero}). The bias was flat across $K$ for every estimator, confirming that it reflects specification rather than sample size: adding sites does not remove it. By borrowing the interaction-including donor from another site, the calibrated estimator recovered all four coefficients that site-specific estimation alone could not. 

\section{Application to Intrapleural Enzyme Therapy}
\label{sec:application}

Pleural infection (PI) leads to substantial morbidity and mortality despite advances in management. For patients who fail initial therapy, intrapleural enzyme therapy (IET) using tissue plasminogen activator and DNase has been shown to improve outcomes, though heterogeneity in patient response persists. Using a multicenter cohort of patients treated with IET, we evaluate baseline albumin as a clinical factor associated with 90-day mortality. Albumin is a prognostic marker previously validated in patients with confirmed PI but not yet evaluated specifically among IET recipients \citep{white2015predicting}. 

Data for this study were obtained from two sources: a retrospective multicenter cohort conducted across 23 medical centers in the United States and the United Kingdom (2014--2020; IRB \#23-2802) hosted by the University of North Carolina (UNC) at Chapel Hill, and a cohort from five Johns Hopkins University (JHU) Health System hospitals (2018--2025; IRB \#00453058). The analytic sample comprised 1,948 individuals (UNC: 1,494; JHU: 454), of whom approximately 88\% ($n = 1{,}710$) had complete data. Albumin was missing for approximately 14\% of UNC individuals and approximately 2\% of JHU individuals. We used logistic regression to estimate the association between 90-day mortality and low albumin ($<$2.7 g/dL), adjusting for sex. The weighting model for albumin was parameterized using logistic regression with fully observed baseline covariates: age, purulence, sex, and BUN. The outcome was excluded from the weighting model since it was measured post-baseline and its inclusion could open a collider structure (Section~\ref{sec:supp-consistency-ipw}).

The JHU and UNC cohorts were broadly similar, with the majority of patients 
being male (75.8\% and 70.0\%, respectively) and presenting with low albumin (56.9\% and 49.9\%). However, 90-day mortality was slightly higher at JHU (18.3\% vs.\ 14.4\%), and JHU individuals were generally older (aged 50 or older: 59\% vs.\ 31\%). Low albumin was associated with increased odds of 90-day mortality across all estimators. Under the CC estimator, the odds ratio for albumin was 1.83 (95\% CI: 1.40, 2.38), and under both the site-specific and calibrated IPW estimators it was 1.85 (95\% CI: 1.42, 2.41)(Table~\ref{tab:unc_jhu_results}). The close agreement between the CC and IPW estimators is consistent with our theoretical results: when the probability of a complete observation does not depend on the outcome, both estimators remain consistent, and with the low missingness rate the estimated weights were close to one for most combinations, further narrowing any difference between them.

The near-identical estimates from the site-specific and calibrated IPW approaches reflect the stability of site-specific weighting in this setting. With a binary outcome and two binary covariates, only eight unique covariate profiles exist; local data at each site suffice to estimate the weighting model without the need of borrowing strength (Table~\ref{tab:jhu_unc_counts}). This is consistent with our simulation results in Section~\ref{sec:supp-simulations}, where site-specific and calibrated weights performed similarly under a shared missingness mechanism with a limited number of covariate combinations. More results can be found in Section~\ref{sec:supp-application} with a worked step-by-step process of applying the CC and IPW algorithms in Section~\ref{sec:step-by-step}.

\section{Discussion}
\label{sec:fl-discussion}

Federated learning has largely developed under the assumption of complete data \citep{jordan2019communication, duan2021leverage}. Multiple imputation can require thousands of communication rounds to converge across sites \citep{chang2020federated}, a cost that limits collaboration. This work resolves that tension by situating federated estimation with incomplete data within the $M$-estimation framework. Expressing the CC and IPW estimators as solutions to estimating equations allows both the estimator and its sandwich variance estimator to be decomposed into site-specific contributions, making explicit which summary quantities each site transmits and bounding the communications required while adjusting for missing data.

The CC estimator is consistent whenever the probability of a complete observation does not depend on $Y$---which is achievable under MCAR, many MAR, and certain MNAR mechanisms. The framing also clarifies a distinction important in applied practice, where MNAR is routinely equated with \textit{non-ignorability}  \citep{heitjan1991ignorability} and, by extension, with bias 
in analyses that involve only complete cases. What determines consistency is not the missingness label but dependence on the outcome \citep{hughes2019accounting, little2024comparison}. The classical question of \citet{heitjan1991ignorability}---when can a coarsening mechanism be ignored?---warrants revisiting here, because in the federated setting ignorability can be framed into a site-level property: mechanisms may differ across sites, and the CC estimator remains consistent so long as each site satisfies the conditions enumerated in Table~\ref{tab:missingness-mcar-mar-mnar}. We further identify when the mechanism \emph{must} be left unadjusted, rather than treated as a free choice between adjusting and not: forcing $Y$ into an unnecessary weighting model can instead introduce bias through a collider structure which our simulations show erodes coverage as the number of sites grows.

When the CC estimator is biased, we provide two IPW estimator variants that differ in their weighting approach. The site-specific approach estimates the weighting model locally, while the calibrated approach  combines candidate models across sites---allowing small sites to supplement their local weighting models with models estimated elsewhere. By situating the problem within the $M$-estimation framework, we derive a stacked sandwich variance estimator that accounts for the estimation of both the weighting model and the calibration of the weights; our simulations show that ignoring either source of estimation uncertainty fails to attain nominal coverage. We further find that the gains from calibration are bounded---pooling too many candidate models degrades the calibrated estimator---so the number of models combined must be chosen with care.

The pleural infection study, a setting with a finite number of outcome-covariate combinations, illustrates the site-level consistency conditions of Proposition~\ref{prop:fl-consistency}(i). With mortality excluded from the weighting model on temporality grounds, the CC and IPW estimators agreed, and the CC estimator was the natural choice, being one-shot. The main limitation of our $M$-estimation framework is its reliance on parametric models, which yield closed-form sandwich estimators; flexible weight models, like random forest, would require bootstrapping across the network which then restores the communication burden. Investigating strategies for selecting among the candidate weighting models---rather than pooling all available ones or those from the largest sites---is a natural extension, since redundant candidates ill-condition the calibration and bias the IPW estimator. 

\section*{Acknowledgments}

{\it Ethics}: The University of North Carolina at Chapel Hill cohort was approved under IRB \#23-2802 and the Johns Hopkins Health System cohort under IRB \#00453058. {\it Funding}: This work was supported by the Johns Hopkins Provost's Postdoctoral Fellowship Program (JEV). {\it Data Availability}: The data that support the findings of this study were obtained from the University of North Carolina at Chapel Hill and the Johns Hopkins Health System under data use agreements and are not publicly available; they may be available from the corresponding author upon permission. R code implementing the proposed estimators is available at \url{https://github.com/jesusepfvazquez/Federated-Learning-Missing-Data}. {\it Conflict of Interest}: None declared.

\printbibliography[notkeyword={bibonly}]

\end{document}


\setcounter{section}{0}
\setcounter{table}{0}
\setcounter{figure}{0}
\setcounter{algorithm}{0}
\renewcommand{\thesection}{S.\arabic{section}}
\renewcommand{\thefigure}{S.\arabic{figure}}
\renewcommand{\thetable}{S.\arabic{table}}
\renewcommand{\thealgorithm}{S.\arabic{algorithm}}
\renewcommand{\theequation}{S.\arabic{equation}}

\spacingset{1} 

\begin{center}
{\LARGE{\bf Supplementary Material to \textit{Federated Learning with Incomplete Data: When to Use Complete Cases and When to Weight}}}
\end{center}

\baselineskip=12pt

\vskip 2mm
\begin{center}
Jesus E. Vazquez$^{*1}$,
Yicheng Shen$^{1}$, 
Jason Akulian$^2$,
Chad Hochberg$^3$,
Theodore J. Iwashyna$^3$,
Elizabeth A. Stuart$^{1}$, 
Jiayi Tong$^{1}$ \\ \vspace{0.2in}
$^1$Department of Biostatistics, Johns Hopkins University, Baltimore, MD 21231\\
$^2$Division of Pulmonary and Critical Care, University of North Carolina, \\ Chapel Hill, NC 27516 \\
$^3$Pulmonary and Critical Care Medicine, Johns Hopkins University, Baltimore, MD 21231\\
\vspace{1em}
*jvazqu18@jh.edu
\end{center}

\section*{Overview}
This supplement provides: (i) algorithms for federated estimation under complete-case and inverse probability weighting (Section~\ref{sec:supp-algorithms}); (ii) proof of consistency of the CC and IPW estimators (Section~\ref{sec:supp-proposition-proof}); (iii) an alternative representation of the asymptotic normality of the estimators under nuisance parameter estimation (Section~\ref{sec:supp-normality}); (iv) additional simulation results (Section~\ref{sec:supp-simulations}); and (v) additional pleural infection application results and a step-by-step example (Sections~\ref{sec:supp-application}--\ref{sec:step-by-step}).

Throughout this \textit{Supplementary Material} we assume that data are partitioned across $K$ sites with $n = \sum_{k=1}^K n^{(k)}$ total observations. Let $Y$ denote the outcome, $X$ a covariate of interest, and $\bZ$ fully observed covariates. The observed data at site $k$ is $\bO^{(k)}$ of which $Y$, $X$, or $(Y,X)$ may be subject to missingness; let $R$ denote the indicator of complete data. We write $f_{Y|X,\bZ}(y_i,x_i,\bz_i;\btheta)$ for the conditional density of $Y$ given $(X,\bZ)$ evaluated at $(y_i,x_i,\bz_i)$. Let $v^{\otimes 2} = vv^\top$ for a column vector $v$.

\section{Algorithms for Federated Learning with Missing Data}
\label{sec:supp-algorithms}

We present algorithms for the CC and IPW estimators (site-specific and calibrated) under two estimation approaches: sufficient information for linear regression models (Algorithms~\ref{alg:fl-cc}--\ref{alg:fl-mr-ipw}) and count aggregation for generalized linear models (Algorithms~\ref{alg:fl-cc-glm}--\ref{alg:fl-ipw-glm}). Table~\ref{tab:alg-summary} summarizes the communication requirements for each. A worked example applying Algorithms~\ref{alg:fl-cc-glm}--\ref{alg:fl-ipw-glm} 
to the pleural infection data is provided in Section~\ref{sec:step-by-step}. 

\begin{table}[h]
\centering
\small
\begin{tabular}{llccc}
\toprule
Approach & Estimator & Rounds & One-shot & Algorithm \\
\midrule
\multirow{3}{*}{Sufficient information} 
  & CC & 3 & No & \ref{alg:fl-cc} \\
  & IPW (site-specific) & 3 & No & \ref{alg:fl-local-ipw} \\
  & IPW (calibrated) & 4 & No & \ref{alg:fl-mr-ipw} \\
\addlinespace
\multirow{3}{*}{Count aggregation} 
  & CC & 1 & Yes & \ref{alg:fl-cc-glm} \\
  & IPW (site-specific) & 3 & No & \ref{alg:fl-ipw-glm} \\
  & IPW (calibrated) & 4 & No & \ref{alg:fl-ipw-glm} \\
\bottomrule
\end{tabular}
\caption{\textbf{Summary of algorithm communication requirements.} ``One-shot" indicates whether both point and variance estimation can be completed in a single round of communication. Under count aggregation, the CC estimator achieves one-shot estimation because variance components can be computed from the transmitted summaries; IPW estimators require additional rounds for weight estimation and sharing variance components.}
\label{tab:alg-summary}
\end{table}

\subsection{Sufficient information approach}

Under a linear regression model with normally distributed errors, the CC and IPW estimators can be computed via the sufficient information approach. Each site transmits (weighted) cross-product matrices, and the coordinating site aggregates these to obtain $\wh\btheta$ via ordinary least squares (OLS) or weighted OLS. Variance estimation requires an additional round in which sites compute local contributions to the sandwich estimator after receiving $\wh\btheta$. This yields a three-round algorithm for CC (Algorithm~\ref{alg:fl-cc}) and site-specific IPW (Algorithm~\ref{alg:fl-local-ipw}), and a four-round algorithm for calibrated IPW (Algorithm~\ref{alg:fl-mr-ipw}), which requires an additional round for sharing nuisance parameter estimates across sites.

\begin{algorithm}
\caption{Complete Case Estimator via Sufficient Information for Federated Learning}
\label{alg:fl-cc}
\begingroup
\small
\setlength{\baselineskip}{11pt}
\begin{algorithmic}

\Statex \textbf{Input:} $K$ sites with observed data $\bO^{(k)}=(Y^{(k)},R^{(k)} X^{(k)},R^{(k)},\bZ^{(k)})$
\Statex \textbf{Output:} $\wh\btheta=(\wh\bbeta,\wh\sigma)$ and $\wh{\mathrm{Var}}(\wh\btheta)$

\Statex
\Statex \textbf{Step 1: Compute sufficient statistics}
\For{each site $k=1,\ldots,K$}
    \State Let $X_R^{(k)} = \diag\{R^{(k)}\}X^{(k)}$
    \State Transmit $\{X_R^{(k)},\bZ^{(k)}\}^{\top} \diag\{R^{(k)}\} \{X_R^{(k)},\bZ^{(k)}\}$ and $\{X_R^{(k)},\bZ^{(k)}\}^{\top} \diag\{R^{(k)}\} Y^{(k)}$
\EndFor

\Statex
\Statex \underline{Coordinating site:} Estimate $\bbeta$
\State Compute $\wh\bbeta_\text{CC} = \left[\sum_{k=1}^{K}\{X_R^{(k)},\bZ^{(k)}\}^{\top} \diag\{R^{(k)}\} \{X_R^{(k)},\bZ^{(k)}\}\right]^{-1} \left[\sum_{k=1}^{K} \{X_R^{(k)},\bZ^{(k)}\}^{\top} \diag\{R^{(k)}\} Y^{(k)} \right]$

\Statex
\Statex \textbf{Step 2: Estimate $\sigma$}
\State Coordinating site distributes $\wh\bbeta_\text{CC}$ to all sites
\For{each site $k=1,\ldots,K$}
    \State Compute $\text{RSS}^{(k)} = \left[Y^{(k)}-\{X_R^{(k)},\bZ^{(k)}\} \wh\bbeta_\text{CC} \right]^\top \diag\{R^{(k)}\} \left[Y^{(k)}-\{X_R^{(k)},\bZ^{(k)}\} \wh\bbeta_\text{CC} \right]$
    \State Transmit $\text{RSS}^{(k)}$ and $n^{(k)}_R$ is equal to the number of complete cases.
\EndFor

\Statex
\Statex \textbf{Coordinating site: Estimate $\sigma^2$}
\State Compute $\wh\sigma^2 = \left(\sum_{k=1}^{K} \text{RSS}^{(k)}\right) / \left(\sum_{k=1}^{K} n^{(k)}_{R} - p\right)$. Take square-root to obtain  $\wh\sigma$

\Statex
\Statex \textbf{Step 3: Variance estimation}
\State Coordinating site distributes $\wh\btheta_\text{CC} = (\wh\bbeta_\text{CC}^\top, \wh\sigma)^\top$ to all sites
\For{each site $k=1,\ldots,K$}
    \State Compute local contributions $\wh\bA_\text{CC}^{(k)}$ and $\wh\bB_\text{CC}^{(k)}$ as in Section 4.3
    \State Transmit $\wh\bA_\text{CC}^{(k)}$ and $\wh\bB_\text{CC}^{(k)}$
\EndFor

\Statex
\Statex \underline{Coordinating site:} Compute robust sandwich variance estimator
\State Aggregate $\wh\bA_\text{CC} = \sum_{k=1}^{K} \wh\bA_\text{CC}^{(k)}$ and $\wh\bB_\text{CC} = \sum_{k=1}^{K} \wh\bB_\text{CC}^{(k)}$
\State Compute $\wh{\mathrm{Var}}(\wh\btheta_\text{CC}) = \wh\bA_\text{CC}^{-1} \wh\bB_\text{CC} \wh\bA_\text{CC}^{-\top}$

\Statex
\State \Return $\wh\btheta_\text{CC}$ and $\wh{\mathrm{Var}}(\wh\btheta_\text{CC})$

\end{algorithmic}
\endgroup
\end{algorithm}

\clearpage

\begin{algorithm}
\caption{Site-specific Inverse Probability Weighting Estimator via Sufficient Information for Federated Learning}
\label{alg:fl-local-ipw}
\begingroup
\small
\setlength{\baselineskip}{11pt}
\begin{algorithmic}

\Statex \textbf{Input:} $K$ sites with observed data $\bO^{(k)}=(Y^{(k)},R^{(k)} X^{(k)},R^{(k)},\bZ^{(k)})$
\Statex \textbf{Output:} $\wh\btheta=(\wh\bbeta,\wh\sigma)$ and $\wh{\mathrm{Var}}(\wh\btheta)$

\Statex
\Statex \textbf{Step 1: Site-specific nuisance estimation and weighted sufficient statistics}
\For{each site $k=1,\ldots,K$}
    \State Solve the local nuisance estimating equation for $\wh\balpha_k$
    \State Let $X_R^{(k)} = \diag\{R^{(k)}\}X^{(k)}$
    \State Let $\bW^{(k)} = \diag\{r^{(k)}_i/\pi_{Y,X,Z}(y_i^{(k)}, r_i^{(k)} x_i^{(k)}, \bz_i^{(k)})\}$
    \State Transmit $\{X_R^{(k)},\bZ^{(k)}\}^{\top} \bW^{(k)} \{X_R^{(k)},\bZ^{(k)}\}$ and $\{X_R^{(k)},\bZ^{(k)}\}^{\top} \bW^{(k)} Y^{(k)}$
\EndFor

\Statex
\Statex \underline{Coordinating site:} Estimate $\bbeta$
\State Compute $\wh\bbeta_{\text{IPW}} = \left[\sum_{k=1}^{K}\{X_R^{(k)},\bZ^{(k)}\}^{\top} \bW^{(k)} \{X_R^{(k)},\bZ^{(k)}\}\right]^{-1} \left[\sum_{k=1}^{K}\{X_R^{(k)},\bZ^{(k)}\}^{\top} \bW^{(k)} Y^{(k)}\right]$

\Statex
\Statex \textbf{Step 2: Estimate $\sigma$}
\State Coordinating site distributes $\wh\bbeta_{\text{IPW}}$ to all sites
\For{each site $k=1,\ldots,K$}
    \State Compute $\text{RSS}^{(k)} = \left[Y^{(k)}-\{X_R^{(k)},\bZ^{(k)}\} \wh\bbeta_{\text{IPW}} \right]^\top \bW^{(k)} \left[Y^{(k)}-\{X_R^{(k)},\bZ^{(k)}\} \wh\bbeta_{\text{IPW}} \right]$
    \State Transmit $\text{RSS}^{(k)}$ and $n^{(k)}_R$ is equal to the number of complete cases.
\EndFor

\Statex
\Statex \underline{Coordinating site:} Estimate $\sigma^2$
\State Compute $\wh\sigma^2 = \left(\sum_{k=1}^{K} \text{RSS}^{(k)}\right) / \left(\sum_{k=1}^{K} n^{(k)}_{R} - p\right)$, where $p$ is the length of $\wh\bbeta$. Take square root to obtain $\wh\sigma$

\Statex
\Statex \textbf{Step 3: Variance estimation}
\State Coordinating site distributes $\wh\btheta_{\text{IPW}} = (\wh\bbeta_{\text{IPW}}^\top, \wh\sigma)^\top$ to all sites.
\For{each site $k=1,\ldots,K$}
    \State Compute local block contributions: $\wh\bA^{(k)}_{\btheta\btheta}$, $\wh\bA^{(k)}_{\btheta\balpha_k}$, $\wh\bA^{(k)}_{\balpha_k\balpha_k}$, $\wh\bB^{(k)}_{\btheta\btheta}$, $\wh\bB^{(k)}_{\btheta\balpha_k}$, $\wh\bB^{(k)}_{\balpha_k\balpha_k}$ as in Section 4.3
    \State Transmit these block matrices to the coordinating site
\EndFor

\Statex
\Statex \underline{Coordinating site:} Assemble and compute robust sandwich variance estimator
\State Aggregate $\btheta$-only blocks: $\wh\bA_{\btheta\btheta} = \sum_{k=1}^{K} \wh\bA^{(k)}_{\btheta\btheta}$ and $\wh\bB_{\btheta\btheta} = \sum_{k=1}^{K} \wh\bB^{(k)}_{\btheta\btheta}$
\State Assemble $\wh\bA_{\text{Stacked}}$ and $\wh\bB_{\text{Stacked}}$ by placing site-specific nuisance blocks $\wh\bA^{(k)}_{\balpha_k\balpha_k}$, $\wh\bA^{(k)}_{\btheta\balpha_k}$, $\wh\bB^{(k)}_{\balpha_k\balpha_k}$, $\wh\bB^{(k)}_{\btheta\balpha_k}$ in their respective positions
\State Compute $\wh{\mathrm{Var}}(\wh\bxi_\text{IPW}) =  \wh\bA_{\text{Stacked}}^{-1} \wh\bB_{\text{Stacked}} \wh\bA_{\text{Stacked}}^{-\top}$

\Statex
\State \Return $\wh\btheta_{\text{IPW}}$ and $\wh{\mathrm{Var}}(\wh\btheta_\text{IPW})$; equal to the upper-left $(p+1) \times (p+1)$ block of $\wh{\mathrm{Var}}(\wh\bxi_\text{IPW})$

\end{algorithmic}
\endgroup
\end{algorithm}

\clearpage

\begin{algorithm}
\caption{Calibrated Inverse Probability Weighting Estimator via Sufficient Information for Federated Learning}
\label{alg:fl-mr-ipw}
\begingroup
\footnotesize
\setlength{\baselineskip}{11pt}
\begin{algorithmic}

\Statex \textbf{Input:} $K$ sites with observed data $\bO^{(k)}=(Y^{(k)},R^{(k)} X^{(k)},R^{(k)},\bZ^{(k)})$
\Statex \textbf{Output:} $\wh\btheta=(\wh\bbeta,\wh\sigma)$ and $\wh{\mathrm{Var}}(\wh\btheta)$
\Statex
\Statex \textbf{Step 1: Site-specific nuisance estimation}
\For{each site $k=1,\ldots,K$}
    \State Solve the local nuisance estimating equation for $\wh\balpha_k$ and transmit $\wh\balpha_k$ to the coordinating site
\EndFor
\Statex
\Statex \textbf{Step 2: Distribute nuisance parameters and compute calibrated weights}
\State Coordinating site distributes $\wh\balpha=(\wh\balpha_1,\ldots,\wh\balpha_J)$ to all sites
\For{each site $k=1,\ldots,K$}
    \State Let $X_R^{(k)} = \diag\{R^{(k)}\}X^{(k)}$ and $\bgamma_i^{(k)} = \left(\pi_{Y,X,Z}(\cdot; \wh\balpha_1), \ldots, \pi_{Y,X,Z}(\cdot; \wh\balpha_J)\right)^\top$, the candidate probabilities for observation $i$
    \State Compute $\wh\btau_k = \left[\sum_{i=1}^{n^{(k)}} \bgamma_i^{(k)} \bgamma_i^{(k)\top}\right]^{-1} \left[\sum_{i=1}^{n^{(k)}} \bgamma_i^{(k)} r_i^{(k)}\right]$ and retain it for variance estimation
    \State Let $\pi_i^{*(k)} = \wh\btau_k^{\top} \bgamma_i^{(k)}$ and $\bW^{(k)} = \diag\{r_i^{(k)}/\pi_i^{*(k)}\}$
\EndFor
\State \underline{Coordinating site:} $\wh\bbeta_{\text{IPW}} = \left[\sum_{k=1}^{K}\{X_R^{(k)},\bZ^{(k)}\}^{\top} \bW^{(k)} \{X_R^{(k)},\bZ^{(k)}\}\right]^{-1} \left[\sum_{k=1}^{K}\{X_R^{(k)},\bZ^{(k)}\}^{\top} \bW^{(k)} Y^{(k)}\right]$
\Statex
\Statex \textbf{Step 3: Estimate $\sigma$}
\State Coordinating site distributes $\wh\bbeta_{\text{IPW}}$ to all sites
\For{each site $k=1,\ldots,K$}
    \State Compute $\text{RSS}^{(k)} = \left[Y^{(k)}-\{X_R^{(k)},\bZ^{(k)}\} \wh\bbeta_{\text{IPW}} \right]^\top \bW^{(k)} \left[Y^{(k)}-\{X_R^{(k)},\bZ^{(k)}\} \wh\bbeta_{\text{IPW}} \right]$ and transmit $\text{RSS}^{(k)}$ and $n^{(k)}_R$, the number of complete cases
\EndFor
\State \underline{Coordinating site:} $\wh\sigma^2 = \left(\sum_{k=1}^{K} \text{RSS}^{(k)}\right) / \left(\sum_{k=1}^{K} n^{(k)}_{R} - p\right)$, $\ p = \dim(\wh\bbeta)$; set $\wh\sigma=\sqrt{\wh\sigma^2}$
\Statex
\Statex \textbf{Step 4: Variance estimation}
\State Coordinating site distributes $\wh\btheta_{\text{IPW}} = (\wh\bbeta_{\text{IPW}}^\top, \wh\sigma)^\top$ to all sites
\For{each site $k=1,\ldots,K$}
    \State Compute the local sandwich blocks of $\wh\bA^{(k)}$ and $\wh\bB^{(k)}$ for $(\btheta,\balpha_1,\ldots,\balpha_J,\btau_k)$ following Section~\ref{sec:supp-variance-calibrated}, and transmit them
\EndFor
\State \underline{Coordinating site:} assemble $\wh\bA_{\text{Stacked}}$ and $\wh\bB_{\text{Stacked}}$---summing the cross-site $\btheta$- and $\balpha_j$-blocks and placing the site-specific $\balpha_j$- and $\btau_k$-blocks (Section~\ref{sec:supp-variance-calibrated})
\State Compute $\wh{\mathrm{Var}}(\wh\bxi) = \wh\bA_{\text{Stacked}}^{-1} \wh\bB_{\text{Stacked}} \wh\bA_{\text{Stacked}}^{-\top}$, $\ \bxi = (\btheta^\top, \balpha_1^\top,\ldots,\balpha_J^\top, \btau_1^\top,\ldots,\btau_K^\top)^\top$
\State \Return $\wh\btheta_{\text{IPW}}$ and $\wh{\mathrm{Var}}(\wh\btheta_\text{IPW})$, the upper-left $(p+1)\times(p+1)$ block of $\wh{\mathrm{Var}}(\wh\bxi)$

\end{algorithmic}
\endgroup
\end{algorithm}

\subsection{Count aggregation approach}

When a GLM is of interest and the outcome-covariate space is discrete with a finite number of unique combinations, the count aggregation approach enables a more communication-efficient estimation. Under this approach, each site identifies the unique outcome-covariate combinations $u_j = (y_j, x_j, \bz_j)$ among complete cases and transmits these combinations along with their corresponding (weighted) counts.

For the CC estimator, count aggregation yields a \emph{one-shot} algorithm (Algorithm~\ref{alg:fl-cc-glm}): once the coordinating site receives the combinations and counts, it can solve the estimating equation \emph{and} compute the variance estimate without additional communication rounds. This is because the variance components $\wh\bA_{\text{CC}}$ and $\wh\bB_{\text{CC}}$ depend only on the score function $\bS_\btheta(u_j; \wh\btheta)$ and its derivatives, which can be evaluated from the transmitted counts. For the IPW estimators, count aggregation reduces communication but does not achieve one-shot estimation (Algorithm~\ref{alg:fl-ipw-glm}). Weight estimation must occur before aggregation, and for calibrated weights, an additional round is needed to share nuisance parameter estimates. Variance estimation for IPW requires the stacked $M$-estimating equation components, which depend on both the outcome model and the nuisance model; these cannot be fully reconstructed from the aggregated counts alone and require a separate communication rounds.

We note that under count aggregation, combinations with fewer than $T$ observations (typically $T = 6$ or $T = 11$) must be suppressed or perturbed to comply with disclosure control policies \citep{malpetti2025technical}; we recommend readers to check with their institutions before sharing any count data. This constraint may limit applicability when the covariate space is large relative to site sample size(s). A worked example using pleural infection data is provided in Section~\ref{sec:step-by-step} showcasing the information transmitted to each of the sites.

\begin{algorithm}
\caption{Complete Case via Count Aggregation for Federated Learning}
\label{alg:fl-cc-glm}
\begingroup
\small
\setlength{\baselineskip}{11pt}
\begin{algorithmic}
\Statex \textbf{Input:} $K$ sites with observed data $\bO^{(k)}=(Y^{(k)}, R^{(k)} X^{(k)}, R^{(k)}, \bZ^{(k)})$; minimum cell threshold $T$.
\Statex \textbf{Output:} $\wh\btheta_{\text{CC}}$ and $\wh{\mathrm{Var}}(\wh\btheta_{\text{CC}})$
\Statex
\Statex \textbf{Step 1: Aggregation and estimation}
\For{each site $k=1,\ldots,K$}
    \State Among complete cases ($r_i^{(k)}=1$), identify unique combinations $u_j = (y_j, x_j, \bz_j)$.
    \State Compute counts $n_j^{(k)} = \sum_{i \in u_j} 1$ for each combination $j$.
    \State Suppress combinations with $n_j^{(k)} < T$ per cell suppression policy.
    \State Transmit each combination and corresponding count $(u_j, n_j^{(k)})$ to coordinating site.
\EndFor

\Statex
\State \underline{Coordinating site:}
\State Solves $\sum_{k=1}^{K} \sum_{j=1}^{q^{(k)}} w_j^{(k)} \bS_\btheta(u_j; \btheta) = \bzero$  where $w_j^{(k)} = n_j^{(k)}$. 
\State Compute $\wh\bA_{\text{CC}}$, $\wh\bB_{\text{CC}}$ evaluated at $\wh\btheta_{\text{CC}}$ using aggregated counts.
\State  Evaluate $\wh{\mathrm{Var}}(\wh\btheta_{\text{CC}}) = \wh\bA_{\text{CC}}^{-1} \wh\bB_{\text{CC}} \wh\bA_{\text{CC}}^{-\top}$.
\Statex
\State \Return $\wh\btheta_{\text{CC}}$, $\wh{\mathrm{Var}}(\wh\btheta_{\text{CC}})$
\end{algorithmic}
\endgroup
\end{algorithm}

\begin{algorithm}
\caption{Inverse Probability Weighting Estimator (Site-Specific or Calibrated Weights) via Count Aggregation for Federated Learning}
\label{alg:fl-ipw-glm}
\begingroup
\small
\setlength{\baselineskip}{11pt}
\begin{algorithmic}
\Statex \textbf{Input:} $K$ sites with observed data $\bO^{(k)}$; weight strategy (site-specific or calibrated); minimum cell threshold $T$.
\Statex \textbf{Output:} $\wh\btheta_{\text{IPW}}$ and $\wh{\mathrm{Var}}(\wh\btheta_{\text{IPW}})$
\Statex
\Statex \textbf{Step 1: Weight estimation}
\For{each site $k=1,\ldots,K$}
    \State Estimate local nuisance parameters $\wh\balpha_k$ via logistic regression on $(R^{(k)}, Y^{(k)}, \bZ^{(k)})$.
\EndFor
\If{strategy = calibrated}
    \State Selected sites broadcast $\wh\balpha_j$ to all sites. \textbf{Note that this incurs in an additional round.}
    \State Each site constructs calibrated weights $\pi^*_i$ as in Section~\ref{sec:fl-mr}.
\EndIf
\Statex
\Statex \textbf{Step 2: Aggregation and estimation}
\For{each site $k=1,\ldots,K$}
    \State Among complete cases, identify unique combinations $u_j = (y_j, x_j, \bz_j)$.
    \State Compute $w_j^{(k)} = \sum_{i \in u_j} 1/\wh\pi_i^{(k)}$ for each combination.
    \State Suppress combinations with fewer than $T$ observations per cell suppression policy.
    \State Transmit each combination and corresponding weight $(u_j, w_j^{(k)})$ to coordinating site.
\EndFor
\Statex 
\State \underline{Coordinating site:} Solve $\sum_{k=1}^{K} \sum_{j=1}^{q^{(k)}} w_j^{(k)} \bS_\btheta(u_j; \btheta) = \bzero$.
\Statex

\Statex \textbf{Step 3: Variance estimation}
\State Coordinating site distributes $\wh\btheta_{\text{IPW}}$ to all sites
\For{each site $k=1,\ldots,K$}
    \State Compute $\wh\bA^{(k)}_{\btheta\btheta}$, $\wh\bA^{(k)}_{\btheta\balpha_j}$, the nuisance diagonal for any model site $k$ contributed, and the analogous $\wh\bB^{(k)}$ blocks
    \If{strategy = calibrated}
        \State Additionally compute $\wh\bA^{(k)}_{\btheta\btau_k}$, $\wh\bA^{(k)}_{\btau_k\balpha_j}$ for $j=1,\ldots,J$, $\wh\bA^{(k)}_{\btau_k\btau_k}$, and the analogous $\wh\bB^{(k)}$ blocks (Section~\ref{sec:supp-variance-calibrated})
    \EndIf
    \State Transmit these block matrices to the coordinating site
\EndFor
\Statex 
\State \underline{Coordinating site:} Assemble $\wh\bA_{\text{Stacked}}$ and $\wh\bB_{\text{Stacked}}$---summing cross-site blocks and placing site-specific blocks---and compute $\wh{\mathrm{Var}}(\wh\bxi) = \wh\bA_{\text{Stacked}}^{-1} \wh\bB_{\text{Stacked}} \wh\bA_{\text{Stacked}}^{-\top}$
\Statex
\State \Return $\wh\btheta_{\text{IPW}}$ and $\wh{\mathrm{Var}}(\wh\btheta_\text{IPW})$, the corresponding upper-left block of $\wh{\mathrm{Var}}(\wh\bxi)$
\Statex
\end{algorithmic}
\endgroup
\end{algorithm}

\clearpage

\section{Proof of Proposition 1}
\label{sec:supp-proposition-proof}

\subsection{Assumptions}
\label{sec:supp-assumptions}

Throughout, we work with data partitioned across $K$ sites, where site $k$ 
contributes $n^{(k)}$ observations independently sampled from a 
site-specific distribution. We assume:

\begin{itemize}

\item[(A1)] \textbf{Within-site independence:} At each site $k$, 
observations $\bO_i^{(k)}$, $i=1,\ldots,n^{(k)}$, are independent and 
identically distributed (iid).

\item[(A2)] \textbf{Across-site independence:} Observations at site $k$ 
are independent of observations at site $k'$ for $k \ne k'$.

\item[(A3)] \textbf{Positivity in complete cases:} At each site $k$, 
$\Pr(R=1) > 0$, ensuring that complete cases exist at every site.

\item[(A4)] \textbf{Positivity (IPW):} At each site $k$, there exists a 
constant $c > 0$ such that the oracle probability of a complete observation 
satisfies $\Pr(R=1 \mid Y, X, \bZ) \geq c$ almost surely. The same bound is required almost surely of the estimated probabilities: 
for the site-specific approach, $\wh{\pi}(\cdot; \wh{\balpha}_k) 
\geq c$; for the calibrated approach, $\pi^*(\cdot; \wh{\balpha}) \geq c$.

\item[(A5)] \textbf{Regularity for $M$-estimation:} The parameter space 
$\bXi$ is compact; $\bxi_0 = (\btheta_0^\top, \balpha_0^\top)^\top \in 
\bXi$ is the unique solution to $E\{\bPhi_\text{Stacked}(\bO; \bxi_0)\} 
= \bzero$; the estimating function $\bPhi_\text{Stacked}(\bO; \bxi)$ is 
twice continuously differentiable in $\bxi$ with integrable derivatives; 
and \newline $E\{\partial \bPhi_\text{Stacked}(\bO; \bxi)/\partial \bxi^\top \mid_{\bxi=\bxi_0}\}$ 
is nonsingular. For the CC estimator, $\bxi = \btheta$ and A5 reduces to 
the standard regularity conditions for $M$-estimation.

\end{itemize}

Assumptions (A1)--(A2) formalize the federated data structure: observations within each site are iid draws from a site-specific distribution, and sites contribute data independently of one another. The site-specific distributions share a common outcome model $E_{Y \mid X, \bZ}\{g(Y)\} = m(X, \bZ; \btheta_0)$ but may differ in their covariate distributions and missingness mechanisms, reflecting the heterogeneity typical of multi-site studies. Assumption (A3) ensures the estimating equations for the CC and IPW estimators are well-defined at each site (i.e., sufficient complete cases). Assumption (A4) prevents inverse probability weights from becoming unbounded, which is required for both valid estimation and asymptotic inference. Assumption (A5) assume standard regularity 
conditions for $M$-estimation \citep{tsiatis2006semiparametric}; these are generally 
satisfied when parametric models are used for both the outcome model and the weighting model, as assumed.

\subsection{Overview of the proof for Proposition~\ref{prop:fl-consistency}}
\label{sec:proof-overview}

For reference, we restate the proposition below.

\begin{Prop}
\label{prop:fl-consistency-supp}
Assume that $Y$, $X$, or $(Y,X)$ are subject to missingness but $\bZ$ is 
fully observed, and that standard regularity conditions for M-estimation hold 
(Supplementary Material, Sections~\ref{sec:supp-assumptions}). 
Under the site-level conditions below, the federated estimator is consistent. 
\begin{enumerate}
    \item[(i)] The complete case estimator is consistent if, at each site, 
    the oracle probability of a complete observation does not depend on 
    the outcome $Y$. This condition may be satisfied by different 
    missingness mechanisms across sites---including MCAR, MAR, and certain 
    MNAR mechanisms---so long as it holds site-by-site 
    (Table~\ref{tab:missingness-mcar-mar-mnar}). 
    \item[(ii)] The inverse probability weighting estimator is consistent if,  at each site, the weighting model is correctly specified. When the oracle probability does not depend on $Y$, correct specification is not required, but $Y$ must be excluded from the weighting model when $X$ is missing to avoid bias.
    \item[(iii)] The calibrated inverse probability weighting estimator is consistent if, at each site, at least one of the $J$ candidate weighting models is correctly specified. The correctly specified model need not be the one estimated at that site.
\end{enumerate}
\end{Prop}

The proof proceeds by establishing that if certain conditions hold at each individual site, the federated estimator inherits consistency. The key step is recognizing that both the CC and IPW estimating equations decompose as sums of site-specific contributions
\begin{equation}
\label{eq:site-decomposition}
\sum_{i=1}^n \bPhi(\bO_i;\bxi_0) = \sum_{k=1}^K \underbrace{\sum_{i=1}^{n^{(k)}} \bPhi^{(k)}(\bO_i^{(k)};\bxi_0)}_{\text{site $k$ contribution}},
\end{equation}
where $\bxi_0 = \btheta_0$ for the CC estimator. Under assumptions (A1)--(A2), site-specific contributions are independent across sites. If each site-specific contribution has expectation zero at the true parameter value $\bxi_0$, then linearity of expectation implies that the expectation of the sum also equals zero:
\begin{equation}
\label{eq:linearity-expectation}
E\left\{ \sum_{k=1}^K \sum_{i=1}^{n^{(k)}} \bPhi^{(k)}(\bO_i^{(k)};\bxi_0) \right\}
= \sum_{k=1}^K E\left\{ \sum_{i=1}^{n^{(k)}} \bPhi^{(k)}(\bO_i^{(k)};\bxi_0) \right\}
= \sum_{k=1}^K \bzero = \bzero.
\end{equation}
Therefore, \emph{consistency at the individual site level is sufficient for consistency of the federated estimator}. The remainder of this section establishes conditions under which each site-specific contribution has mean zero. The proof proceeds in three parts. First, in Section~\ref{sec:supp-pr-complete}, we define what the probability of a complete observation is, which is then used to establish the consistency of the CC and IPW estimators. Then, in Sections~\ref{sec:supp-consistency-cc} and~\ref{sec:supp-consistency-ipw}, we establish the consistency of the CC and IPW estimators, respectively.

\begin{remark}[Site-level conditions are sufficient but not necessary]
\label{rem:site-sufficient}
The conditions in Proposition~\ref{prop:fl-consistency-supp} are \emph{sufficient} for consistency of the federated estimator, but not necessary. It is possible for site-specific biases could cancel in aggregate, yielding a consistent pooled estimator even when individual sites are biased. However, such cancellation would require precise knowledge of the bias structure (how and where it arises) across sites and is difficult to verify from the observed data alone. We therefore focus on site-level conditions because they are they are robust to changes in the composition or number of participating sites in the network.
\end{remark}

\subsubsection{Probability of a complete observation}
\label{sec:supp-pr-complete}

How we define the missingness mechanism (i.e., MCAR, MAR, MNAR) depends on both (i) what the probability of a complete observation depends on and (ii) the missing variable(s). For example, when missingness is only on the outcome $Y$, our observed data is $\bO=(RY,R,X,\bZ)$ with the observed data density defined by $f_{RY,R,X,\bZ}$. When the missingness mechanism is MCAR for missingness in $Y$, the probability of missingness can be defined as 
\bse
&& E_{RY, R, X, \bZ} (R) \\
&& =\int r f_{RY,Y,X,\bZ} (ry,y,x,\bz) \, dry \, dr \, dx \, d\bz \\
&& = \int r \left\{ \int f_{Y,R,X,\bZ} (y,r,x,\bz) \, dr \right\}^r \left\{ \int f_{Y,R,X,\bZ} (y,r,x,\bz) \, dy \right\}^{1-r} d ry \, dr \, dx \, d\bz. 
\ese
The first part is the contribution in which we observe $Y$ and the second is the contribution in which we do not observe $Y$. Then it follows that after we integrate over $R$ we obtain an expectation with respect to the full $(Y,R,X,\bZ)$ distribution
\bse
E_{RY, R, X, \bZ} (R) &=& \int I(R=1) f_{Y,R,X,\bZ} (y,r,x,\bz) \, dr \, dy \, dx \, d\bz \\
&=& \int I(R=1) f_{R|Y,X,\bZ} (r,y,x,\bz) f_{Y,X,\bZ} (y,x,\bz) \, dr \, dy \, dx \, d\bz \\
&=& \int I(R=1) f_{R} (r) \, dr \int f_{Y,X,\bZ} (y,x,\bz) \, dy \, dx \, d\bz \\
&=& \int I(R=1) f_{R} (r) \, dr \\
&=& \Pr(R=1),
\ese
where the second to third line follows from the definition of MCAR, which states that $R \independent (Y,X,\bZ)$. We generalize this argument to define the probability of a complete observation depending on (i) what explains the missingness and (ii) whether the missingness is only $Y$, $X$, or both. Through this process we obtain Table~\ref{tab:missingness-mcar-mar-mnar}, which enumerates various forms of MCAR, MAR, and MNAR that depend on the missing variable(s).

\subsection{Consistency of the complete case estimator}
\label{sec:supp-consistency-cc}

Having defined the probability of a complete observation in Table~\ref{tab:missingness-mcar-mar-mnar}, we now formally establish when the CC estimator is consistent. In the absence of missingness, consistency follows directly since $E_{Y|X,\bZ}\{\bS_{\btheta}(Y,X,\bZ;\btheta)\} = \bzero$. Using a similar argument as \cite{tsiatis2006semiparametric}, this statement follows since $\int f_{Y|X,\bZ}(y,x,\bz;\btheta)\,dy=1$ for all $(x,\bz)$, and differentiating this identity with respect to $\btheta^\top$ yields 
\bse
\bzero &=& \partial /\partial \btheta^\top \int  f_{Y|X,\bZ}(y,x,\bz;\btheta) \, dy \\
&=& \int \partial /\partial \btheta^\top  f_{Y|X,\bZ}(y,x,\bz;\btheta) \, dy \\
&=& \int \left\{ \partial /\partial \btheta^\top \log f_{Y|X,\bZ}(y,x,\bz;\btheta) \right\}  f_{Y|X,\bZ}(y,x,\bz;\btheta) \, dy \\
&=& \int \bS_{\btheta}(y,x,\bz;\btheta) f_{Y|X,\bZ}(y,x,\bz;\btheta) \, dy  \\
&=& E_{Y|X,\bZ}\{\bS_{\btheta}(Y,X,\bZ;\btheta)\}.
\ese
With this property, we show that the CC estimator is consistent as long as the probability of observing a complete case does not depend on $Y$. Let $\pi(\cdot)$ denote the probability of a complete observation, which in this case may depend on $X$, $\bZ$, or both, but not on $Y$. That is, we assume
\[
\Pr(R=1 \mid Y, X, \bZ) = \pi(X, \bZ),
\]
for some function $\pi$ that does not involve $Y$. Under this condition, the CC estimator is consistent since
\bse
E \{R\bS_{\btheta}(Y,X,\bZ;\btheta)\} 
&=& E_{Y,X,\bZ} \{ \Pr(R=1|Y,X,\bZ) \bS_{\btheta}(Y,X,\bZ;\btheta)\}  \\
&=& E_{Y,X,\bZ} \{ \pi(X,\bZ) \bS_{\btheta}(Y,X,\bZ;\btheta) \}  \\
&=& E_{X,\bZ}\left[\pi(X,\bZ) \, E_{Y|X,\bZ} \{ \bS_{\btheta}(Y,X,\bZ;\btheta) \}\right] \\
&=& E_{X,\bZ}\{\pi(X,\bZ) \times \bzero \} \\
&=& \bzero,
\ese
where the fourth equality follows from $E_{Y|X,\bZ}\{\bS_{\btheta}(Y,X,\bZ;\btheta)\} = \bzero$. The \textit{key step} here is that the missingness probability $\pi(X,\bZ)$ can be factored out of the conditional expectation over $Y$ because it does not depend on $Y$. This result holds regardless of the missingness mechanism---MCAR, MAR, or MNAR---provided the probability of a complete case is independent of $Y$. 

\subsubsection{Site-specific consistency implies federated consistency}

The preceding argument establishes consistency for a single sample. In the federated setting, the CC estimating equation decomposes as a sum of $K$ site-specific contributions. For the missing covariate problem, the CC estimator can be decomposed as
\bse
\sum_{i=1}^n \bPhi_\text{CC}(\bO_i;\btheta_0) 
&=& \sum_{k=1}^K \underbrace{\sum_{i=1}^{n^{(k)}} r_i^{(k)} \bS_\btheta(y_i^{(k)}, r_i^{(k)} x_i^{(k)}, \bz_i^{(k)};\btheta_0)}_{\text{site $k$ contribution}}=\bzero.
\ese
Suppose that at each site $k$, the oracle probability of a complete observation does not depend on $Y$---that is, $\pi_{X,\bZ}(x^{(k)}, \bz^{(k)})$, $\pi_{\bZ}(\bz^{(k)})$, $\pi_{X}(x^{(k)})$, or some random function $\pi(\cdot)$ under MCAR. The argument above then implies that each site-specific contribution has expectation $\bzero$:
\bse
E\left\{ \sum_{i=1}^{n^{(k)}} r_i^{(k)} \bS_\btheta(y_i^{(k)}, r_i^{(k)} x_i^{(k)}, \bz_i^{(k)};\btheta_0) \right\} = \bzero, \quad k = 1, \ldots, K.
\ese
Under assumptions (A1)--(A2), the site-specific contributions are independent across sites. By linearity of expectation, the expectation of the sum equals the sum of the expectations:
\bse
E\left\{ \sum_{k=1}^K \sum_{i=1}^{n^{(k)}} r_i^{(k)} \bS_\btheta(\cdot;\btheta_0) \right\}
&=& \sum_{k=1}^K E\left\{ \sum_{i=1}^{n^{(k)}} r_i^{(k)} \bS_\btheta(\cdot;\btheta_0) \right\} \\
&=& \sum_{k=1}^K \bzero \\
&=& \bzero.
\ese
Therefore, if the site-level condition holds at every site---possibly under different missingness mechanisms across sites---the federated CC estimator is consistent. Importantly, this result does \emph{not} require that all sites share the same missingness mechanism; each site need only satisfy the condition that its probability of a complete observation is independent of $Y$.

\subsection{Consistency of the inverse probability weighting estimator}
\label{sec:supp-consistency-ipw}

The proof of consistency of the IPW estimator proceeds in two parts, corresponding to the two cases in Proposition~\ref{prop:fl-consistency-supp}(ii):
\begin{enumerate}
    \item[(i)] When the oracle probability of a complete observation 
    depends on $Y$, the IPW estimator is consistent if the weighting 
    model is correctly specified at each site.
    \item[(ii)] When the oracle probability does not depend on $Y$, 
    correct specification is not required; however, when $X$ is missing, 
    $Y$ must be excluded from the weighting model to avoid bias.
\end{enumerate}
As with the CC estimator, site-level consistency of each contribution 
implies consistency of the federated IPW estimator under 
assumptions~(A1)--(A2) and~(A5).

\subsubsection{Case (i): Probability of a complete observation depends on $Y$}
\label{sec:supp-ipw-casei}

Let $\pi(\cdot)$ denote the true probability of a complete observation, which depends on $Y$ and may also depend on $(X, \bZ)$. That is, we assume
\[
\Pr(R=1 \mid Y=y, X=x, \bZ=\bz) = \pi(y, \bw),
\]
where $\bW \subseteq (X, \bZ)$ and $\bw$ represents its realization. Let $\pi^*(y, \bw)$ denote the specified weight. The IPW estimator is consistent if $\pi^*(y, \bw) = \Pr(R=1 \mid Y=y, \bW=\bw)$.

To establish this, we show that the weighted estimating equation has mean zero:
\bse
&& E \left\{ \frac{R \, \bS_{\btheta}(Y, X, \bZ; \btheta)}{\pi^*(Y, \bW)} \right\} \\
&& = E_{Y, X, \bZ} \left\{ \frac{\Pr(R=1 \mid Y, X, \bZ) \, \bS_{\btheta}(Y, X, \bZ; \btheta)}{\pi^*(Y, \bW)} \right\} \\
&& = E_{Y, X, \bZ} \left\{ \frac{\Pr(R=1 \mid Y, \bW) \, \bS_{\btheta}(Y, X, \bZ; \btheta)}{\pi^*(Y, \bW)} \right\} \\
&& = E_{Y, X, \bZ} \left\{ \bS_{\btheta}(Y, X, \bZ; \btheta) \right\} \\
&& = E_{X, \bZ} \left[ E_{Y \mid X, \bZ} \left\{ \bS_{\btheta}(Y, X, \bZ; \btheta) \right\} \right] \\
&& = \bzero,
\ese
where the second equality follows from the assumption that $\Pr(R=1 \mid Y, X, \bZ) = \Pr(R=1 \mid Y, \bW)$, the third equality holds when $\pi^*(Y, \bW) = \Pr(R=1 \mid Y, \bW)$, and the final equality follows from the property $E_{Y \mid X, \bZ}\{\bS_{\btheta}(Y, X, \bZ; \btheta)\} = \bzero$ established in Section~\ref{sec:supp-consistency-cc}. However, when $\pi^*(Y, \bW) \ne \Pr(R=1 \mid Y, \bW)$, then 
\bse
E_{Y, X, \bZ} \left\{ \frac{\Pr(R=1 \mid Y, W) \, \bS_{\btheta}(Y, X, \bZ; \btheta)}{\pi^*(Y, W)} \right\} \ne E_{X, \bZ} \left[ E_{Y \mid X, \bZ} \left\{ \bS_{\btheta}(Y, X, \bZ; \btheta) \right\} \right],
\ese
and the resulting expectation cannot be guaranteed to equal zero because the $\pi(\cdot)$ in the numerator and denominator do not cancel. Correct specification, $\pi^*(y, \bw) = \Pr(R=1 \mid Y=y, \bW=\bw)$, 
is therefore a sufficient condition for consistency of the IPW estimator. Proving necessity is difficult and generally not tractable; we therefore focus on establishing sufficiency here.

\paragraph{Site-specific consistency implies federated consistency.}
As in the CC proof, the federated IPW estimating equation decomposes as 
\bse
\sum_{i=1}^n \bPhi_\text{IPW}(\bO_i;\btheta_0,\balpha_0) 
&=& \sum_{k=1}^K 
\underbrace{\sum_{i=1}^{n^{(k)}} \frac{r_i^{(k)} \bS_\btheta(y_i^{(k)}, r_i^{(k)}x_i^{(k)}, \bz_i^{(k)};\btheta_0)}{\pi_{Y,\bZ}(y_i^{(k)}, \bz_i^{(k)}; \balpha^{(k)}_{0})}}_{\text{site $k$ contribution}}.
\ese
Under (A1)--(A2) and (A4), if $\pi_{Y,\bZ}(\cdot; \balpha^{(k)}_{0}) = 
\Pr(R=1 \mid Y, \bW)$ at each site $k$, then each site-specific 
contribution has expectation $\bzero$, and by linearity of expectation:
\bse
E\left\{ \sum_{k=1}^K \sum_{i=1}^{n^{(k)}} \frac{r_i^{(k)} 
\bS_\btheta(y_i^{(k)}, r_i^{(k)}x_i^{(k)}, \bz_i^{(k)};\btheta_0)}
{\pi_{Y,\bZ}(y_i^{(k)}, \bz_i^{(k)}; \balpha_{0,k})} \right\}
&=& \sum_{k=1}^K E\left\{ \sum_{i=1}^{n^{(k)}} \frac{r_i^{(k)} 
\bS_\btheta(y_i^{(k)}, r_i^{(k)}x_i^{(k)}, \bz_i^{(k)};\btheta_0)}
{\pi_{Y,\bZ}(y_i^{(k)}, \bz_i^{(k)}; \balpha_{0,k})} \right\} \\
&=& \sum_{k=1}^K \bzero \\
&=& \bzero.
\ese
Therefore, correct specification of the weighting model at each site is 
sufficient for consistency of the federated IPW estimator. As established 
above, proving necessity is not tractable in general; we therefore focus 
on sufficiency throughout.

\subsubsection{Case (ii): Probability of a complete observation does not 
depend on $Y$}

Let $\pi(\cdot)$ denote the oracle probability of a complete observation, 
which depends on $\bW \subseteq (X, \bZ)$ but not on $Y$:
\[
\Pr(R=1 \mid Y, X, \bZ) = \pi(\bw),
\]
where $\bw$ denotes the realization of $\bW$. Let $\pi^*(\bw)$ denote the specified probability. The IPW estimator is consistent regardless of whether $\pi^*(\bw) = \Pr(R=1 \mid \bW=\bw)$:
\bse
E \left\{ \frac{R \, \bS_{\btheta}(Y, X, \bZ; \btheta)}{\pi^*(\bW)} \right\} 
&=& E_{Y, X, \bZ} \left\{ \frac{\Pr(R=1 \mid Y, X, \bZ) \, 
    \bS_{\btheta}(Y, X, \bZ; \btheta)}{\pi^*(\bW)} \right\} \\
&=& E_{Y, X, \bZ} \left\{ \frac{\Pr(R=1 \mid \bW) \, 
    \bS_{\btheta}(Y, X, \bZ; \btheta)}{\pi^*(\bW)} \right\} \\
&=& E_{X, \bZ} \left[ \frac{\Pr(R=1 \mid \bW)}{\pi^*(\bW)} 
    E_{Y \mid X, \bZ} \left\{ \bS_{\btheta}(Y, X, \bZ; \btheta) 
    \right\} \right] \\
&=& \bzero,
\ese
where the second to third line uses the fact that $\Pr(R=1 \mid \bW)$ 
and $\pi^*(\bW)$ are both functions of $\bW$ alone and can therefore be 
factored out of $E_{Y \mid X, \bZ}\{\cdot\}$. However, if $Y$ forced into the weighting model when the oracle 
probability does not depend on $Y$, then this may result in  bias:
\bse
 E_{Y, X, \bZ} \left\{ \frac{\Pr(R=1 \mid \bW) \, 
    \bS_{\btheta}(Y, X, \bZ; \btheta)}{\pi^*(Y, \bW)} \right\} \ne E_{X, \bZ} \left[ \frac{\Pr(R=1 \mid \bW)}{\pi^*(\bW)} 
    E_{Y \mid X, \bZ} \left\{ \bS_{\btheta}(Y, X, \bZ; \btheta) 
    \right\} \right],
\ese
since $\pi^*(Y, \bW)$ depends on $Y$ and cannot be factored out of 
$E_{Y \mid X, \bZ}\{\cdot\}$.

\paragraph{Site-specific consistency implies federated consistency.}
If at each site $k$ the oracle probability does not depend on $Y$ and 
the site-specific weighting model $\pi(\bw; \balpha_0^{(k)})$ excludes 
$Y$, then each site-specific IPW contribution has expectation $\bzero$. 
Under assumptions (A1)--(A2), the expectation of the federated IPW 
estimating equation equals $\bzero$ by linearity, yielding a consistent 
federated estimator.

\paragraph{When IPW introduces bias:}

This phenomenon has been noted by \citet{bartlett2014improving} and \citet{vazquez2026estimators}, who investigate settings where the outcome $Y$ appears significant in the weighting model for an unobserved covariate $X$. In both cases, the apparent association arises because missingness depends on the missing covariate $X$, which is itself associated with $Y$---a relationship guaranteed by the outcome model of interest (i.e., $g\{E_{Y|X,\bZ}(Y)\} = m(X,\bZ;\btheta)$). However, this empirical association (i.e., signal) does not imply that $Y$ should be included in the weighting model.

\begin{figure}[ht!]
    \centering
    \includegraphics[width=0.8\textwidth]{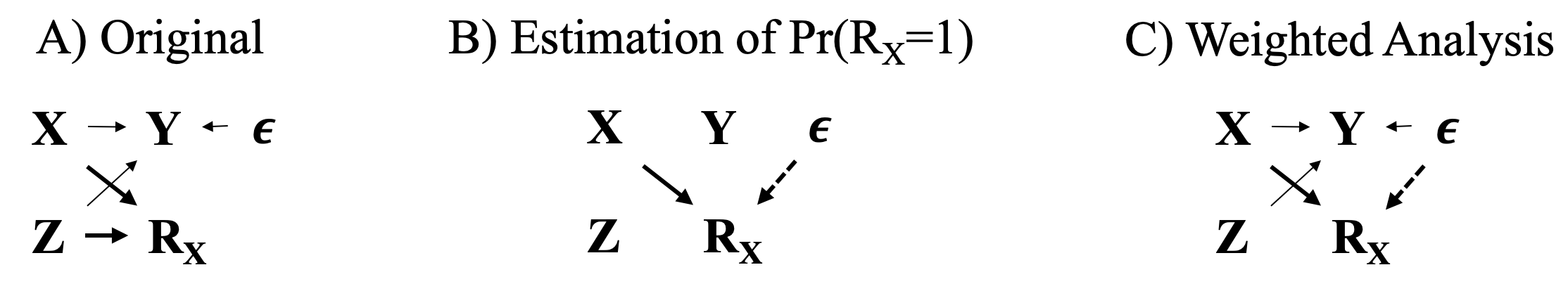}
    \caption{\textbf{Illustration of bias introduced by inverse probability weighting (IPW).}
        (A)~\textit{Original:} Missingness in $X$, denoted by $R_X$, depends solely on $(X, \bZ)$. 
        (B)~\textit{Estimation of $\Pr(R_X=1)$:} When the probability of a complete case is modeled using $(Y, \bZ)$, a non-causal path opens between the outcome error term $\epsilon$ and $R_X$ (dashed arrow). 
        (C)~\textit{Weighted analysis:} Although weighting removes arrows into $R_X$, the open path $\epsilon \dashrightarrow R_X$ persists, inducing bias in the estimated effect $\btheta$.
    }
    \label{fig:dag-ipw}
\end{figure}

To illustrate how the IPW estimator may introduce bias, we extend a missing data directed acyclic graph with weighting (mDAG; \citealp{mohan2021graphical}). In Figure~\ref{fig:dag-ipw}, Panel~A shows the original data-generating process, where missingness in $X$, denoted by $R_X$, depends only on $(X, \bZ)$ and is independent of the outcome $Y$ and its error term $\epsilon$. Under this mechanism, the CC estimator is unbiased because no path connects $Y$ and $R_X$. Panel~B depicts the misspecification scenario in which the probability of a complete case is modeled using $(Y, \bZ)$ rather than $(X, \bZ)$. This opens a non-causal path from $\epsilon$ to $R_X$ through $Y$, inducing dependence between $Y$ and $R_X$ via a backdoor path (dashed arrow). Panel~C shows that after reweighting, the path $\bZ \rightarrow R_X$ is closed; however, the open backdoor path $\epsilon \dashrightarrow R_X$ remains, leading to biased estimation of $\btheta$.

This example illustrates that the IPW estimator can introduce, rather than mitigate, bias when the weighting model opens a non-causal path through a collider structure. In such settings, when missingness is not driven by $Y$, the CC estimator is the preferred estimator---a finding explored by \citet{bartlett2014improving} and extended by \citet{vazquez2026estimators} to other forms of data coarsening. As these authors note, this setting arises when $Y$ is measured during follow-up while $X$ and $\bZ$ are recorded at study entry. By temporality, missingness in $X$ cannot depend on $Y$ and is instead driven by $(X, \bZ)$ or unmeasured baseline factors.

\paragraph{Practical guidance.}
In cases where the CC estimator fails to remain consistent---corresponding to the unshaded cells in Table~\ref{tab:missingness-mcar-mar-mnar}---the 
IPW estimator may achieve consistency under correct specification of the weighting model. A practical strategy is therefore to use the IPW estimator when the oracle probability depends on $Y$, and the CC estimator when it does not. In the latter case, the IPW estimator 
requires additional steps to estimate the weights yet provides no bias correction, and may even introduce bias if $Y$ is forced in the weighting model. An additional difficulty arises in settings where the missingness distribution (and by extension $\balpha$) is not identifiable from the observed data alone (Table~\ref{tab:missingness-mcar-mar-mnar}). In these cases, regardless of the estimation approach used---parametric, semiparametric, or nonparametric---the weighting model cannot be consistently estimated, and the probability of a complete observation will be misspecified \citep{molenberghs2008every}. For example, the distribution for missingness in $Y$ where the probability of a complete observation depends on $(Y,X,\bZ)$, a MNAR case, is not identifiable from the observed data alone.

\subsection{Consistency of the calibrated estimator}
\label{sec:supp-consistency-calibrated}

The calibration step is the multiply-robust construction of \citet{han2014multiply} and \citet{chuen2014mr}, in the least squares (projection) form of \citet{chen2023unified}; its defining property is that the combined model is correctly specified whenever at least one candidate is. Our contribution is to establish this property \emph{site by site} in the federated setting---where the correct candidate may be contributed locally or from another site---so that consistency follows from the site decomposition of Proposition~\ref{prop:fl-consistency-supp}. We restate the recovery argument here for completeness.

Proposition~\ref{prop:fl-consistency-supp}(iii) follows by reducing the calibrated weighting model to the correctly specified case of Section~\ref{sec:supp-consistency-ipw}. At site $k$, let $\bgamma(\bO; \balpha_0) = \{\pi(\bO; \balpha_{0,1}), \ldots, \pi(\bO; \balpha_{0,J})\}^\top$ collect the $J$ candidate probabilities of a complete observation, each a function of the weighting-model covariates $(Y, \bW)$, and let the calibrated probability be $\pi^\text{cal}(\bO; \balpha_0) = \btau_k^\top \bgamma(\bO; \balpha_0)$. The site-$k$ calibration coefficients solve the population moment $E^{(k)}[\bgamma\{R - \bgamma^\top \btau_k\}] = \bzero$, where $E^{(k)}$ denotes expectation under the site-$k$ distribution. Solving for $\btau_k$ gives
\bse
\btau_k^* = \left\{ E^{(k)}(\bgamma \bgamma^\top) \right\}^{-1} E^{(k)}(\bgamma R).
\ese
Equivalently, $\pi^\text{cal} = \btau_k^{*\top}\bgamma$ is the least-squares fit of $R$ on the candidate probabilities: among all linear combinations of $\bgamma$, the one closest to $R$ in mean square (its $L_2$ projection onto their span). The moment condition above is the orthogonality that defines the fit; the residual $R - \pi^\text{cal}$ is uncorrelated with each candidate.

We now show that, when at least one candidate is correctly specified, the calibration step recovers the true probability of a complete observation. Assume the candidate probabilities are not collinear, so that $E^{(k)}(\bgamma\bgamma^\top)$ is nonsingular and $\btau_k$ is identifiable (the calibration analogue of the nonsingularity required in (A5)). Suppose that for some $j$,
\bse
\pi(\bO; \balpha_{0,j}) = \Pr(R=1 \mid Y, \bW) = E^{(k)}(R \mid Y, \bW) \quad \text{almost surely,}
\ese
so the $j^{th}$ candidate model equals the conditional mean of the complete-observation indicator $R$ at site $k$ (true model). Since every component of $\bgamma$ is a function of $(Y, \bW)$, the rules of conditional expectation give
\bse
E^{(k)}(\bgamma R) = E^{(k)}\{\bgamma\, E^{(k)}(R \mid Y, \bW)\} = E^{(k)}\{\bgamma\, \pi(\bO; \balpha_{0,j})\},
\ese
which is the $j^{th}$ column of $E^{(k)}(\bgamma\bgamma^\top)$, because $\pi(\bO; \balpha_{0,j})$ is the $j^{th}$ entry of $\bgamma$. Multiplying by $\{E^{(k)}(\bgamma\bgamma^\top)\}^{-1}$ selects that column---the inverse of a matrix applied to its own $j^{th}$ column returns the unit vector $\mathbf{e}_j$---so that
\bse
\btau_k^* = \{E^{(k)}(\bgamma\bgamma^\top)\}^{-1} E^{(k)}(\bgamma R) = \mathbf{e}_j,
\ese
the unit vector placing all weight on the correctly specified candidate. The calibrated probability therefore equals the truth,
\bse
\pi^\text{cal}(\bO; \balpha_0) = \btau_k^{*\top}\bgamma = \pi(\bO; \balpha_{0,j}) = \Pr(R=1 \mid Y, \bW) \quad \text{almost surely.}
\ese

Recovery places the calibrated weighting model in the correctly specified case of Section~\ref{sec:supp-consistency-ipw}: with $\pi^\text{cal} = \Pr(R=1 \mid Y, \bW)$, the site-$k$ IPW estimating equation contribution has expectation $\bzero$ by the argument of Section~\ref{sec:supp-ipw-casei}. By the site decomposition~\eqref{eq:site-decomposition} and assumptions~(A1)--(A2), if every site has at least one correctly specified candidate, the federated calibrated estimating equation has expectation $\bzero$, and the calibrated estimator is consistent. The correctly specified candidate need not originate at site $k$: the projection uses only the shared candidate parameters $\balpha_0$ and site $k$ data, so a model contributed by any site recovers $\Pr(R=1 \mid Y, \bW)$ wherever it is correct.

\begin{remark}[Conditioning and pool size]
The identifying condition weakens as candidates grow redundant or highly correlated: the condition number $\kappa$ of $E^{(k)}(\bgamma\bgamma^\top)$ rises (Section~\ref{sec:supp-kappa}), inflating the finite-sample variance of $\wh\btau_k$, and the regularization used to stabilize the near-singular inverse introduces a small finite-sample bias. Both raise the mean squared error as the pool grows, so a small, well-conditioned pool is preferred (Section~\ref{sec:sim-pool}).
\end{remark}

\begin{remark}[Constrained parametrization]
The non-negative, normalized parametrization used in the simulations \citep{chen2023unified} preserves recovery, since the unit-weight solution $\mathbf{e}_{j}$ is feasible on the simplex and attains the same projection. The construction is the multiple-robustness principle of \citet{han2014multiply} and \citet{chuen2014mr}, applied here site-by-site across the network.
\end{remark}

\section{Asymptotic normality}
\label{sec:supp-normality} 

Following a similar argument as \citet{tsiatis2006semiparametric} and \citet{vazquez2026estimators}, the asymptotic variance of the IPW estimator, conditional that we have a consistent estimator of $\balpha$, is equal to 
\bse
n^{1/2}(\wh{\btheta}_\text{IPW}-\btheta_0) \rightarrow_d \Normal(\bzero, \bA_\text{IPW}^{-1}\bB_\text{IPW}\bA_\text{IPW}^\top), 
\ese
where 
\bse
&& \bA_\text{IPW} = E \left\{  \partial \bPhi_{\rm IPW}(\bO; \btheta,\balpha_0)/ 
\partial \btheta^\top |_{\btheta = \btheta_0} \right\}, \\
&& \bB_\text{IPW} =  E\biggr\{ \bigr(\bPhi_{\rm IPW}(\bO; \btheta_0,\balpha_0) \\
&& \quad - \underbrace{E\{\partial\bPhi_{\rm IPW}(\bO;\btheta_0,\balpha)/\partial\balpha^T 
|_{\balpha = \balpha_0} \}[E\{\partial\bPhi_\text{Nuisance}(\bO;\balpha)/\partial\balpha^T 
|_{\balpha = \balpha_0}\}]^{-1}\bPhi_\text{Nuisance}(\bO;\balpha_0)}_{\text{adjustment for \balpha\ estimation}} \bigr)^{\otimes2}\biggr\}.
\ese
For which the second line is the contribution of unknown nuisance parameters $\balpha$. The robust sandwich estimator for $\var(\wh\btheta)$ may be obtained by replacing $(\bA_\text{IPW},\bB_\text{IPW})$ with their empirical analogues $(\wh\bA_\text{IPW},\wh\bB_\text{IPW})$, and inference for $\btheta$ follows from the corresponding empirical form. Under a pooled setting, empirical versions of these matrices are computed from the combined sample. In a federated setting, however, the same quantities can be obtained by aggregating site-specific summary contributions, allowing estimation without sharing individual-level data. 

We can express these empirical components as sums of site-specific contributions similar to Section~\ref{sec:robust-sandwhich-estimator}:
\bse
\wh\bA_\text{IPW} &=& E \left\{  \partial \bPhi_{\rm IPW}(\bO; \btheta, \balpha)/ \partial \btheta^\top |_{\btheta = \wh\btheta, \balpha = \wh\balpha}
\right\} = n^{-1} \sum_{k=1}^K \underbrace{\sum_{i=1}^{n^{(k)}} \left\{  \partial \bPhi_{\rm IPW}(\bO_i^{(k)}; \btheta, \wh\balpha) / \partial \btheta^\top |_{\btheta = \wh\btheta} \right\}}_{\text{site $k$ contribution}}.
\ese
Each site can compute its local contribution and transmit to a coordinating site. It follows that we can expand the expression for $\bB_\text{IPW}$, of which the following components can be empirically estimated as
\bse
&& E\{(\bPhi_{\rm IPW}(\bO; \btheta_0,\balpha_0) \bPhi_{\rm IPW}(\bO; \btheta_0,\balpha_0)^\top \}  = n^{-1} \sum_{k=1}^K \sum_{i=1}^{n^{(k)}} \left\{ \bPhi_{\rm IPW}(\bO_i^{(k)}; \wh\btheta, \wh\balpha)\bPhi_{\rm IPW}(\bO_i^{(k)}; \wh\btheta, \wh\balpha)^\top \right\}, \\
&& E \{ \bPhi_{\rm IPW}(\bO; \btheta_0,\balpha_0) \bPhi_\text{Nuisance}(\bO;\balpha_0) \} = n^{-1} \sum_{k=1}^K \sum_{i=1}^{n^{(k)}} \left\{ \bPhi_{\rm IPW}(\bO_i^{(k)}; \wh\btheta, \wh\balpha)\bPhi_{\rm Nuisance}(\bO_i^{(k)}; \wh\balpha)^\top \right\}, \\
&& E\{\partial\bPhi_{\rm IPW}(\bO;\btheta_0,\balpha)/\partial\balpha^T |_{\balpha = \balpha_0} \} = n^{-1} \sum_{k=1}^K   \partial  \left\{\sum_{i=1}^{n^{(k)}} \bPhi_{\rm IPW}(\bO_i^{(k)}; \wh\btheta, \balpha) \right\} / \partial \balpha^\top |_{\balpha = \wh\balpha}, \\
&& E\{\partial\bPhi_\text{Nuisance}(\bO;\balpha)/\partial\balpha^T |_{\balpha = \balpha_0}\} = n^{-1} \sum_{k=1}^K   \partial \left\{ \sum_{i=1}^{n^{(k)}} \bPhi_{\rm Nuisance}(\bO_i^{(k)}; \balpha) \right\} / \partial \balpha^\top |_{\balpha = \wh\balpha}, \\
&&  E\left\{ \bPhi_\text{Nuisance}(\bO;\balpha_0)\bPhi_\text{Nuisance}(\bO;\balpha_0)^\top \right\}  = n^{-1} \sum_{k=1}^K \sum_{i=1}^{n^{(k)}} \left\{ \bPhi_{\rm Nuisance}(\bO_i^{(k)}; \wh\balpha) \bPhi_{\rm IPW}(\bO_i^{(k)}; \wh\balpha)^\top \right\}. 
\ese

\subsection{Calibrated weight estimation}
\label{sec:supp-variance-calibrated}

Under calibrated weight estimation, the stacked matrices no longer decompose into distinct per-site blocks, because the calibration coefficients $\btau$ constitute an additional estimated component. We append the calibration estimating equation, which at site $k$ takes the form $\sum_i \bgamma(\bO_i; \balpha)\{r_i - \bgamma(\bO_i; \balpha)^\top \btau_k\} = \bzero$, to the stacked system. Let $\btau = (\btau_1^\top, \ldots, \btau_K^\top)^\top$ and $\bxi = (\btheta^\top, \balpha_1^\top, \ldots, \balpha_J^\top, \btau_1^\top, \ldots, \btau_K^\top)^\top$; the stacked estimating equation is
\be
\label{eqn:estimating-equation-stacked-tau}
&& \sumi\bPhi_\text{Stacked}(\bO_i ;\bxi) \nonumber \\
&& =
\sumi
\left\{
\begin{matrix}
\bPhi_\text{IPW}(\bO_i;\btheta,\balpha, \btau) \\
\bPhi_\text{Site 1, Nuisance}(\bO_i;\balpha_1) \\
\vdots \\
\bPhi_\text{Site J, Nuisance}(\bO_i;\balpha_J) \\
\vdots \\ 
\bPhi_\text{Site 1, Calibration}(\bO_i;\btau_1,\balpha) \\
\vdots \\
\bPhi_\text{Site K, Calibration}(\bO_i;\btau_K, \balpha) \\
\end{matrix}
\right\} 
=
\sumi
\left\{
\begin{matrix}
\bPhi_\text{IPW}(\bO_i;\btheta, \balpha, \btau) \\
\bPhi_\text{Nuisance}(\bO_i; \balpha) \\
\bPhi_\text{Calibration}(\bO_i; \btau, \balpha)
\end{matrix}
\right\}
= \bzero 
\ee
Let $J$ denote the number of candidate weighting models used for calibration, where each $\balpha_j$ ($j = 1, \ldots, J$) indexes a distinct model contributed by a participating site and $\btau_k$ ($k = 1, \ldots, K$) the calibration occurring at each site. As noted in Section~\ref{sec:fl-mr}, $J$ may differ from $K$: it may exceed $K$ if any site contributes more than one candidate model, or be smaller than $K$ if only a subset of sites share their coefficients.

For pedagogical purposes, consider the case in which each site contributes a single candidate model, so that $J = K$ and site $k$ contributes $\balpha_k$, with all $K$ estimates shared and used at the calibration step of every site. Let $\bA^{(k)}_{\btheta\btheta}$, $\bA^{(k)}_{\btheta\balpha_k}$, $\bA^{(k)}_{\balpha_k\btheta}$, and $\bA^{(k)}_{\balpha_k\balpha_k}$ take the same form as in Section~\ref{sec:robust-sandwhich-estimator}. Let the calibration blocks be $\bA^{(k)}_{\btheta\btau_k} = E\{ \partial \bPhi_\text{IPW}(\bO; \btheta, \balpha, \btau_k) / \partial \btau_k^\top \}$, $\bA^{(k)}_{\btau_k\balpha_j} = E\{ \partial \bPhi_\text{Calibration}(\bO; \btau_k, \balpha) / \partial \balpha_j^\top \}$, and $\bA^{(k)}_{\btau_k\btau_k} = E\{ \partial \bPhi_\text{Calibration}(\bO; \btau_k, \balpha) / \partial \btau_k^\top \}$. The stacked matrix $\wh\bA_{\text{Stacked}}$ takes the form:
\bse
&& \wh\bA_{\text{Stacked}} = \\
&& \begin{pmatrix}
\sum_{k=1}^K \wh\bA^{(k)}_{\btheta\btheta}
& \sum_{k=1}^K \wh\bA^{(k)}_{\btheta\balpha_1}
& \cdots
& \sum_{k=1}^K \wh\bA^{(k)}_{\btheta\balpha_K}
& \bA^{(1)}_{\btheta\btau_1}
& \cdots
& \bA^{(K)}_{\btheta\btau_K}
\\
\bzero
& \wh\bA^{(1)}_{\balpha_1\balpha_1}
& \cdots
& \bzero
& \bzero
& \cdots
& \bzero
\\
\vdots
& \vdots
& \ddots
& \vdots
& \vdots
& \ddots
& \vdots
\\
\bzero
& \bzero
& \cdots
& \wh\bA^{(K)}_{\balpha_K\balpha_K}
& \bzero
& \cdots
& \bzero
\\
\bzero
& \bA^{(1)}_{\btau_1\balpha_1}
& \cdots
& \bA^{(1)}_{\btau_1\balpha_K}
& \bA^{(1)}_{\btau_1\btau_1}
& \cdots
& \bzero
\\
\vdots
& \vdots
& \ddots
& \vdots
& \vdots
& \ddots
& \vdots
\\
\bzero
& \bA^{(K)}_{\btau_K\balpha_1}
& \cdots
& \bA^{(K)}_{\btau_K\balpha_K}
& \bzero
& \cdots
& \bA^{(K)}_{\btau_K\btau_K}
\end{pmatrix} 
\ese
The first column-block corresponds to $\btheta$, the next $K$ to the candidate weighting model parameters $\balpha_1, \ldots, \balpha_K$, and the final $K$ to the site-specific calibration coefficients $\btau_1, \ldots, \btau_K$. The $\btheta\btheta$ block and the $\btheta\balpha_j$ blocks are summed across the $K$ sites, because every site's calibrated weight uses all $K$ candidate models, so the IPW estimating function differentiates with respect to each $\balpha_j$ at every site. The $\balpha$- and $\btau$-blocks are not summed, since each candidate model is fit at a single site and each calibration moment is evaluated at a single site. The cross-derivatives $\wh\bA^{(k)}_{\btau_k\balpha_j}$ are nonzero for all $j$, since the calibration moment at site $k$ depends on every candidate model through $\bgamma$. By contrast, $\wh\bA^{(k)}_{\balpha_j\balpha_{j'}} = \bzero$ for $j \ne j'$: even when a site contributes multiple weighting models, each is fit independently, so the estimating equation for one does not depend on the parameters of another.

The score covariance matrix $\wh\bB_{\text{Stacked}}$ has a similar but sparser structure. The superscript $(k)$ denotes expectation under the data distribution at site $k$. Under one model per site, the nonzero component blocks are $\bB^{(k)}_{\btheta\btheta} = E^{(k)}\{\bPhi_\text{IPW}(\bO; \btheta, \balpha, \btau_k)^{\otimes 2}\}$, $\bB^{(k)}_{\btheta\balpha_k} = E^{(k)}\{\bPhi_\text{IPW}(\bO; \btheta, \balpha, \btau_k)\,\bPhi_\text{Nuisance}(\bO; \balpha_k)^\top\}$, $\bB^{(k)}_{\btheta\btau_k} = E^{(k)}\{\bPhi_\text{IPW}(\bO; \btheta, \balpha, \btau_k)\,\bPhi_\text{Calibration}(\bO; \btau_k, \balpha)^\top\}$, $\bB^{(k)}_{\balpha_k\balpha_k} = E^{(k)}\{\bPhi_\text{Nuisance}(\bO; \balpha_k)^{\otimes 2}\}$, $\bB^{(k)}_{\balpha_k\btau_k} = E^{(k)}\{\bPhi_\text{Nuisance}(\bO; \balpha_k)\,\bPhi_\text{Calibration}(\bO; \btau_k, \balpha)^\top\}$, and $\bB^{(k)}_{\btau_k\btau_k} = E^{(k)}\{\bPhi_\text{Calibration}(\bO; \btau_k, \balpha)^{\otimes 2}\}$. Every block pairing components from different sites vanishes, because the nuisance score for $\balpha_k$ and the calibration moment for $\btau_k$ are nonzero only at site $k$: $\wh\bB_{\balpha_j\balpha_m} = \bzero$ for $j \ne m$, $\wh\bB_{\btau_k\balpha_j} = \bzero$ for $j \ne k$, and $\wh\bB_{\btau_k\btau_{k'}} = \bzero$ for $k \ne k'$. The stacked matrix is:
\bse
&& \wh\bB_{\text{Stacked}} = \\
&&
\begin{pmatrix}
\sum_{k=1}^K \wh\bB^{(k)}_{\btheta\btheta}
& \wh\bB^{(1)}_{\btheta\balpha_1} & \cdots & \wh\bB^{(K)}_{\btheta\balpha_K}
& \wh\bB^{(1)}_{\btheta\btau_1} & \cdots & \wh\bB^{(K)}_{\btheta\btau_K} \\
\wh\bB^{(1)}_{\balpha_1\btheta}
& \wh\bB^{(1)}_{\balpha_1\balpha_1} & \cdots & \bzero
& \wh\bB^{(1)}_{\balpha_1\btau_1} & \cdots & \bzero \\
\vdots & \vdots & \ddots & \vdots & \vdots & \ddots & \vdots \\
\wh\bB^{(K)}_{\balpha_K\btheta}
& \bzero & \cdots & \wh\bB^{(K)}_{\balpha_K\balpha_K}
& \bzero & \cdots & \wh\bB^{(K)}_{\balpha_K\btau_K} \\
\wh\bB^{(1)}_{\btau_1\btheta}
& \wh\bB^{(1)}_{\btau_1\balpha_1} & \cdots & \bzero
& \wh\bB^{(1)}_{\btau_1\btau_1} & \cdots & \bzero \\
\vdots & \vdots & \ddots & \vdots & \vdots & \ddots & \vdots \\
\wh\bB^{(K)}_{\btau_K\btheta}
& \bzero & \cdots & \wh\bB^{(K)}_{\btau_K\balpha_K}
& \bzero & \cdots & \wh\bB^{(K)}_{\btau_K\btau_K}
\end{pmatrix}.
\ese
The first row and column correspond to $\btheta$, the next $K$ to the candidate weighting model parameters $\balpha_1, \ldots, \balpha_K$, and the final $K$ to the site-specific calibration coefficients $\btau_1, \ldots, \btau_K$. Unlike $\wh\bA_{\text{Stacked}}$, only the $\btheta\btheta$ block is summed across sites. The reason is that $\wh\bB_{\text{Stacked}}$ is a covariance of estimating functions rather than a matrix of derivatives: a pair of components contributes only at observations where both are nonzero. The IPW estimating function is nonzero at every observation, whereas the nuisance score for $\balpha_k$ and the calibration moment for $\btau_k$ are nonzero only at site $k$. Functions tied to different sites therefore never share an observation, and their cross-covariances vanish. The matrix is therefore block-arrowhead: a dense $\btheta$ border, with an interior that is block-diagonal in the per-site pairs $\{\balpha_k, \btau_k\}$. Equivalently, $\wh\bB_{\text{Stacked}} = \sum_{k=1}^K \wh\bB^{(k)}_{\text{Stacked}}$, where each site-$k$ term is zero outside the rows and columns indexed by $\{\btheta, \balpha_k, \btau_k\}$. The asymmetry between $\wh\bA_{\text{Stacked}}$ and $\wh\bB_{\text{Stacked}}$ reflects the difference between functional dependence, captured by derivatives, and shared support, captured by covariances.

The two matrices also differ in what each site transmits. For $\wh\bB_{\text{Stacked}}$, site $k$ forms only the outer products among its own scores $\{\btheta, \balpha_k, \btau_k\}$, a block of dimension $[(p + \dim(\balpha_k) + \dim(\btau_k))]$, and sends that single block; no cross-site terms exist. For $\wh\bA_{\text{Stacked}}$, site $k$ additionally forms the cross-derivative blocks $\wh\bA^{(k)}_{\btheta\balpha_j}$ and $\wh\bA^{(k)}_{\btau_k\balpha_j}$ for every $j$, because its calibrated weight depends on all $K$ candidate models. The coordinating site sums the $\btheta\btheta$ blocks (and, for $\wh\bA_{\text{Stacked}}$, the $\btheta\balpha_j$ blocks) across sites, places the remaining site-specific blocks of both matrices in their positions, and forms $\wh{\mathrm{Var}}(\wh\bxi) = \wh\bA_{\text{Stacked}}^{-1} \wh\bB_{\text{Stacked}} \wh\bA_{\text{Stacked}}^{-\top}$ (Algorithms~\ref{alg:fl-mr-ipw} and~\ref{alg:fl-ipw-glm}); the variance of $\wh\btheta$ is the leading $(p+1) \times (p+1)$ block.

\section{Additional simulation details and results} 
\label{sec:supp-simulations}

This section provides supplementary simulation details and results that complement the main text. We first give the data-generating details and the condition-number definition for the calibration pool-selection study (Section~\ref{sec:supp-pool}). We then present extended results for the linear regression model under shared and heterogeneous missingness mechanisms (Section~\ref{sec:supp-linear}), and results for logistic regression using only aggregated counts (Section~\ref{sec:supp-logistic}).

\subsection{Calibration pool selection}
\label{sec:supp-pool}

This section gives the data-generating details for the pool-selection study of Section~\ref{sec:sim-pool} and defines the condition number used to diagnose the calibration step.

\subsubsection{Candidate missingness library}
\label{sec:supp-pool-forms}

Each site was assigned a missingness mechanism for $X$ drawn at random from a library of ten logistic forms for the probability of a complete observation under MAR. Writing $\tilde Y = Y - \bar Y$ for the centered outcome, every form shares the linear predictor
\bse
&& \eta(Y,\bZ) = \alpha_0 + \alpha_Y \tilde Y + \alpha_{Z_1} Z_1 + \alpha_{Z_2} Z_2 + (\text{interaction terms}) \\
&& \Pr(R=1\mid Y,\bZ) = \text{expit}\{\eta(Y,\bZ)\},
\ese
with base coefficients $(\alpha_0, \alpha_Y, \alpha_{Z_1}, \alpha_{Z_2}) = (-0.7, 0.08, 0.3, 0.2)$, interaction coefficients $0.05$ for any $\tilde Y$-by-$Z$ term and $0.10$ for the $Z_1 Z_2$ term, and independent $\Normal(0, 0.02^2)$ perturbations added to every coefficient at each site. Centering $Y$ and keeping the coefficients small holds the probability of a complete observation away from $0$ and $1$, yielding a subject-level missingness rate in $X$ of roughly $60\%$. The ten forms differ only in their interaction structure:

\begin{table}[h!]
\centering
\begin{tabular}{cl}
\toprule
\textbf{Form} & \textbf{Structure (added to the main effects $\tilde Y, Z_1, Z_2$)} \\
\midrule
1  & main effects only \\
2  & $+\,\tilde Y Z_1$ \\
3  & $+\,\tilde Y Z_2$ \\
4  & $+\,Z_1 Z_2$ \\
5  & $+\,\tilde Y Z_1 + \tilde Y Z_2$ \\
6  & $+\,\tilde Y Z_1 + Z_1 Z_2$ \\
7  & $+\,\tilde Y Z_2 + Z_1 Z_2$ \\
8  & $Z_2$ omitted (depends on $\tilde Y, Z_1$ only) \\
9  & $Z_1$ omitted (depends on $\tilde Y, Z_2$ only) \\
10 & $+\,\tilde Y Z_1 + \tilde Y Z_2 + Z_1 Z_2$ (full two-way) \\
\bottomrule
\end{tabular}
\caption{Library of candidate missingness forms used in the pool-selection study. Each site's true mechanism, and its working weighting model, were drawn independently and uniformly from these ten forms.}
\label{tab:supp-pool-forms}
\end{table}

Each site fit a working weighting model whose form was likewise drawn uniformly from the same library, so the working model coincided with the data-generating mechanism only by chance. We varied the number of sites $K$ from $10$ to $50$ and the number of donor models from one to nine, taking the donors to be the largest sites by sample size; the candidate pool at each site comprised the donor models together with the site's own working model.

\subsubsection{Conditioning of the calibration step}
\label{sec:supp-kappa}

Let $\bG^{(k)}$ denote the $n^{(k)} \times m$ matrix whose columns are the predicted probabilities of a complete observation from the $m$ pooled candidate models, evaluated at the observations of site $k$. The calibration coefficients $\wh\btau$ solve the least-squares projection of the missingness indicator onto the columns of $\bG^{(k)}$, so their stability is governed by the condition number
\bse
\kappa\big(\bG^{(k)}\big) = \frac{\sigma_{\max}\big(\widetilde\bG^{(k)}\big)}{\sigma_{\min}\big(\widetilde\bG^{(k)}\big)},
\ese
the ratio of the largest to the smallest singular value of the column-standardized matrix $\widetilde\bG^{(k)}$, equivalently the square root of the corresponding eigenvalue ratio of its Gram matrix. A value near one indicates near-orthogonal candidates that each contribute independent information; a large value indicates near-collinear candidates, for which the projection is unstable and the variance of $\wh\btau$ inflates. We report the network average
\bse
\bar\kappa = \frac{1}{K}\sum_{k=1}^{K} \kappa\big(\bG^{(k)}\big)
\ese
for each simulation, then averaged across the $2{,}000$ simulations.

\subsection{Extended results for the linear regression model with random errors}
\label{sec:supp-linear}

We further examine two scenarios: (i) a homogeneous missingness mechanism across all sites, and (ii) heterogeneous missingness mechanisms across sites.

\subsubsection{Homogeneous missingness}

\begin{figure}[h!]
    \centering
    \includegraphics[width=0.6\linewidth]{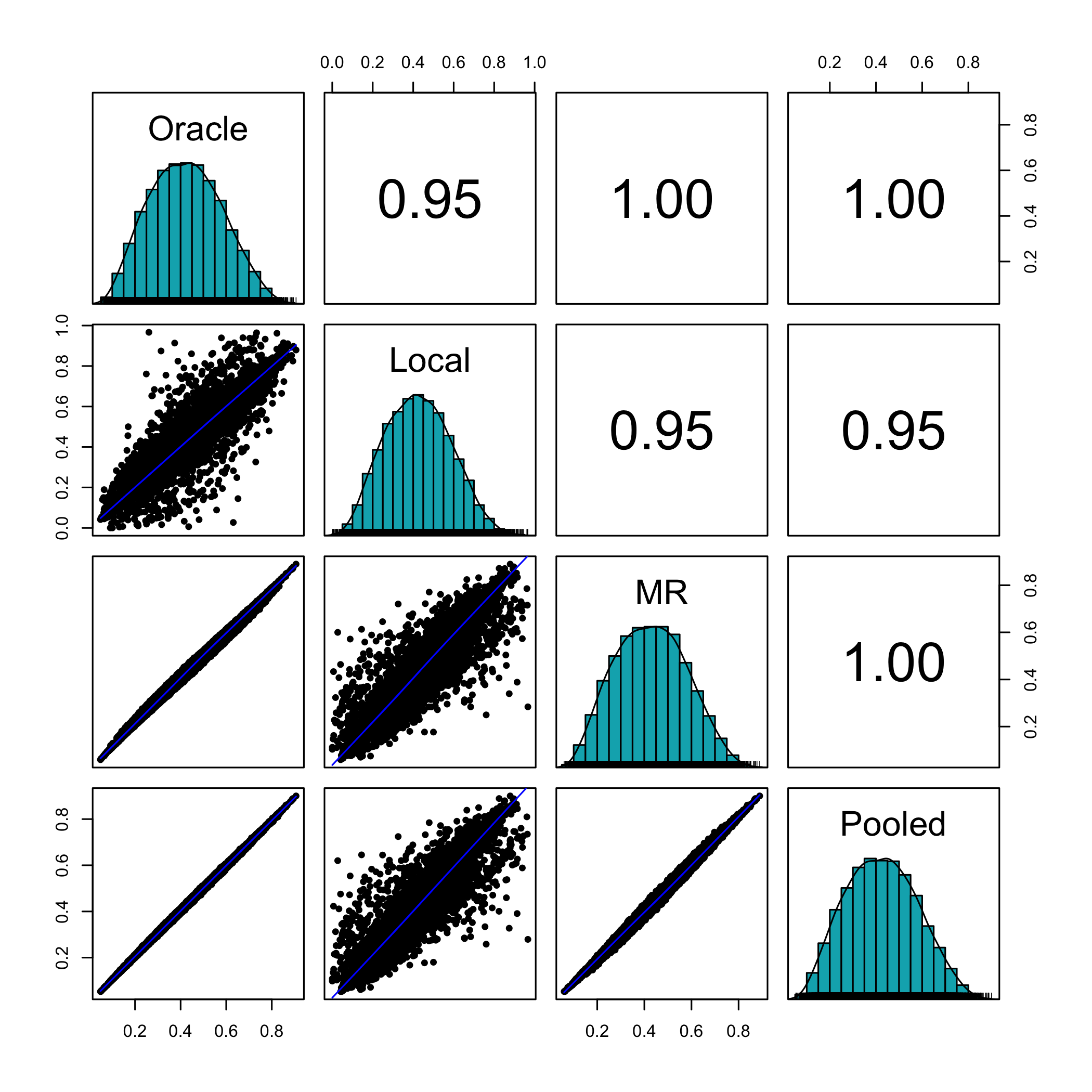}
    \caption{\textbf{Scatter plots and Pearson correlation estimates of oracle and estimated probabilities of a complete observation.} \textit{Oracle} denotes the true probability of complete observation; \textit{Site-specific} denotes site-specific estimates; \textit{MR} denotes the calibrated weights estimated by sharing nuisance parameters across sites; and \textit{Pooled} denotes estimates obtained using pooled data. Results shown are for the case of data MAR.}
    \label{fig:weight-scatter}
\end{figure}

Figure~\ref{fig:weight-scatter} presents scatterplots, marginal distributions, and Pearson correlation coefficients comparing the $\Pr(R = 1 \mid Y, \bZ)$ obtained from four approaches: true values are known (Oracle), local estimation at each site only (Site-specific; 30 sites), calibrated (two largest site shared $\balpha_k$), and estimation using pooled data (Pooled). As expected, the pooled estimator provides the closest approximation to the oracle probabilities. In contrast, when estimation is performed locally, the resulting probabilities exhibit a weaker, yet strong, linear relationship with the oracle values (correlation $\approx0.95$). Under the calibrated weight estimation framework, sharing the estimated nuisance parameters $\balpha_k$ across sites improves weight estimation, yielding probabilities that closely resemble that of the pooled estimates, (correlation $\approx1.0$). These numerical results signal that sharing $\balpha_k$ between sites may lead to better specification of the weights.

\subsubsection{Heterogeneous missingness}

\begin{figure}[h!]
    \centering
    \includegraphics[width=0.6\linewidth]{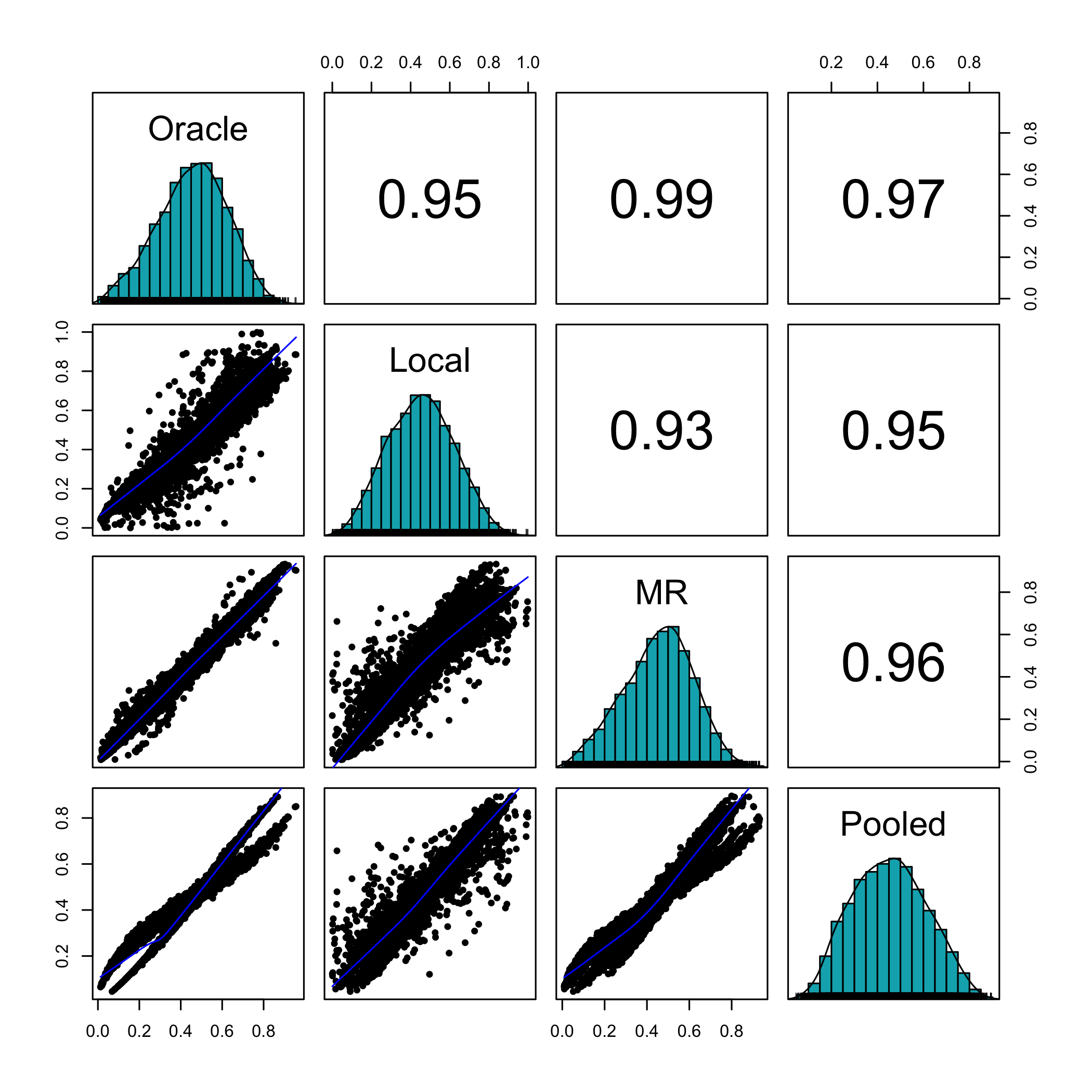}
    \caption{\textbf{Scatter plots and Pearson correlation estimates of oracle and estimated probabilities of a complete observation, different MAR across sites.} \textit{Oracle} denotes the true probability of complete observation; \textit{Site-specific} denotes site-specific estimates; \textit{MR} denotes multiply robust estimates obtained by sharing nuisance parameters across sites; and \textit{Pooled} denotes estimates obtained using pooled data. Results shown are for the case of data MAR.}
    \label{fig:weight-scatter-diff}
\end{figure}

We next examined a setting in which the missingness mechanism varied across sites. Figure~\ref{fig:weight-scatter-diff} displays the pairwise correlations between the oracle probabilities of observing a complete case and their estimated counterparts. Weights computed under the calibrated weight estimation framework exhibited the strongest linear association with the oracle probabilities (correlation $\approx 0.99$), followed by the pooled estimator. In contrast, probabilities estimated using local models showed the weakest agreement with the oracle probabilities. These results were expected, as the working model specification was incorrectly specified for both the pooled and locally estimated weights in a subset of sites. However, under the calibrated weight estimation framework, $\balpha$ from selected sites were shared across, allowing the weights to be adjusted. 

The sharing of $\balpha$ improved the recovery of the oracle probabilities and, by extension, the estimation of $\btheta$. Under calibrated weights, the IPW estimator achieved near-zero bias, performing comparably to the oracle-weight case (Tables~\ref{tab:supp-numerical-results-hetero}). When weights were instead estimated by pooled or site-specific approaches, the IPW estimator remained biased regardless of $K$. When correct site-specific specification is not possible, calibration stabilizes the weights and, in turn, the estimation of the regression parameters.

\begin{table}[h!]
\centering
\resizebox{0.76\linewidth}{!}{
\begin{tabular}[t]{llrrrrrr}
\toprule
\textbf{Estimator} &  & \textbf{Bias} & \textbf{SD} & \textbf{Bias} & \textbf{SD} & \textbf{Bias} & \textbf{SD} \\
\midrule
\addlinespace[0.3em]
\addlinespace
\multicolumn{2}{l}{Parameter: Intercept ($\beta_0$)} & \multicolumn{2}{c}{\textbf{K = 10}} & \multicolumn{2}{c}{\textbf{K = 30}} & \multicolumn{2}{c}{\textbf{K = 50}} \\
\addlinespace
\hspace{1em}Oracle &  & 0.12 & 12.77 & -0.24 & 6.82 & 0.04 & 5.10 \\
\addlinespace
\hspace{1em}CC &  & -112.90 & 16.82 & -113.09 & 9.08 & -112.75 & 6.96 \\
\addlinespace
\hspace{1em}IPW & Oracle & -0.40 & 19.05 & -0.33 & 10.23 & -0.01 & 7.95 \\
 & Pooled & 26.13 & 20.64 & 25.17 & 11.06 & 25.53 & 8.31 \\
 & Site-specific & 28.06 & 21.30 & 27.58 & 11.54 & 27.91 & 8.64 \\
 & Calibrated & 0.41 & 19.23 & 1.01 & 15.55 & 1.47 & 15.28 \\
 & Uniform & -113.09 & 20.39 & -113.10 & 11.03 & -112.48 & 8.36 \\
\addlinespace
\multicolumn{2}{l}{Parameter: $X$ ($\beta_1$)} & \multicolumn{6}{c}{} \\
\addlinespace
\hspace{1em}Oracle &  & 0.02 & 8.09 & 0.02 & 4.36 & 0.02 & 3.35 \\
\addlinespace
\hspace{1em}CC &  & -8.95 & 12.03 & -8.54 & 6.46 & -8.63 & 4.79 \\
\addlinespace
\hspace{1em}IPW & Oracle & -0.54 & 16.43 & -0.13 & 8.96 & -0.06 & 7.01 \\
 & Pooled & -1.24 & 14.91 & -0.70 & 8.09 & -0.70 & 6.01 \\
 & Site-specific & -2.87 & 15.18 & -2.30 & 8.07 & -2.21 & 6.29 \\
 & Calibrated & -0.52 & 16.25 & -0.28 & 8.95 & -0.30 & 6.76 \\
 & Uniform & -8.86 & 14.25 & -8.62 & 7.70 & -8.74 & 5.76 \\
\addlinespace
\multicolumn{2}{l}{Parameter: $Z_1$ ($\beta_2$)} & \multicolumn{6}{c}{} \\
\addlinespace
\hspace{1em}Oracle &  & -0.13 & 20.54 & 0.08 & 10.83 & 0.07 & 8.21 \\
\addlinespace
\hspace{1em}CC &  & -65.53 & 36.80 & -64.08 & 19.32 & -64.16 & 14.56 \\
\addlinespace
\hspace{1em}IPW & Oracle & 0.37 & 37.27 & 0.56 & 19.79 & 0.34 & 15.12 \\
 & Pooled & -58.48 & 43.74 & -56.14 & 23.26 & -56.31 & 17.36 \\
 & Site-specific & -66.76 & 44.56 & -65.26 & 23.90 & -65.53 & 17.71 \\
 & Calibrated & -0.17 & 37.79 & -2.56 & 24.05 & -2.98 & 21.28 \\
 & Uniform & -65.15 & 41.41 & -64.18 & 22.25 & -64.36 & 16.24 \\
\addlinespace
\multicolumn{2}{l}{Parameter: $Z_2$ ($\beta_3$)} & \multicolumn{6}{c}{} \\
\addlinespace
\hspace{1em}Oracle &  & 0.23 & 10.03 & 0.12 & 5.18 & -0.10 & 4.04 \\
\addlinespace
\hspace{1em}CC &  & -12.17 & 14.08 & -12.26 & 7.39 & -12.48 & 5.71 \\
\addlinespace
\hspace{1em}IPW & Oracle & -0.62 & 18.58 & -0.23 & 9.98 & -0.30 & 7.57 \\
 & Pooled & 0.63 & 17.44 & 0.76 & 9.28 & 0.81 & 6.98 \\
 & Site-specific & 0.21 & 17.77 & 0.02 & 9.53 & -0.02 & 7.21 \\
 & Calibrated & -0.78 & 18.76 & -0.22 & 10.36 & -0.29 & 8.00 \\
 & Uniform & -12.10 & 16.98 & -12.15 & 9.05 & -12.39 & 6.84 \\ 
\addlinespace
\bottomrule
\end{tabular}}
\caption{\textbf{Simulation results of regression analysis with a missing covariate under heterogeneous missingness.} Percent bias (\%) and empirical standard deviation (SD) for each regression coefficient ($\beta_0$--$\beta_3$), by number of sites ($K = 10, 30, 50$). Half the sites follow a main-effects MAR mechanism and half a MAR mechanism carrying a $YZ_1$ interaction; the pooled and site-specific estimators use a working model that omits the interaction, whereas the calibrated estimator borrows a correctly specified candidate from another site. \textit{Oracle} denotes the no-missingness benchmark; the IPW rows use true (Oracle), pooled, site-specific, calibrated, and uniform weights. \label{tab:supp-numerical-results-hetero}}
\end{table}

\clearpage
\subsection{Results for the logistic regression model}
\label{sec:supp-logistic}

We conducted an additional simulation study to evaluate the performance of the proposed estimators under a logistic regression model. Data were generated across 2,000 iterations under the model:
\bse
\text{logit}\{\Pr(Y=1 \mid X=x, \bZ=\bz)\} = \beta_0 + \beta_1 X + \beta_2 Z_1 + \beta_3 Z_2,
\ese
where $\{\beta_0, \beta_1, \beta_2, \beta_3\} = \{1, 1, 1, 1\}$. Covariates $(X, Z_1, Z_2)$ were each generated as $\text{Bernoulli}(0.5)$. Therefore, the number of $u_j=(y,x,z_1,z_2)$ combinations was $2^4=16$. As before, the number of sites $K$ varied from 5 to 50, with site-specific sample sizes $n^{(k)}$ randomly drawn from $\{30, 100, 1000\}$ with equal probability.

We considered the case in which the missingness mechanism was the same across sites under MAR. Subject-level missingness in $X$ was approximately 60\%, introduced according to $\Pr(R_X=1 \mid Y, \bZ) = \text{expit}(-0.1 + 0.1Y + 0.2Z_1 + 0.2Z_2)$. We compared four weighting strategies: (i) oracle (true weights), (ii) site-specific (site-specific estimation), (iii) calibrated (sharing $\wh{\balpha}$ from the two largest sites), and (iv) random (generated from 1/uniform(0.1, 0.9) distribution). The probability of a complete observation was modeled using logistic regression with covariates $(Y, Z_1, Z_2)$. For local estimation, no parameters were shared across sites; for the calibrated weights, only the largest site shared its $\balpha$ estimates. As a benchmark, we also evaluated the case with no missingness.

Most of the bias was observed in the intercept coefficient when bias was expected (Table~\ref{tab:glm-numerical-results}). As expected, the CC estimator exhibited approximately 12\% bias under MAR, with coverage deteriorating from 72\% to 13\% as the number of sites increased from 10 to 50 (Table~\ref{tab:glm-numerical-results}). In contrast, the IPW estimator, using either local or calibrated weights, achieved near-zero bias and nominal coverage probabilities close to the 95\% level when using the corrected sandwich variance estimator. Without variance correction, standard errors were overestimated, resulting in higher nominal coverage probabilities for both the IPW estimators using the site-specific and calibrated weights. The benefits of variance correction, again, were mostly observed for the intercept coefficient than for the coefficients for $(X,Z_1,Z_2)$.

Site-specific and calibrated weights performed similarly under a shared missingness mechanism; the advantages of calibration are expected to emerge under heterogeneous mechanisms across sites. With a binary outcome and binary covariates, there were only 16 unique combinations of $(Y, X, Z_1, Z_2)$ for the outcome model and 16 unique combinations of $(R, Y, Z_1, Z_2)$ for the missingness model. Thus, the much of the information about both models can be recovered at the local level without the need to share information across sites or without the need of large sample sizes. When the standard errors were calculated, the calibrated weights yielded slightly larger estimates than the local weights; reflecting that the calibration provides limited information in this setting. Instead, it only introduces additional model uncertainty through the inclusion of nuisance parameters from other sites in the weight estimation as well as in the calibration coefficients $\btau$.

\begin{table}[h!]
\centering
\resizebox{0.9\linewidth}{!}{
\begin{tabular}[t]{llrrrrrrrrrrrr}
\toprule
\textbf{Estimator} &  & \textbf{Bias} & \textbf{SE} & \textbf{SD} & \textbf{Cov} & \textbf{Bias} & \textbf{SE} & \textbf{SD} & \textbf{Cov} & \textbf{Bias} & \textbf{SE} & \textbf{SD} & \textbf{Cov} \\
\midrule
\addlinespace[0.3em]
\addlinespace
\multicolumn{2}{l}{Parameter: Intercept ($\beta_0$)} & \multicolumn{4}{c}{\textbf{K=10}} & \multicolumn{4}{c}{\textbf{K=30}} & \multicolumn{4}{c}{\textbf{K=50}}\\
\addlinespace
\hspace{1em}Oracle &  & 0.06 & 5.52 & 5.54 & 94.38 & -0.08 & 3.18 & 3.08 & 95.95 & 0.03 & 2.47 & 2.50 & 94.79 \\
\addlinespace
\hspace{1em}CC &  & 12.27 & 8.82 & 8.55 & 72.62 & 11.92 & 5.08 & 4.96 & 34.92 & 12.05 & 3.93 & 3.95 & 13.26 \\
\addlinespace
\hspace{1em}IPW & Oracle & 0.30 & 8.82 & 8.55 & 95.24 & -0.06 & 5.08 & 4.96 & 95.50 & 0.07 & 3.94 & 3.96 & 95.04 \\
 & Site-specific & 0.18 & 8.88 & 7.61 & 97.77 & -0.06 & 5.12 & 4.31 & 97.72 & 0.05 & 3.97 & 3.48 & 97.77 \\
 & Calibrated & 0.18 & 8.88 & 7.61 & 97.77 & -0.06 & 5.12 & 4.31 & 97.72 & 0.05 & 3.97 & 3.48 & 97.77 \\
 & Uniform & 12.68 & 5.33 & 10.44 & 39.93 & 11.99 & 3.06 & 6.18 & 15.64 & 12.08 & 2.37 & 4.80 & 5.72 \\
\addlinespace
 & & \multicolumn{12}{c}{\textit{Robust sandwich estimator}}\\
\addlinespace
 & Site-specific & 0.18 & 7.70 & 7.61 & 95.50 & -0.06 & 4.43 & 4.31 & 95.50 & 0.05 & 3.43 & 3.48 & 95.09 \\
 & Calibrated & 0.17 & 7.97 & 7.62 & 96.10 & -0.07 & 4.53 & 4.33 & 95.85 & 0.05 & 3.50 & 3.48 & 95.55 \\
\addlinespace
\multicolumn{2}{l}{Parameter: $X$ ($\beta_1$)} & \multicolumn{12}{c}{}\\
\addlinespace
\hspace{1em}Oracle &  & 0.32 & 7.46 & 7.48 & 94.48 & 0.05 & 4.30 & 4.41 & 94.28 & -0.04 & 3.33 & 3.30 & 95.14 \\
\addlinespace
\hspace{1em}CC &  & 0.69 & 11.87 & 11.86 & 95.04 & 0.11 & 6.83 & 6.82 & 95.34 & 0.08 & 5.29 & 5.23 & 94.99 \\
\addlinespace
\hspace{1em}IPW & Oracle & 0.70 & 11.87 & 11.87 & 95.09 & 0.12 & 6.83 & 6.83 & 95.34 & 0.08 & 5.29 & 5.23 & 95.14 \\
 & Site-specific & 0.72 & 11.97 & 11.97 & 94.94 & 0.14 & 6.89 & 6.87 & 95.45 & 0.10 & 5.33 & 5.29 & 95.04 \\
 & Calibrated & 0.72 & 11.97 & 11.97 & 94.94 & 0.14 & 6.89 & 6.87 & 95.45 & 0.10 & 5.33 & 5.29 & 95.04 \\
 & Uniform & 0.64 & 7.17 & 14.29 & 67.56 & 0.28 & 4.12 & 8.19 & 68.07 & 0.06 & 3.19 & 6.37 & 66.60 \\
\addlinespace
 & & \multicolumn{12}{c}{\textit{Robust sandwich estimator}}\\
\addlinespace
 & Site-specific & 0.72 & 11.87 & 11.97 & 94.79 & 0.14 & 6.83 & 6.87 & 95.34 & 0.10 & 5.29 & 5.29 & 94.84 \\
 & Calibrated & 0.72 & 11.90 & 11.97 & 94.79 & 0.14 & 6.84 & 6.87 & 95.29 & 0.10 & 5.30 & 5.29 & 94.84 \\
\addlinespace
\multicolumn{2}{l}{Parameter: $Z_1$ ($\beta_2$)} & \multicolumn{12}{c}{}\\
\addlinespace
\hspace{1em}Oracle &  & 0.17 & 7.46 & 7.42 & 95.14 & 0.09 & 4.30 & 4.35 & 94.94 & -0.06 & 3.33 & 3.28 & 95.39 \\
\addlinespace
\hspace{1em}CC &  & -0.29 & 11.73 & 11.66 & 95.65 & -0.31 & 6.75 & 6.87 & 93.72 & -0.44 & 5.23 & 5.12 & 95.19 \\
\addlinespace
\hspace{1em}IPW & Oracle & 0.20 & 11.73 & 11.66 & 95.55 & 0.17 & 6.75 & 6.87 & 93.72 & 0.04 & 5.23 & 5.12 & 95.39 \\
 & Site-specific & 0.18 & 11.83 & 11.75 & 95.70 & 0.17 & 6.81 & 6.93 & 93.72 & 0.04 & 5.28 & 5.17 & 95.14 \\
 & Calibrated & 0.18 & 11.83 & 11.75 & 95.70 & 0.17 & 6.81 & 6.93 & 93.72 & 0.04 & 5.28 & 5.17 & 95.14 \\
 & Uniform & -0.37 & 7.09 & 14.11 & 67.56 & -0.31 & 4.08 & 8.44 & 65.94 & -0.46 & 3.16 & 6.29 & 67.97 \\
\addlinespace
 & & \multicolumn{12}{c}{\textit{Robust sandwich estimator}}\\
\addlinespace
 & Site-specific & 0.18 & 11.73 & 11.75 & 95.45 & 0.17 & 6.75 & 6.93 & 93.57 & 0.04 & 5.23 & 5.17 & 95.09 \\
 & Calibrated & 0.18 & 11.76 & 11.75 & 95.50 & 0.17 & 6.77 & 6.93 & 93.62 & 0.04 & 5.24 & 5.16 & 95.24 \\
\addlinespace
\multicolumn{2}{l}{Parameter: $Z_2$ ($\beta_3$)} & \multicolumn{12}{c}{}\\
\addlinespace
\hspace{1em}Oracle &  & 0.27 & 7.46 & 7.48 & 95.04 & 0.12 & 4.30 & 4.21 & 96.05 & -0.01 & 3.33 & 3.34 & 94.94 \\
\addlinespace
\hspace{1em}CC &  & -0.28 & 11.73 & 11.38 & 95.60 & -0.31 & 6.75 & 6.61 & 95.14 & -0.66 & 5.23 & 5.23 & 94.84 \\
\addlinespace
\hspace{1em}IPW & Oracle & 0.21 & 11.73 & 11.38 & 95.60 & 0.18 & 6.75 & 6.61 & 95.39 & -0.17 & 5.23 & 5.24 & 95.34 \\
 & Site-specific & 0.23 & 11.83 & 11.45 & 95.70 & 0.17 & 6.81 & 6.67 & 95.29 & -0.17 & 5.27 & 5.25 & 95.60 \\
 & Calibrated & 0.23 & 11.83 & 11.45 & 95.70 & 0.17 & 6.81 & 6.67 & 95.29 & -0.17 & 5.27 & 5.25 & 95.60 \\
 & Uniform & -0.23 & 7.09 & 14.00 & 69.08 & -0.28 & 4.08 & 8.22 & 66.80 & -0.68 & 3.15 & 6.48 & 65.38 \\
\addlinespace
 & & \multicolumn{12}{c}{\textit{Robust sandwich estimator}}\\
\addlinespace
 & Site-specific & 0.23 & 11.73 & 11.45 & 95.70 & 0.17 & 6.75 & 6.67 & 95.14 & -0.17 & 5.23 & 5.25 & 95.34 \\
 & Calibrated & 0.23 & 11.76 & 11.44 & 95.70 & 0.17 & 6.77 & 6.67 & 95.14 & -0.17 & 5.24 & 5.25 & 95.39 \\
\addlinespace
\bottomrule
\end{tabular}}
\caption{\textbf{Simulation results under logistic regression model with a missing covariate by number of sites under MAR.} Percent bias (\%), mean estimated standard errors (SE), empirical standard deviations (SD), and 95\% confidence interval coverage (Cov). Results shown are for the intercept coefficient ($\beta_0$) under a shared MAR mechanism across sites. \label{tab:glm-numerical-results}}
\end{table}

\clearpage
\section{Additional data application results}
\label{sec:supp-application}

Data from the University of North Carolina (UNC) at Chapel Hill included 1,829 individuals from 24 sites in the United States of America and United Kingdom (2014-2020). One of the sites included individuals from the affiliated hospitals across the Johns Hopkins University (JHU) network. Information from JHU was therefore removed from the analysis, as it risked double counting individuals already present in the separately obtained from JHU (2018-2025).

\subsection{Study characteristics}

Overall, the total sample size from UNC included 1,494 individuals, of whom 14.4\% experienced 90-day mortality, compared with 18.3\% of the 454 individuals from JHU (Table~\ref{tab:descriptive-unc-hopkins}). The two analytical samples were similar in the proportion of males, individuals with low albumin levels (albumin $<$ 2.7 g/dL), and those with high BUN levels (BUN $\geq$ 14 mg/dL). However, two  differences were present: the proportion of individuals aged at least 50 years was higher for JHU than for UNC (59\% vs. 31\%), and the proportion of missing albumin values was much higher for UNC than for JHU (15\% vs.\ 2\%). 

\begin{table}[h!]
\centering
\begin{tabular}{lcc}
\toprule
 & \textbf{JHU (n=454)} & \textbf{UNC (n=1,494)} \\
\midrule
\textbf{90-day mortality} & 83 (18.3\%) & 215 (14.4\%) \\
\textbf{Sex (Male)} & 344 (75.8\%) & 1046 (70.0\%) \\
\textbf{Albumin (g/dL) $<$ 2.7} & 253 (56.9\%) & 631 (49.9\%) \\
\quad Missing & 10 (2.2\%) & 228 (15.3\%) \\
\textbf{Age (years) $\geq$ 50} & 267 (58.8\%) & 463 (31.0\%) \\
\textbf{BUN (mg/dL) $\geq$ 14} & 180 (39.6\%) & 500 (33.5\%) \\
\bottomrule
\end{tabular}
\caption{Baseline characteristics of the Johns Hopkins University (JHU; 5 sites) and University of North Carolina (UNC; 23 sites). Counts (percentages) are presented within overall group.}
\label{tab:descriptive-unc-hopkins}
\end{table}

Among the JHU analytic sample, the most common subgroup comprised males who were alive at 90 days and had low albumin ($n=126$, 28\%), followed by males alive at 90 days with high albumin ($n=93$, 20\%)(Table~\ref{tab:jhu_unc_counts}). The same two subgroups were also the largest in the UNC analytic sample, accounting for 21\% ($n=314$) and 24\% ($n=364$), respectively. For JHU, the site-specific and calibrated weighted counts closely matched the crude counts, consistent with the low proportion of missing albumin in the analytic sample (2.2\%). In contrast, the weighted counts for UNC were much higher than the crude counts---reflecting the up-weighting needed to compensate for the 15.3\% of individuals with missing albumin---although the site-specific and calibrated weights produced identical estimates to the first decimal place.

\clearpage

\begin{landscape}

\begin{table}[h!]
\centering
\begin{tabular}{llllcccccc}
\toprule
& & & & \multicolumn{3}{c}{\textbf{JHU}} & \multicolumn{3}{c}{\textbf{UNC}} \\
\cmidrule(lr){5-7} \cmidrule(lr){8-10}
\textbf{90-day} & \textbf{Sex} & \textbf{Albumin} & 
& \textbf{Crude} & \textbf{Site-specific} & \textbf{Calibrated} 
& \textbf{Crude} & \textbf{Site-specific} & \textbf{Calibrated} \\
\midrule
Alive & Female & $\geq$2.7 g/dL && 65 & 67.4 & 67.4 & 203 & 240.1 & 240.1 \\
Alive & Female & $<$2.7 g/dL    && 76 & 78.7 & 78.7 & 190 & 224.0 & 224.0 \\
Alive & Male   & $\geq$2.7 g/dL && 93 & 94.7 & 94.7 & 364 & 431.6 & 431.6 \\
Alive & Male   & $<$2.7 g/dL    && 126 & 127.8 & 127.8 & 314 & 374.9 & 374.9 \\
\addlinespace
Dead  & Female & $\geq$2.7 g/dL && 17 & 17.4 & 17.4 & 30 & 34.6 & 34.6 \\
Dead  & Female & $<$2.7 g/dL    && 23 & 23.6 & 23.6 & 49 & 56.1 & 56.1 \\
Dead  & Male   & $\geq$2.7 g/dL && 16 & 16.2 & 16.2 & 37 & 42.4 & 42.4 \\
Dead  & Male   & $<$2.7 g/dL    && 28 & 28.3 & 28.3 & 78 & 89.3 & 89.3 \\
\bottomrule
\end{tabular}
\vspace{1ex}
\caption{\textbf{Observed and weighted combination counts by outcome, sex, and albumin 
level.} Crude counts reflect complete cases only from Johns Hopkins University (JHU; 5 sites) and University of North Carolina (UNC; 23 sites). Site-specific and calibrated weighted counts up-weight complete cases to represent the full sample under inverse probability weighting.}
\label{tab:jhu_unc_counts}
\end{table}

\end{landscape}

\subsection{Regression Results}

Table~\ref{tab:unc_jhu_results} presents log odds ratio estimates for 90-day mortality using only JHU or UNC data, and from their joint federated analysis, under the CC and IPW estimators. Under the CC estimator, the estimated effect 
of low albumin was 0.20 (SE $= 0.248$) for JHU and 0.76 (SE $= 0.163$) for UNC---the latter being larger and more efficient. The combined estimate fell between the two at 0.60 (SE $= 0.136$), pulled closer to the UNC estimate given its larger sample size. Sex was not significantly associated with 90-day mortality in either sample or in the combined analysis.

For the JHU analysis, all three estimators yielded nearly identical results. This was expected given the low percentage of missing data, which produced crude and weighted counts that were highly similar. For the UNC analysis, the CC and IPW estimates were slightly more different, reflecting the 
higher missingness rate in that sample. Among the IPW approaches, the log-odds estimates were identical, though the standard errors for the intercept differed slightly between the site-specific and calibrated approaches. Under the combined federated analysis, point estimates differed slightly between the CC and IPW estimators, while the two IPW approaches 
agreed closely, with only minor differences in standard errors.

Across all analyses, the agreement between CC and IPW is expected: because the outcome was excluded from the missingness model on temporality grounds, both estimators are consistent in this setting, and their close agreement provides empirical support for the validity of the CC estimator. The near-identical results from the site-specific and calibrated IPW estimator are similarly expected. With a binary outcome and two binary covariates, only eight unique combinations exist, and the local data at each site may have been enough to estimate the weighting model without borrowing strength across sites. The minor differences in standard errors between the two IPW approaches reflect the adjustment for additional nuisance parameters under calibrated weight estimation approach.

\begin{table}[h!]
\centering
\begin{tabular}{lcccccc}
\toprule
 
& \multicolumn{2}{c}{\textbf{JHU}} 
& \multicolumn{2}{c}{\textbf{UNC}}
& \multicolumn{2}{c}{\textbf{Combined}} \\
\midrule
& \underline{Estimate} & \underline{SE} 
& \underline{Estimate} & \underline{SE}
& \underline{Estimate} & \underline{SE} \\
\multicolumn{7}{l}{\textbf{Complete Case}} \\
\hspace{1em} Intercept & -1.3745 & 0.2304 & -2.0345 & 0.1653 & -1.8428 & 0.1342 \\
\hspace{1em} Albumin $<2.7$ g/dL & 0.2042 & 0.2479 & 0.7560 & 0.1630 & 0.6041 & 0.1357 \\
\hspace{1em} Sex (male) & -0.3534 & 0.2440 & -0.1639 & 0.1611 & -0.2313 & 0.1339 \\

\addlinespace
\multicolumn{7}{l}{\textbf{IPW (site-specific)}} \\
\hspace{1em} Intercept & -1.3881 & 0.2304 & -2.0581 & 0.1650 & -1.8808 & 0.1344 \\
\hspace{1em} Albumin $<2.7$ g/dL & 0.2087 & 0.2479 & 0.7498 & 0.1632 & 0.6152 & 0.1361 \\
\hspace{1em} Sex (male) & -0.3474 & 0.2438 & -0.1754 & 0.1605 & -0.2340 & 0.1337 \\
\addlinespace
\multicolumn{7}{l}{\textbf{IPW (calibrated)}} \\
\hspace{1em} Intercept & -1.3881 & 0.2304 & -2.0581 & 0.1649 & -1.8808 & 0.1344 \\
\hspace{1em} Albumin $<2.7$ g/dL & 0.2087 & 0.2479 & 0.7498 & 0.1632 & 0.6152 & 0.1364 \\
\hspace{1em} Sex (male) & -0.3474 & 0.2438 & -0.1754 & 0.1605 & -0.2340 & 0.1340 \\
\addlinespace
\bottomrule
\end{tabular}
\caption{\textbf{Log odds ratio estimates for 90-day mortality.} Estimates are log odds ratios; SE denotes the standard error obtained via the sandwich estimator, adjusted for nuisance parameter estimation uncertainty. Results are presented for the Johns Hopkins University (JHU; 5 sites) and University of 
North Carolina (UNC; 23 sites) samples separately, and for their joint 
federated analysis.}
\label{tab:unc_jhu_results}
\end{table}

\section{Step-by-step example}
\label{sec:step-by-step}

We demonstrate the federated estimation procedure using the pleural infection data from Section~\ref{sec:supp-application}. This section provides a narrative walkthrough; an accompanying R-Markdown tutorial 
with code is available at 
\url{https://github.com/jesusepfvazquez/Federated-Learning-Missing-Data}. Each subsection shows what each site transmits and how the coordinating site computes the final estimates. 

\subsection{Complete case estimator}
\label{sec:example-cc}

Under the CC estimator (Algorithm~\ref{alg:fl-cc-glm}), each site transmits 
the unique outcome-covariate combinations present in its data along with their 
corresponding counts in the single round of communication. In the PI application, JHU and UNC each transmit eight 
such combinations, as illustrated in Table~\ref{tab:jhu_unc_counts}. 
Specifically, each site transmits:

\begin{verbatim}
# JHU transmits: 8 combinations x 3 columns
summary_jhu <- read.csv("results_jhu_counts.csv")
#   dead90  albumin_c2  sex_c2  n
#   0       0           0       65
#   0       0           1       93
#   ...
# UNC transmits: 8 combinations x 3 columns
summary_unc <- read.csv("results_unc_counts.csv")
#   dead90  albumin_c2  sex_c2  n
#   0       0           0       203
#   0       0           1       364
#   ...
\end{verbatim}
The coordinating site pools the transmitted summaries, expands the counts 
into individual-level records, and fits the logistic regression model:
\begin{verbatim}
# Coordinating site: combine transmitted summaries
summary_combined <- rbind(summary_jhu, summary_unc)

# Expand counts to individual counts data
data_expanded <- summary_combined |> tidyr::uncount(weights = n)

# Fit logistic regression
fit_cc <- glm(dead90 ~ albumin_c2 + sex_c2, data = data_expanded, family = "binomial")

# Point estimates
coef(fit_cc)
# (Intercept)  albumin_c2      sex_c2 
#     -1.8428      0.6041     -0.2313 

# Variance via sandwich estimator (computed from same summaries)
sqrt(diag(sandwich::vcovHC(fit_cc, type = "HC0")))
# (Intercept)  albumin_c2      sex_c2 
#      0.1342      0.1357      0.1339
\end{verbatim}
This is a \emph{one-shot} algorithm: all estimates are obtained from a single 
round of transmitted counts. The sandwich variance estimator can be computed via \texttt{vcovHC()} from the \texttt{sandwich} R package, 
though the $\wh\bA_\text{CC}$ and $\wh\bB_\text{CC}$ components may also 
be constructed from the transmitted counts.

\subsection{Inverse probability weighting: site-specific weighting}
\label{sec:example-ipw-site-specific}
The site-specific IPW estimator (Algorithm~\ref{alg:fl-ipw-glm}) requires 
three rounds of communication: weight estimation, outcome model estimation, 
and variance estimation. Each round is described in turn below.

\subsubsection{Step 1: Weight estimation}
In the first round, each site fits a missingness model using its local data 
and computes the probability of a complete observation for each individual. 
Weighted counts are then constructed by summing the inverse probabilities 
over each outcome-covariate combination. Note that the missingness model for UNC includes pairwise interactions, whereas the model for JHU includes only main effects.

\begin{verbatim}
# Estimate missingness model: JHU (main effects only)
fit_miss_jhu <- glm(R ~ age + purulence + sex + bun, 
                    data = data_jhu, family = "binomial")

# Compute weights and aggregate by combination
data_jhu$w <- 1 / fitted(fit_miss_jhu)
summary_jhu_weighted <- data_jhu |>
  filter(R == 1) |>
  group_by(dead90, albumin_c2, sex_c2) |>
  summarise(n_weighted = sum(w))

# Estimate missingness model: UNC (pairwise interactions)
fit_miss_unc <- glm(R ~ (age + purulence + sex + bun)^2, 
                    data = data_unc, family = "binomial")

# Compute weights and aggregate by combination
data_unc$w <- 1 / fitted(fit_miss_unc)
summary_unc_weighted <- data_unc |>
  filter(R == 1) |>
  group_by(dead90, albumin_c2, sex_c2) |>
  summarise(n_weighted = sum(w))
\end{verbatim}

Each site then transmits \texttt{summary\_jhu\_weighted} and 
\texttt{summary\_unc\_weighted} to the coordinating site.

\subsubsection{Step 2: Outcome model estimation}
Upon receiving the weighted counts from each site 
(Table~\ref{tab:jhu_unc_counts}, ``Site-specific'' columns), the coordinating 
center pools the transmitted summaries and fits the outcome model:

\begin{verbatim}
# Coordinating site receives weighted summaries from each site
summary_jhu_weighted <- read.csv("results_jhu_counts.csv")
#   dead90  albumin_c2  sex_c2  n_weighted
#   0       0           0       67.4
#   0       0           1       94.7
#   ...
summary_unc_weighted <- read.csv("results_unc_counts.csv")
#   dead90  albumin_c2  sex_c2  n_weighted
#   0       0           0       240.1
#   0       0           1       431.6
#   ...

# Pool summaries and fit weighted logistic regression
summary_combined <- rbind(summary_jhu_weighted, summary_unc_weighted)
fit_ipw <- glm(dead90 ~ albumin_c2 + sex_c2, 
               data = summary_combined, 
               weights = n_weighted,
               family = "binomial")

coef(fit_ipw)
# (Intercept)  albumin_c2      sex_c2 
#     -1.8808      0.6152     -0.2340 
\end{verbatim}

The pooled estimate $\wh\btheta$ is then transmitted back to each site for 
variance estimation. 

\subsubsection{Step 3: Variance estimation}
Each site computes its local $\wh\bA^{(k)}$ and $\wh\bB^{(k)}$ matrices 
evaluated at the coordinating site's $\wh\btheta$. The dimensions of these 
matrices differ across sites, reflecting the different models of the local 
missingness models. The JHU main-effects model has 6 parameters for 
$\balpha_\text{JHU}$, yielding $9 \times 9$ matrices ($3$ for $\btheta$ and 
$6$ for $\balpha_\text{JHU}$). The UNC interaction model has 16 parameters 
for $\balpha_\text{UNC}$, yielding $19 \times 19$ matrices.

\begin{verbatim}
# JHU transmits: 9 x 9 matrices (3 for theta + 6 for alpha_jhu)
A_jhu <- as.matrix(read.csv("results_jhu_A_mat.csv"))
B_jhu <- as.matrix(read.csv("results_jhu_B_mat.csv"))

# UNC transmits: 19 x 19 matrices (3 for theta + 16 for alpha_unc)
A_unc <- as.matrix(read.csv("results_unc_A_mat.csv"))
B_unc <- as.matrix(read.csv("results_unc_B_mat.csv"))
\end{verbatim}

The coordinating site assembles the stacked matrices following the block 
structure described in Algorithm~\ref{alg:fl-ipw-glm}. The $\btheta$-blocks 
are summed across sites, while the $\balpha$-blocks remain site-specific:

\begin{verbatim}
# Initialize combined matrix: dim = 3 + 16 + 6 = 25
A_combined <- matrix(0, nrow = 25, ncol = 25)

# Place UNC block (theta + alpha_unc)
A_combined[1:19, 1:19] <- A_unc

# Add JHU theta-theta block to existing UNC theta-theta block
A_combined[1:3, 1:3] <- A_combined[1:3, 1:3] + A_jhu[1:3, 1:3]

# Place JHU cross-term (theta x alpha_jhu) and alpha-alpha blocks
A_combined[1:3, 20:25] <- A_jhu[1:3, 4:9]
A_combined[20:25, 20:25] <- A_jhu[4:9, 4:9]
\end{verbatim}

The same assembly applies to $\wh\bB_{\text{Stacked}}$. However, the off-diagonal blocks $\wh\bB^{(k)}_{\balpha_k\btheta}$ are nonzero and must be placed:

\begin{verbatim}
# B matrix assembly (same structure, plus symmetric off-diagonal blocks)
B_combined[1:19, 1:19] <- B_unc
B_combined[1:3, 1:3] <- B_combined[1:3, 1:3] + B_jhu[1:3, 1:3]
B_combined[1:3, 20:25] <- B_jhu[1:3, 4:9]
B_combined[20:25, 1:3] <- t(B_combined[1:3, 20:25])
B_combined[20:25, 20:25] <- B_jhu[4:9, 4:9]
\end{verbatim}

The sandwich variance is then computed and standard errors for $\btheta$ 
extracted from the first three diagonal elements:

\begin{verbatim}
# Sandwich variance estimator
A_inv <- solve(A_combined)
cov_combined <- A_inv %*% B_combined %*% t(A_inv)

# Standard errors for theta (first 3 parameters)
sqrt(diag(cov_combined)[1:3])
# (Intercept)  albumin_c2      sex_c2 
#      0.1344      0.1361      0.1337
\end{verbatim}

\clearpage
\defbibfilter{myfilter}{%
    keyword=bibonly      
    or keyword=mainbib
}

\printbibliography